\documentclass{llncs}
\usepackage[table]{xcolor}
\usepackage{todonotes}
\usepackage[linesnumbered,ruled]{algorithm2e}
\usepackage{array}
\usepackage{booktabs}
\usepackage{amsmath}
\usepackage{xcolor}
\usepackage[square,sort,comma,numbers]{natbib}
\usepackage{caption}
\usepackage{subcaption}
\usepackage{amssymb}
\usepackage{amsmath}
\usepackage{cleveref}

\usepackage{breakcites}

\begin{document}
%colors for prism and mst
\definecolor{colPrism}{HTML}{E41A1C}
\definecolor{colMst}{HTML}{377EB8}
\definecolor{colInit}{HTML}{4DAF4A}
\definecolor{fgcolor}{rgb}{0.345, 0.345, 0.345}

\newenvironment{knitrout}{}{} % an empty environment to be redefined in TeX

\title{Node Overlap Removal by Growing a Tree}
\titlerunning{Hamiltonian Mechanics} 
\author{Lev Nachmanson\inst{1} \and Arlind Nocaj\inst{2}\and Sergey Bereg\inst{3}
\and Leishi Zhang\inst{4} \and Alexander Holroyd\inst{1}}\institute{  
$^1$Microsoft Research, Redmond, USA\\    
$^2$University of Konstanz, Konstanz, Germany\\    
$^3$The University of Texas at Dallas, Richardson, USA\\    
$^4$Middlesex University, London,UK\\    
\email{levnach@microsoft.com, arlind.nocaj@uni-konstanz.de, besp@utdallas.edu, L.X.Zhang@mdx.ac.uk, holroyd@microsoft.com} 
}
\maketitle
\begin{abstract} {Node overlap removal is a necessary step in many scenarios
including laying out a graph, or visualizing a tag cloud. Our
contribution is a new overlap removal algorithm that iteratively
builds a Minimum Spanning Tree on a Delaunay triangulation of the
node centers and removes the node overlaps by "growing" the
tree. The algorithm is simple to implement yet produces high 
quality layouts. According to our experiments it runs several 
times faster than the current state-of-the-art methods.}
\end{abstract}

% Keywords: add 3 to 10 keywords
%\keyword{overlap; removal; layout; algorithm; spanning tree}

% The fields PACS, MSC, and JEL may be left empty or commented out if not applicable
%\PACS{}
%\MSC{}
%\JEL{}

% If this is an expanded version of a conference paper, please cite it here: enter the full citation of your conference paper, and add $^\dagger$ in the end of the title of this article.
%\conference{}

%%%%%%%%%%%%%%%%%%%%%%%%%%%%%%%%%%%%%%%%%%

% \section{Introduction} A configuration with overlapped nodes often appears after applying a graph layout algorithm~\citep{Tamassia:2007:HGD:1202383}, due to the fact that many algorithms treat nodes as points that do not have a size and shape. To remove the overlaps a specialized algorithm is usually applied. Our contribution is such an algorithm that we call Growing Tree, or GTree further on. The basic idea of GTree is to first capture most of the overlap and the local structure with a specific spanning tree on top of a proximity graph, and then resolve the overlap by letting the tree "grow". The algorithm is simple, yet, as our experiments show, usually runs faster and produces outputs of comparable quality with PRISM~\citep{DBLP:journals/jgaa/GansnerH10}, which is widely used for this purpose and considered to be the state-of-the-art. To compare GTree with PRISM we implemented GTree in the open source graph visualization software \href{http://www.graphviz.org/}{Graphviz}, where PRISM
%  is the default overlap removal algorithm. From the other side, GTree is the default in \href{https://github.com/Microsoft/automatic-graph-layout}{MSAGL}.
\section{Introduction} 
Removing node overlap after laying out a graph is a common task in network visualization. Most graph layout algorithms \citep{Tamassia:2007:HGD:1202383} consider nodes as points that do not occupy any geometrical space. In practice, nodes often have shapes, labels, and so on. These shapes and labels may overlap in the visualization and affect the visual readability. To remove such overlaps a specialized algorithm is usually applied. 

The main contribution of this paper is a new node overlap removal algorithm that we call Growing Tree, or GTree further on. The basic idea is to first capture most of the overlap and the local structure with a specific spanning tree on top of a proximity graph, and then resolve the overlap by letting the tree "grow". 

We compare GTree with PRISM~\citep{DBLP:journals/jgaa/GansnerH10},
which is widely used for the same purpose. Needing more area than
PRISM, our method preserves the original layout well and is up to
eight times faster than PRISM. To compare the two algorithms we
implemented GTree in the open source graph visualization software 
%\href{\tt http://www.graphviz.org}, %
Graphviz~\footnote{\tt http://www.graphviz.org/},
where PRISM is the default overlap removal algorithm. On the other
side, GTree is the default in 
%\href{https://github.com/Microsoft/automatic-graph-layout}
MSAGL\footnote{https://github.com/Microsoft/automatic-graph-layout},
where we also have an implementation of PRISM. We ran comparisons by
using both tools.
%%%%%%%%%%%%%%%%%%%%%%%%%%%%%%%%%%%%%%%%%%

\section{Related Work}
There is vast research on node overlap removal. Some methods,
including hierarchical layouts~\citep{DBLP:conf/acsc/FriedrichS04}, incorporate the overlap removal with the layout step. Likewise,
force-directed methods~\citep{fr-gdfdp-91} have been extended to take
the node sizes into account~\citep{DBLP:journals/vlc/LinYC09,DBLP:conf/apvis/LiEN05,DBLP:conf/gd/WangM95},
but it is difficult to guarantee overlap-free layouts without
increasing the repulsive forces extensively.
Dwyer et al.~\cite{DBLP:journals/tvcg/DwyerKM06a} show how to avoid
node overlaps with Stress
Majorization~\citep{DBLP:conf/gd/GansnerKN04}. The method can remove
node overlaps during the layout step, but it needs an initial state
that is overlap free; sometimes such a state is not given.

Another approach, which we also choose, is to use a post-processing step.
In Cluster
Busting~\citep{DBLP:journals/jgaa/LyonsMR98,DBLP:conf/gd/GansnerN98}
the nodes are iteratively moved towards the centers of their Voronoi
cells. The process has the disadvantage of distributing the nodes uniformly in a given bounding box.

Imamichi et al.~\cite{DBLP:conf/gd/ImamichiAGHN08} approximate the node shapes by circles and minimize a function penalizing the circle overlaps.

Starting from the center of a node,
RWorldle~\citep{DBLP:journals/cgf/StrobeltSSKD12} removes the overlaps by discovering the free space around a node by using a spiral curve and then utilizing this space. The approach requires a large number of intersection queries that are time consuming. This idea is extended by Strobelt et al.~\cite{strobelt2012rolled} to discover available space by scanning the plane with a line or a circle.

Another set of node overlap removal algorithms focus on the idea of defining pairwise node
constraints and translating the nodes to satisfy the
constraints~\citep{DBLP:journals/vlc/MisueELS95,
  DBLP:journals/scjapan/HayashiIMF02,
  DBLP:journals/constraints/MarriottSTH03,DBLP:conf/acsc/HuangL03}. These
methods consider horizontal and vertical problems separately, which
often leads to a distorted aspect
ratio~\citep{DBLP:journals/jgaa/GansnerH10}. A Force-transfer-algorithm is introduced by Huang et al.~\cite{huang2007new};
horizontal and vertical scans of overlapped nodes create forces moving
nodes vertically and horizontally; the algorithm takes $\mathcal{O}(n^2)$
steps, where $n$ is the number of the nodes.  
Gomez et al.~\cite{gomez2013mixed} develop Mixed Integer Optimization for Layout
Arrangement to remove overlaps in a set of rectangles. The paper discusses the quality of the layout, which seems to be high, but not the effectiveness of the method, which relies on a mixed integer problem solver. Dwyer et al.~\cite{dwyer2006fast} reduce the overlap removal to a quadratic
problem and solve it efficiently in $\mathcal{O}(n\log n)$ steps. According
to Gansner and Hu~\cite{DBLP:journals/jgaa/GansnerH10}, the quality and the speed of
the method of Dwyer et al.~\cite{dwyer2006fast} is very similar to the ones of PRISM.

The ProjSnippet method~\citep{gomez2013similarity} generates good quality
layouts. The method requires $\mathcal{O}(n^2)$ amount of memory, at least if applied directly as described in the paper, and the usage of a nonlinear problem solver.

In PRISM~\citep{DBLP:journals/jgaa/GansnerH10,DBLP:journals/corr/abs-0911-0626}, a Delaunay triangulation on the node centers is used as the starting point of an iterative step. Then a stress model for node overlap removal is built on the edges of the triangulation and the stress function of the model is minimized.  GTree also starts with building this Delaunay triangulation, but then the algorithms diverge.

\section{GTree Algorithm}\label{sec:gtree}

An input to GTree is a set of nodes $V$, where each node $i \in V$ is
represented by an axis-aligned rectangle $B_i$ with the center
$p_i$. We assume that for different $i,j \in V$ the centers $p_i, p_j$
are different too. If this is not the case, we randomly shift the
nodes by tiny offsets. We denote by $D$ a Delaunay triangulation of
the set $\{p_i: i \in V\}$, and let $E$ be the set of edges of $D$.

On a high level, our method proceeds as follows. First we
calculate the triangulation $D$, then we define a cost function on $E$
and build a minimum cost spanning tree on $D$ for this cost
function. Finally, we let the tree ``grow''. The steps are repeated
until there are no more overlaps. The last several steps are slightly
modified. Now we explain the algorithm in more detail.

\begin{figure}[hbt]
\begin{subfigure}[b]{0.5\textwidth}     
\begin{tikzpicture}
\usetikzlibrary{arrows}
\usetikzlibrary{calc}
\usetikzlibrary{shapes}
\tikzstyle{every node}=[thick, draw=black, rectangle, minimum width=32pt, minimum height=18pt,
    align=center]
\node (a) at (0pt, -3pt){};
\node[draw=none,minimum width=0pt,minimum height=0pt, anchor=north east,yshift=2pt] at (a) {$p_i$};
\node (b) at (13pt,10pt) {};
\node[draw=none,minimum width=0pt,minimum height=0pt, anchor=west,yshift=2pt] at (b) {$p_j$};
\draw (a.center)--(b.center);
%\draw (a.center)--(b.center) node[above] {$\tau_{r}$};
\path[every node/.style={anchor=south,auto=false}] (a.center) edge node{$s$} (b.center);
\fill (a.center) circle [radius=1pt];
\fill (b.center) circle [radius=1pt];
\node[below=-6pt] (a) at (60pt,0pt) {};
\node[below=-6pt] (b) at (78pt,18pt) {};
\draw (a.center)--(b.center);
%\draw (a.center)--(b.center) node[above] {$\tau_{r}$};
\path[every node/.style={anchor=south,auto=false}] (a.center) edge node{$d$} (b.center);
\fill (a.center) circle [radius=1pt];
\fill (b.center) circle [radius=1pt];
\node[draw=none,anchor=west] (textA) at(90pt,0pt) {
$\begin{aligned}
d &=t_{ij}s\\
     c_{ij} &=s-d\\
  \end{aligned}
$};

%caption of subfigure
\node[ draw=none] (captionA) at (30pt,-22pt) {overlapping nodes};
%\node[right=150pt, below=0pt] (a) at (20pt,1pt) {Beta};
%\draw (current bounding box.north east) rectangle (current bounding box.south west);
\end{tikzpicture}
\end{subfigure} 
\begin{subfigure}[b]{0.3\textwidth}%
\begin{tikzpicture}
\usetikzlibrary{arrows}
\usetikzlibrary{shapes}
\tikzstyle{every node}=[thick, draw=black, rectangle, minimum width=32pt, minimum height=18pt,
    align=center]
    
\begin{scope}[shift={(0pt,0)}]  
\node[below=0pt] (a) {$B_i$};
\node[below=0pt] (b) at (25pt,25pt) {$B_j$};
\coordinate (p1) at ([xshift=+3pt, yshift=-6.5pt] b.south west);
\coordinate (p2) at ([xshift=+3pt] b.south west);
\draw [thick] (p1) -- (p2);
%\path[draw=none, anchor=south east,auto=false] (p1) edge node{$dist(B_i,B_j)$} (p2);
\node[draw=none, anchor=west] at (a.north east) {$dist(B_i,B_j)$};
\fill (p1) circle [radius=1pt];
\fill (p2) circle [radius=1pt];
\end{scope}

\node[draw=none,anchor=west] (textB) at(70pt,-12pt) {
$\begin{aligned}
c_{ij} &=\; dist(B_i,B_j)
  \end{aligned}
$};

\node[ draw=none] (captionA) at (25pt,-27pt) {non overlapping nodes};

%\node[right=150pt, below=0pt] (a) at (20pt,1pt) {Beta};
%\draw (current bounding box.north east) rectangle (current bounding box.south west);
\end{tikzpicture}%
\end{subfigure}
\caption{Cost function $c_{ij}$ for edges of the Delaunay
  triangulation. For overlapping nodes $- c_{ij}$ is equal to the
  minimal distance that is necessary to shift the boxes along the edge
  direction so they touch each other.}
\label{fig:costFunction}
\end{figure}
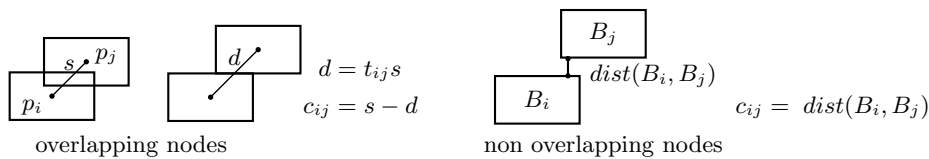

We define the cost function $c$ on $E$ in such a way that the
larger the overlap on an edge becomes, the smaller the cost of this
edge comes to be. Let $(i,j)\in E$. If the rectangles $B_i$ and $B_j$
do not overlap then $c(i,j) = dist(B_i,B_j)$, that is the
distance between $B_i$ and $B_j$. Otherwise, for a real number $t$ let
us denote by $B_j(t)$ a rectangle with the same dimensions as $B_j$
and with the same orientation, but with the center at $p_i + t(p_j - p_i)$. We find $t_{ij} > 1$ such that the
rectangles $B_i$ and $B_j(t_{ij})$ touch each other. Let $s = \|p_j -
p_i\|$, where $\|\|$ denotes the Euclidean norm. We set $c(i,j) =
-(t_{ij}-1)s$. See Figure~\ref{fig:costFunction} for an
illustration. %%We set $t_{ij} = 1$ when $B_i$ and $B_j$ do not overlap.

Having the cost function ready, we compute a minimum spanning tree $T$ on $D$. 
Remember that it is a tree with the set of vertices $V$ for which the cost,
%\begin{align}
\mbox{$\sum_{e \in E'}{c(e)}$}, is minimal, where $E'$ is the set of edges of the tree. %The cost is negative on the edges connecting overlapping nodes and is not negative on the rest of the edges. Therefore, the former are most probably included into $T$.
We use Prim's algorithm to find $T$.

The algorithm proceeds by growing $T$, similar to the
growth of a tree in nature. It starts from the root of $T$. For each child of the root overlapping with the root, it extends the edge connecting the root and the child to remove the overlap. To achieve this, it keeps the root fixed but translates the sub-tree of the child. The edges between the root and other children remain unchanged. The algorithm recursively processes the children of the root in the same manner. This process is described in Algorithm~\ref{algo:growTree}.

\begin{algorithm}
\SetAlgoVlined
\DontPrintSemicolon
\SetKwProg{myalg}{function}{}{}
\SetKwFunction{grow}{GrowAtNode}

\KwIn{Current center positions $p$ and root $r$ }\KwOut {New center positions $p'$}
%\KwResult{how to write algorithm with \LaTeX2e }
$p'_r = p_r$\;
\grow$(r)$\;

\myalg{\grow(i)}{
\ForEach{$j\in \text{Children}(i)$}{
  $p'_j = p'_i + t_{ij}(p_j-p_i)$\;\label{alg:t}
\grow(j);
}
}
\caption[Growing the tree]{Growing $T$}
\label{algo:growTree}
\end{algorithm}

\begin{figure}[b]
        \centering
        \begin{subfigure}[b]{0.3\textwidth}
                \includegraphics[width=\textwidth]{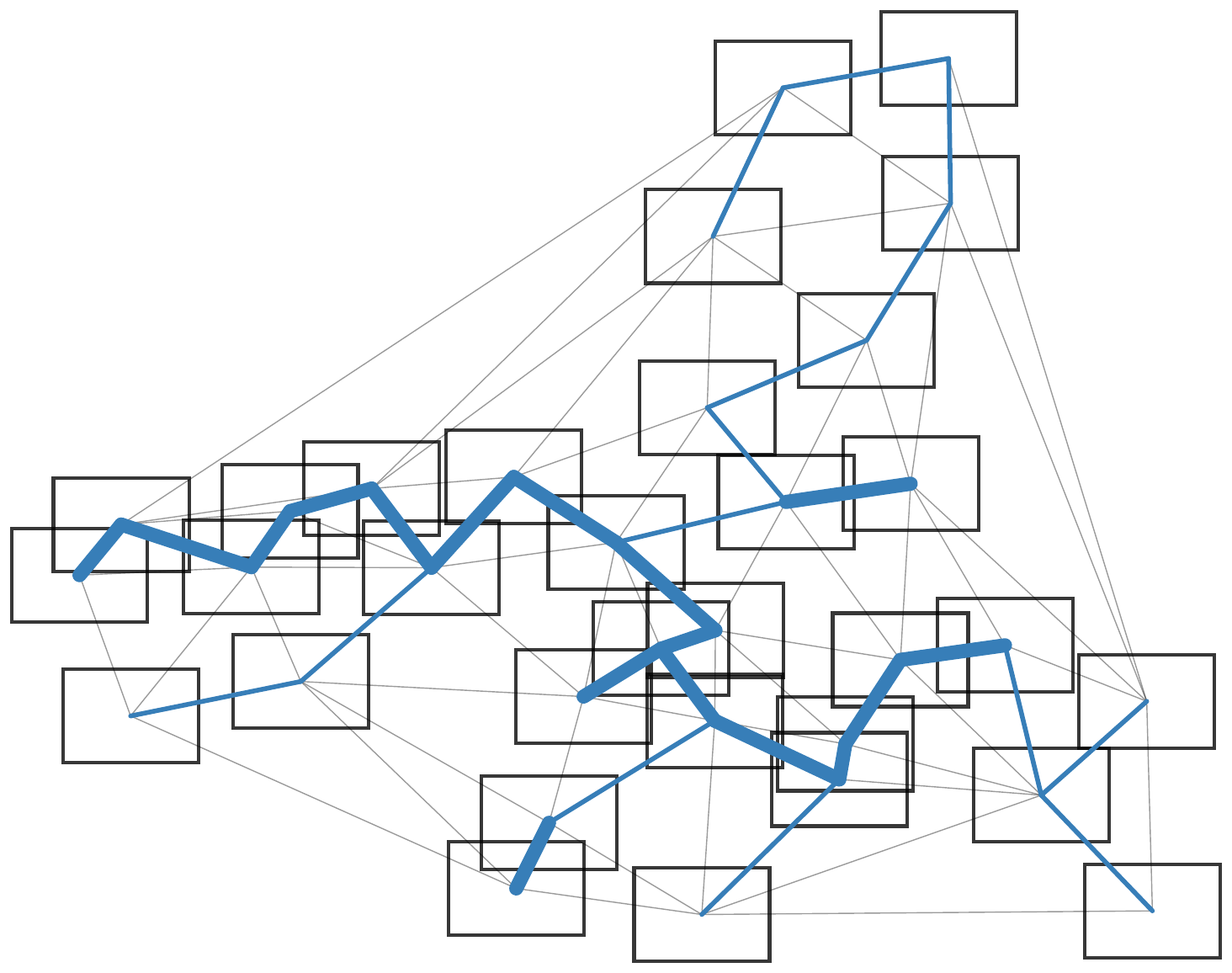}
                \caption{iteration 1}
        \end{subfigure}%
        ~ %add desired spacing between images, e. g. ~, \quad, \qquad etc.
        \begin{subfigure}[b]{0.3\textwidth}
                \includegraphics[width=\textwidth]{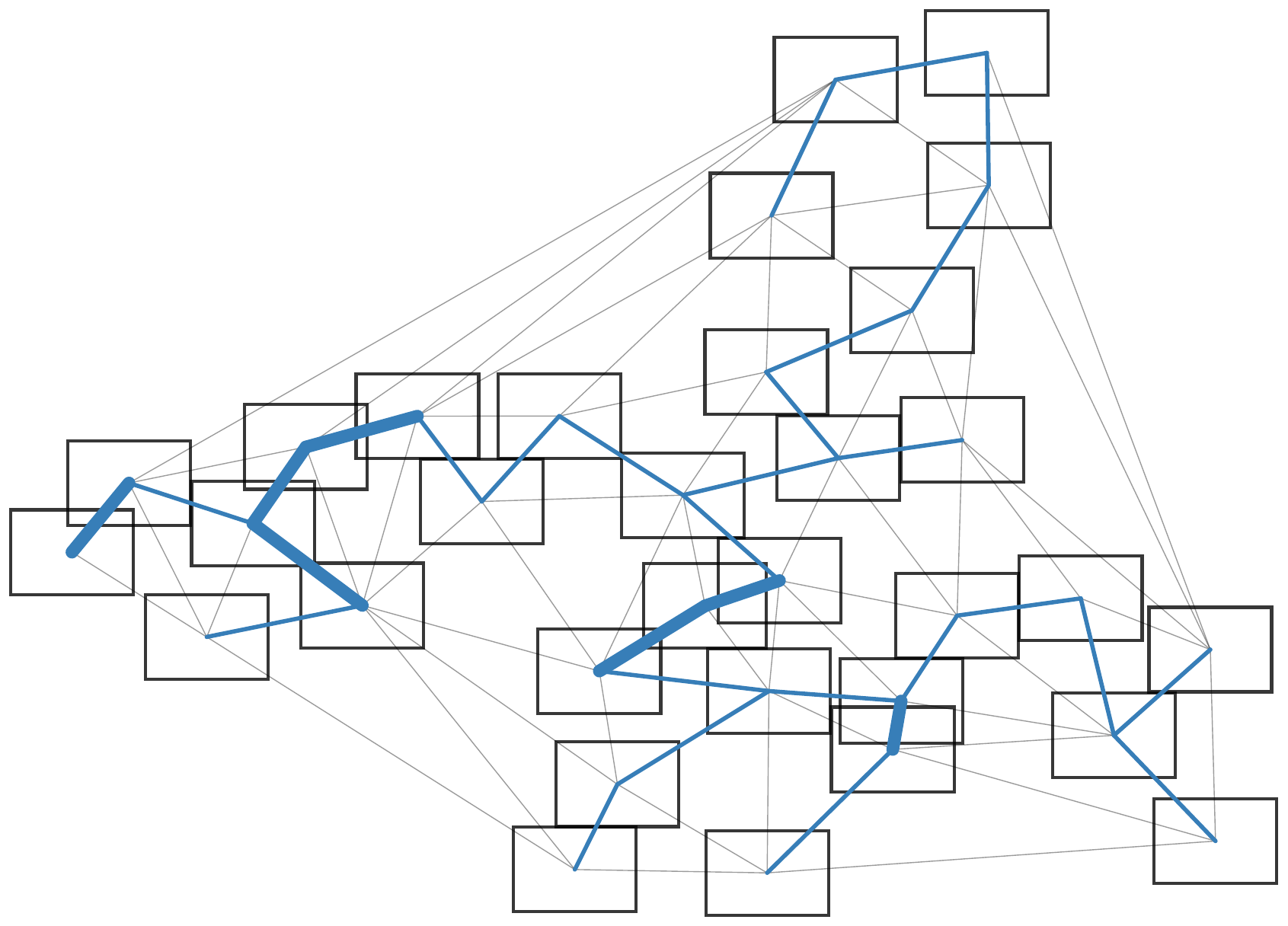}
                \caption{iteration 2}
        \end{subfigure}%
        ~ %add desired spacing between images, e. g. ~, \quad, \qquad etc.
        \begin{subfigure}[b]{0.3\textwidth}
                \includegraphics[width=\textwidth]{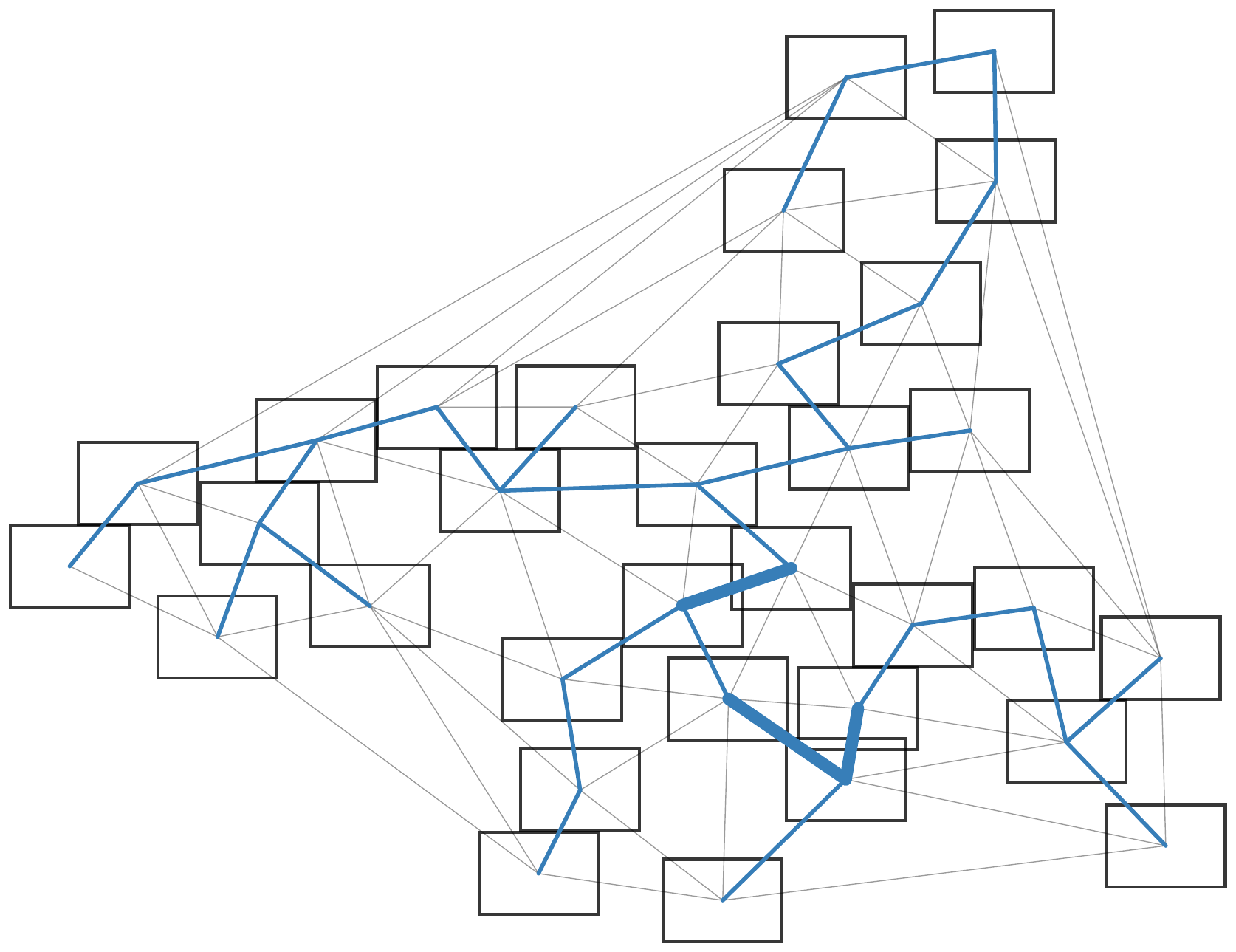}
                \caption{iteration 3}
        \end{subfigure}
        
        \begin{subfigure}[b]{0.3\textwidth}
                        \includegraphics[width=\textwidth]{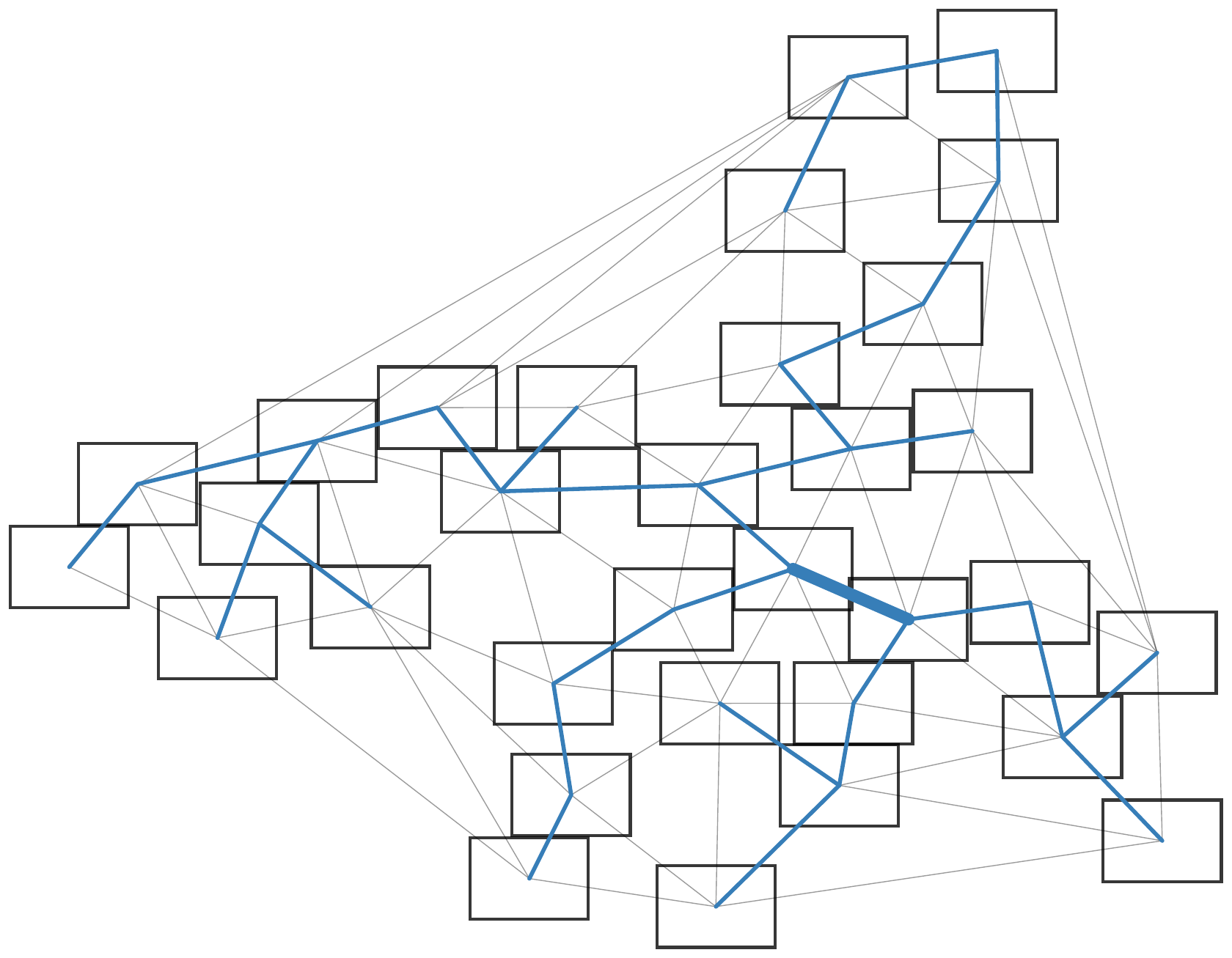}
                        \caption{iteration 4}
		\end{subfigure}%
		\begin{subfigure}[b]{0.3\textwidth}
                                \includegraphics[width=\textwidth]{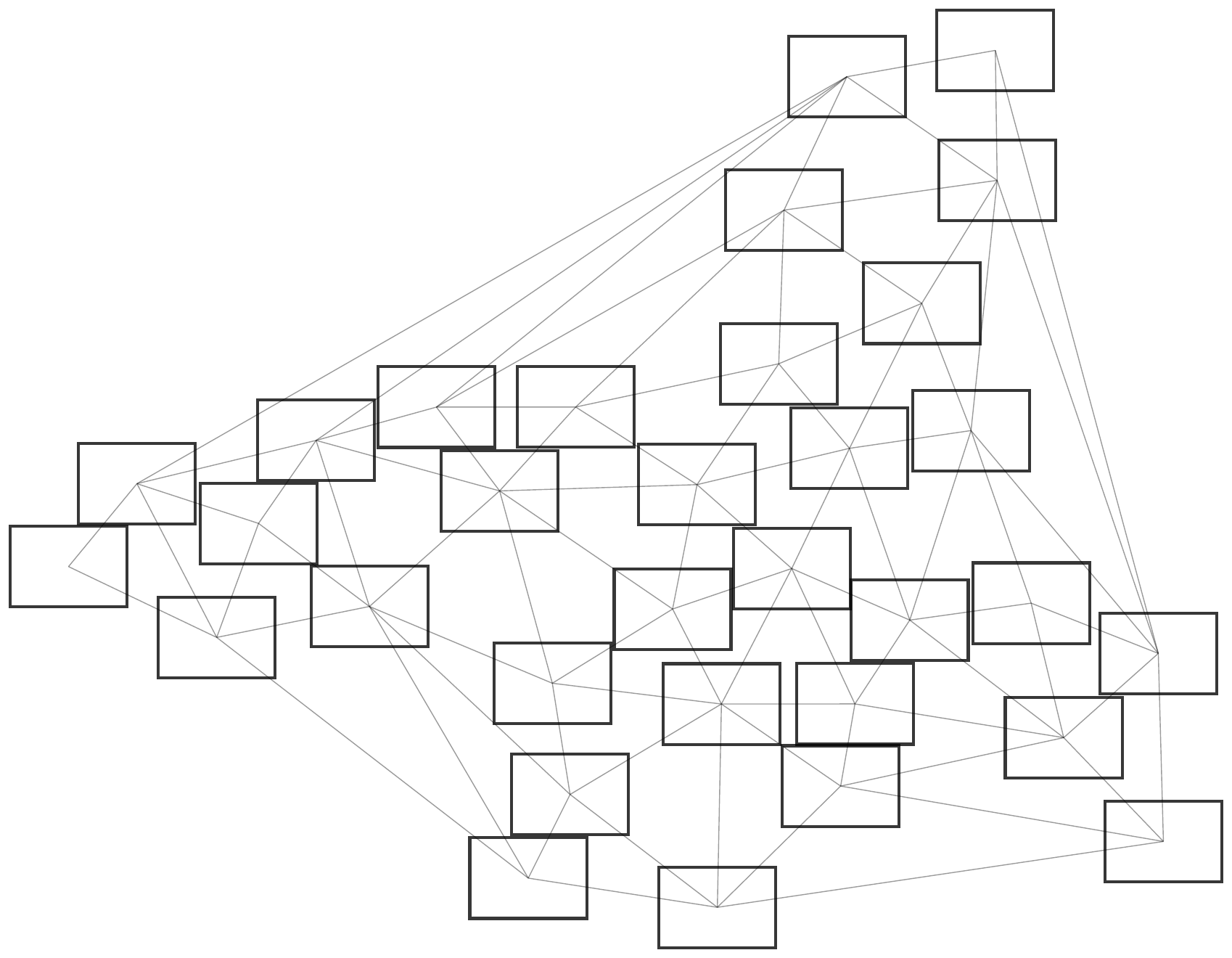}
                                \caption{iteration 5}
		\end{subfigure}
				\begin{subfigure}[b]{0.3\textwidth}
		                                \includegraphics[width=\textwidth]{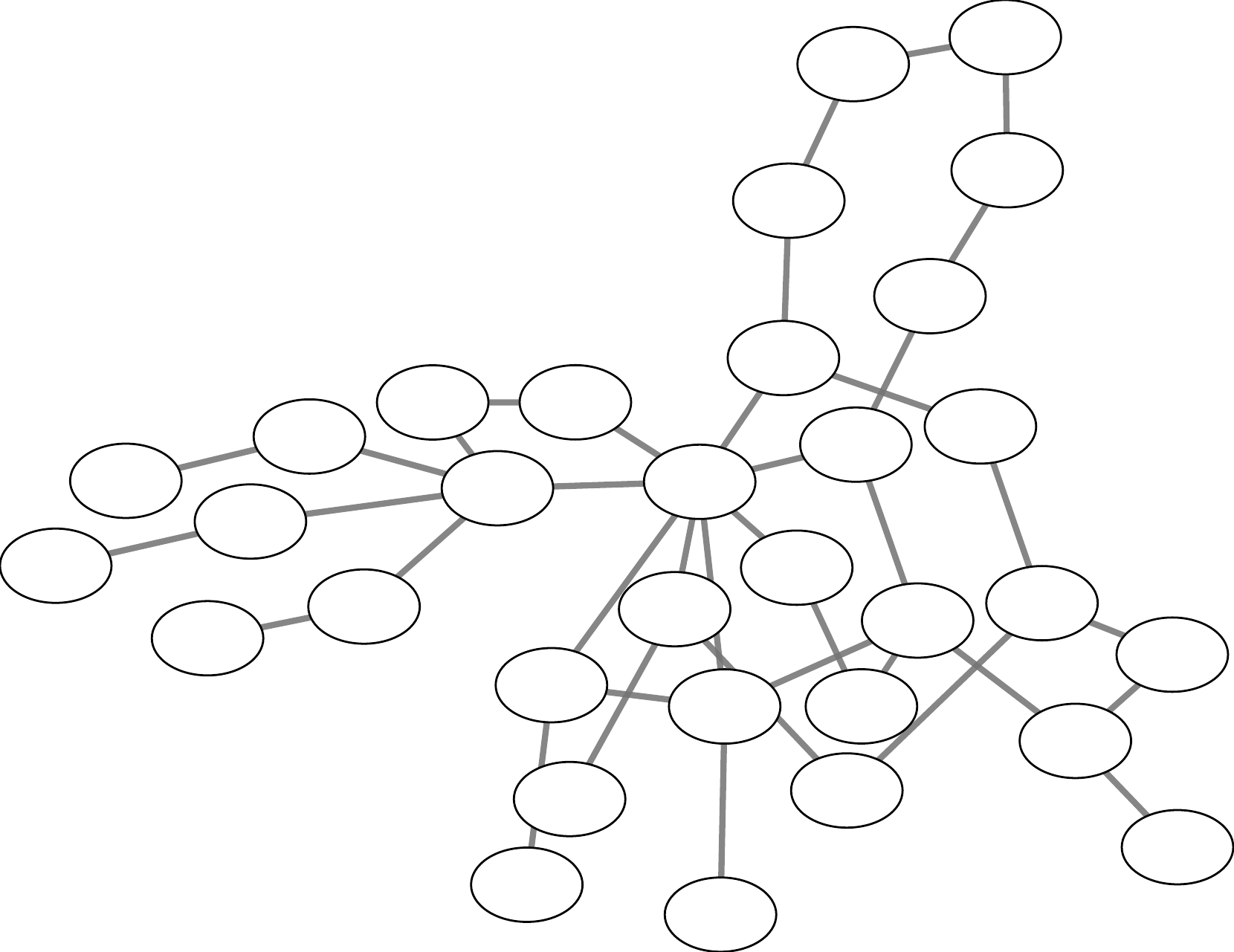}
		                                \caption{final overlap free graph \newline with original shapes}
				\end{subfigure}		
        \caption{Overlap removal process with the minimum spanning tree on the proximity graph, where the latter here is a Delaunay triangulation on rectangle centers. The blue edges form a tree; there are four different trees in the figure. The tree edges connecting overlapped nodes are thick and solid.  In each iteration the thick edges are elongated and the dashed tree edges shift accordingly. Overlap is completely resolved in four iterations. }
        \label{fig:iterationExample}
\end{figure}

The number $t_{ij}$ in line \ref{alg:t} of
Algorithm~\ref{algo:growTree} is the same as in the definition of the cost of the edge $(i,j)$ when $B_i$ and $B_j$ overlap, and is $1$ otherwise. 

The algorithm does not update all positions for the child sub-tree nodes immediately, but updates only the root of the sub-tree. Using the initial positions of a parent and a child, and the new position of the parent, the algorithm obtains the new position of the child in line \ref{alg:t}.   
In total, Algorithm~\ref{algo:growTree} works in $O(|V|)$ steps.
The choice of the root of the tree
does not matter. Different roots produce the same results modulo a
translation of the plane by a vector. Indeed it can be shown that after applying the
algorithm, for any $i,j \in V$ the vector $p'_j - p'_i$ is defined
uniquely by the path from $i$ to $j$ in $T$.  

While an overlap along any edge of the triangulation exists, we iterate, starting from finding a Delaunay triangulation, then
building a minimum spanning tree on it, and finally running Algorithm~\ref{algo:growTree}. See Figure~\ref{fig:iterationExample} for an example.

When there are no overlaps on the edges of the triangulation, as noticed
by Gansner and Hu~\cite{DBLP:journals/jgaa/GansnerH10}, overlaps are still possible. We follow the same idea as PRISM and modify the iteration step. In
addition to calculating the Delaunay triangulation we run a sweep-line
algorithm to find all overlapping node pairs and augment the Delaunay graph
$D$ with each such a pair.  As a consequence, the resulting minimum
spanning tree contains non-Delaunay edges catching the overlaps, and
the rest of the overlaps are removed. This stage usually requires much
less time than the previous one.

It is possible to create an example where the algorithm will not remove all overlaps. However, such examples are extremely rare and have not been seen yet in practice of using MSAGL or in our experiments. MSAGL applies random tiny changes to the initial layout which prevents GTree from cycling.

% Usually, to remove all overlaps with our method no more than $30$ runs of the high level steps is required. For experimental results and timing please see Table~\ref{fig:pdfs} and Table~\ref{fig:runtimeOverview}.

\newcolumntype{C}{m{0.16\linewidth}}

\section{Comparing PRISM and GTree by Measuring Layout Similarity, Quality, and  Run Time}\label{sec:measures}
Our data includes the same set of graphs that was used by the authors
of PRISM to compare it with other
algorithms~\citep{DBLP:journals/jgaa/GansnerH10}. The set is available
in the Graphviz open source package\footnote{http://www.graphviz.org}. We
%\href{http://www.graphviz.org/}{package}
also used a small collection of random graphs and a collection of
about 10,000 files\footnote{https://www.dropbox.com/sh/4q0k89yrv4x3ae3/AAA3xyKFRhLyyHXcG9jpcgata?dl=0}. For 
%residing~\href{https://www.dropbox.com/sh/4q0k89yrv4x3ae3/AAA3xyKFRhLyyHXcG9jpcgata?dl=0}{here}\footnote{https://www.dropbox.com/sh/4q0k89yrv4x3ae3/AAA3xyKFRhLyyHXcG9jpcgata?dl=0}. For
the experiments we use a modified version of Dot, where we can invoke
either GTree or Prism for the overlap removal step, and we also used
MSAGL, where we implemented PRISM and GTree. MSAGL was used only to
obtain the quality measures. We ran the experiments on a PC with
Linux, 64bit and an Intel Core i7-2600K CPU@3.40GHz with 16GB RAM.

Some of resulting layouts can be seen in Figures~\ref{fig:pdfs1}, ~\ref{fig:root},
~\ref{fig:pdfs}.
 \newlength{\mylength}
\setlength{\mylength}{0.22\linewidth}
%\newcolumntype{C}{m{0.22\linewidth}}
\newcolumntype{C}{ >{\centering\arraybackslash} m{0.22\linewidth} }

\setlength{\mylength}{0.22\linewidth}
%\newcolumntype{C}{m{0.22\linewidth}}
\newcolumntype{C}{ >{\centering\arraybackslash} m{0.22\linewidth} }

%\clearpage
\begin{figure}
\centering
\begin{tabular}{CCC}%
\includegraphics[width=\mylength,height=\mylength,keepaspectratio]{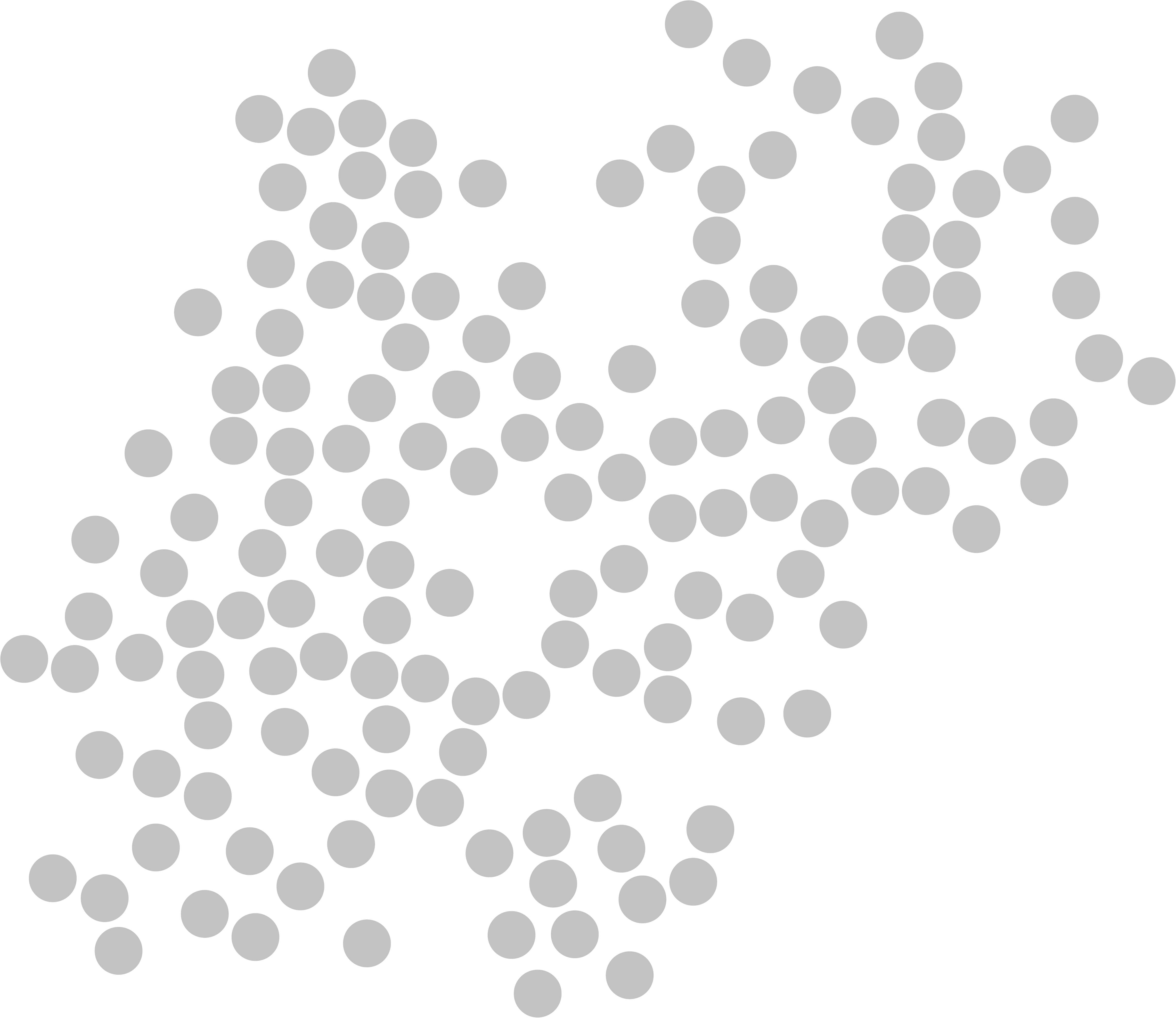}&  \includegraphics[width=\mylength,height=\mylength,keepaspectratio]{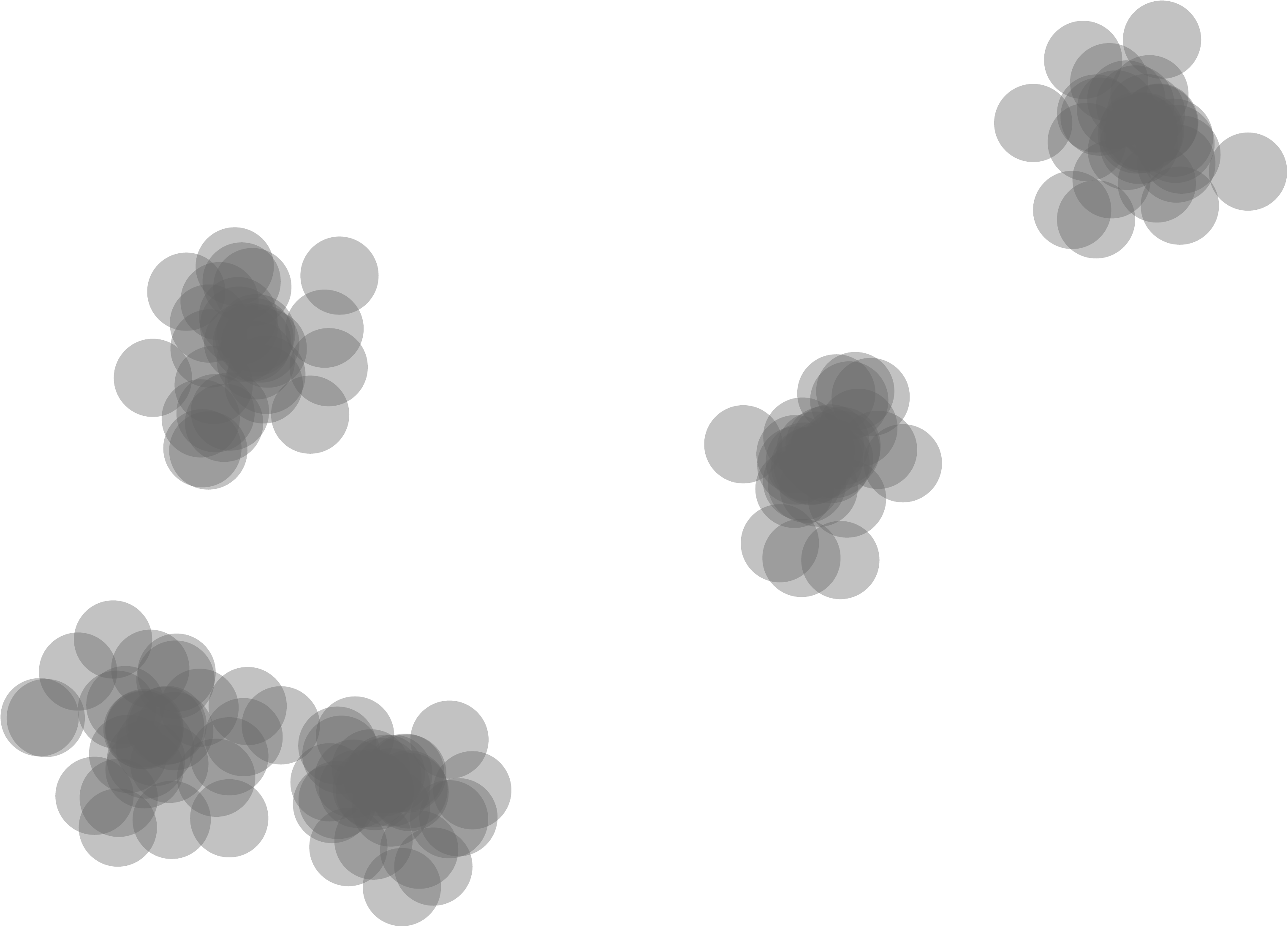}&  \includegraphics[width=2\mylength,height=2\mylength,keepaspectratio]{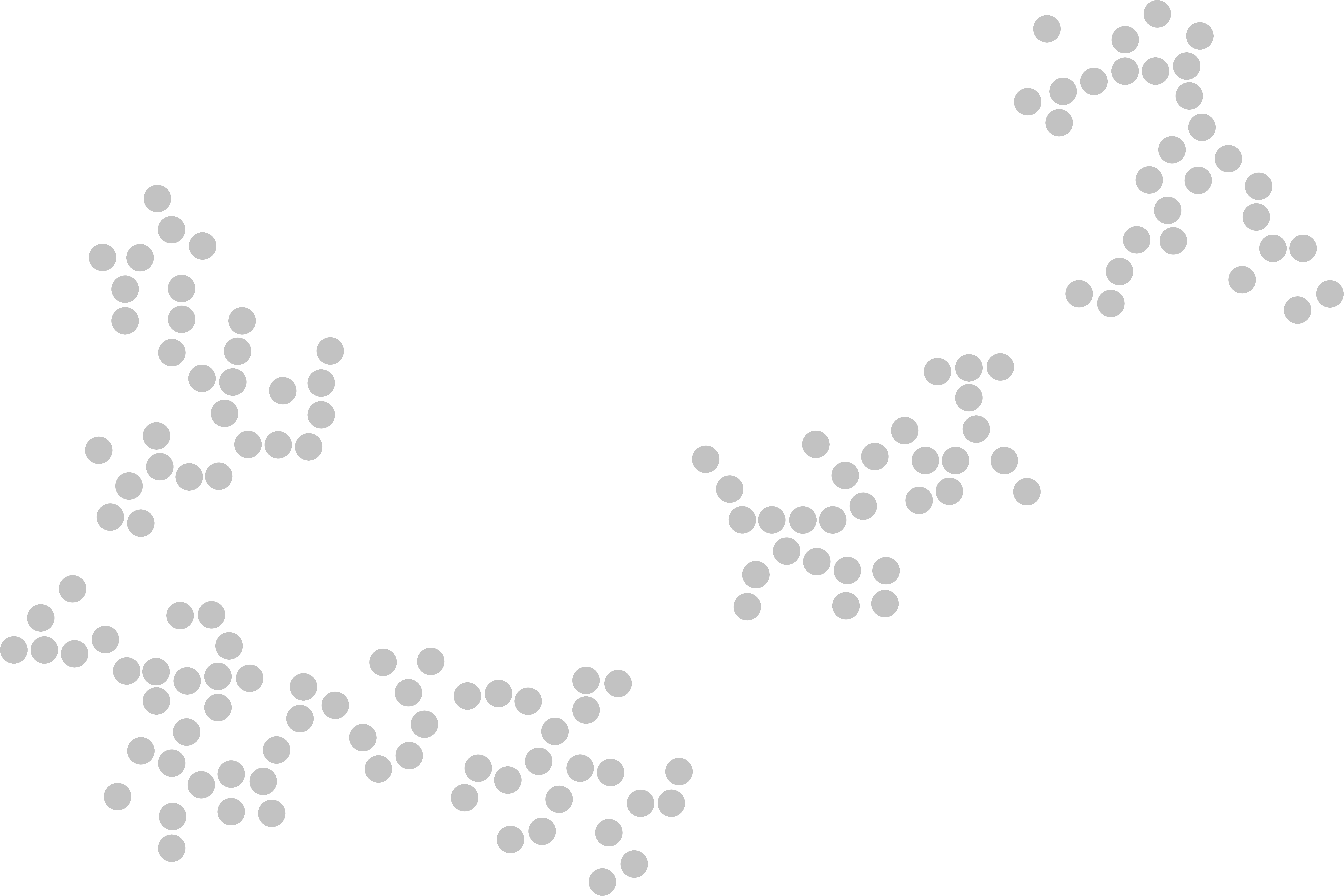}\\%
\includegraphics[width=\mylength,height=\mylength,keepaspectratio]{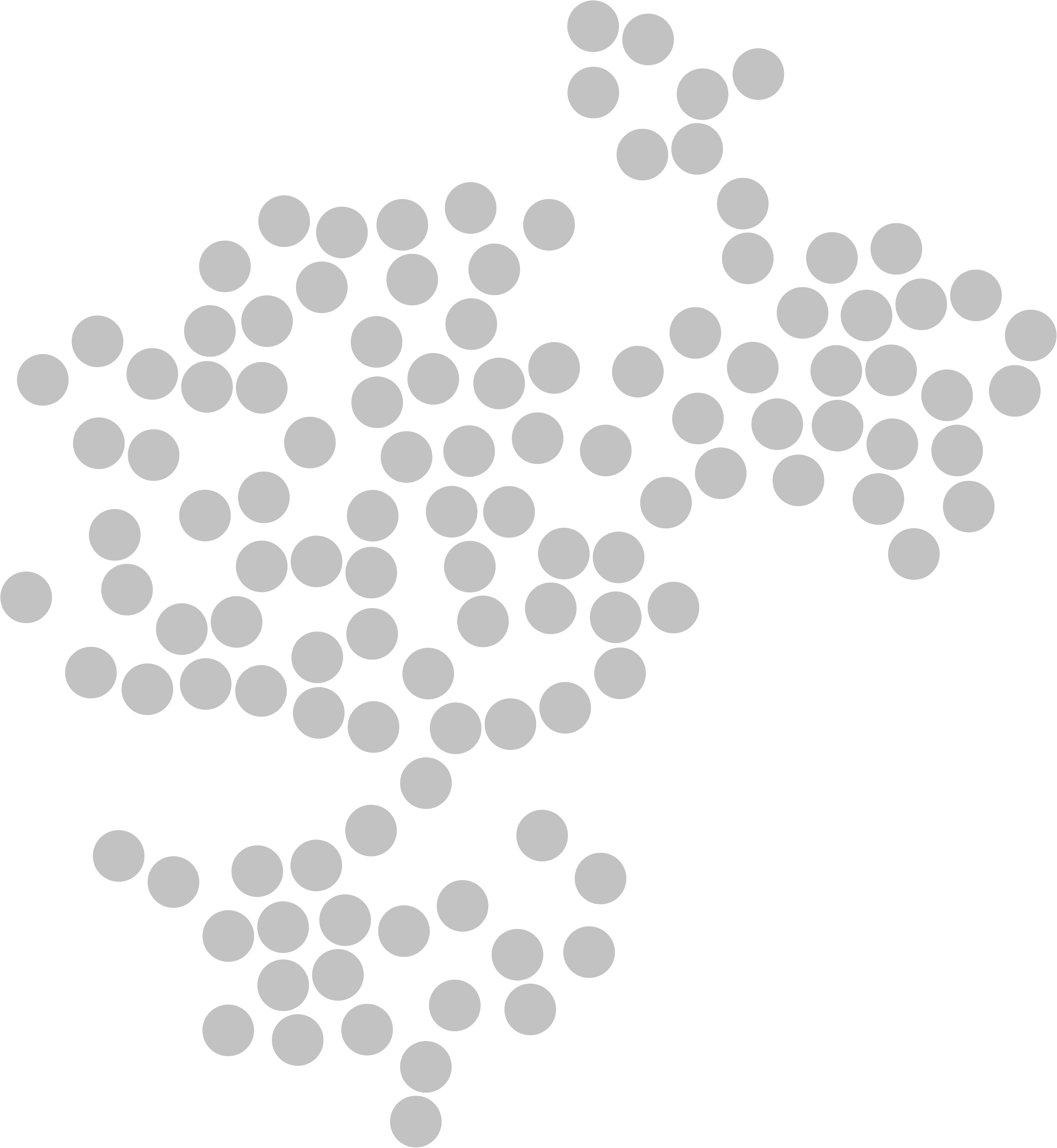}&  \includegraphics[width=\mylength,height=\mylength,keepaspectratio]{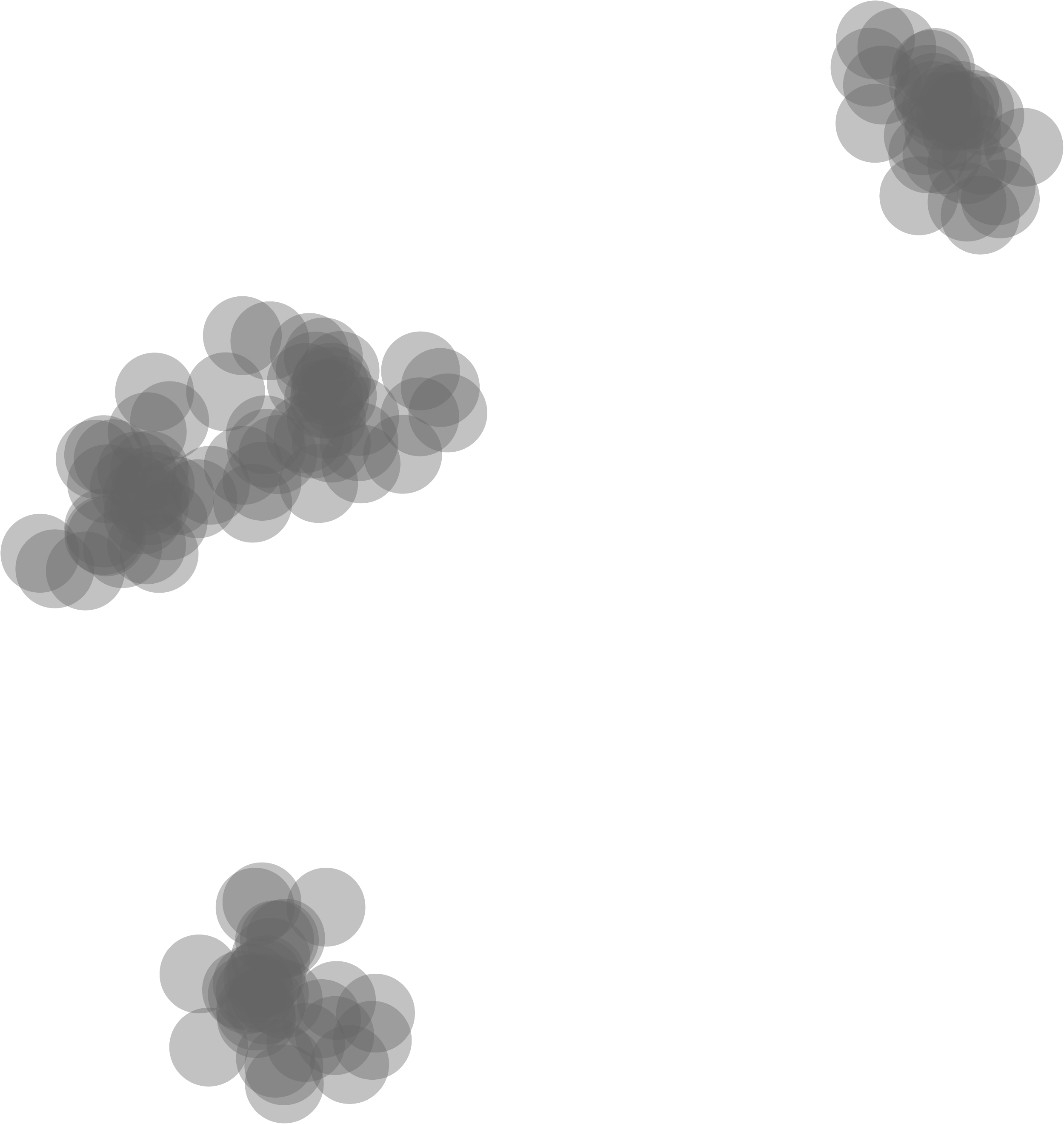}&  \includegraphics[width=1.5\mylength,height=1.5\mylength,keepaspectratio]{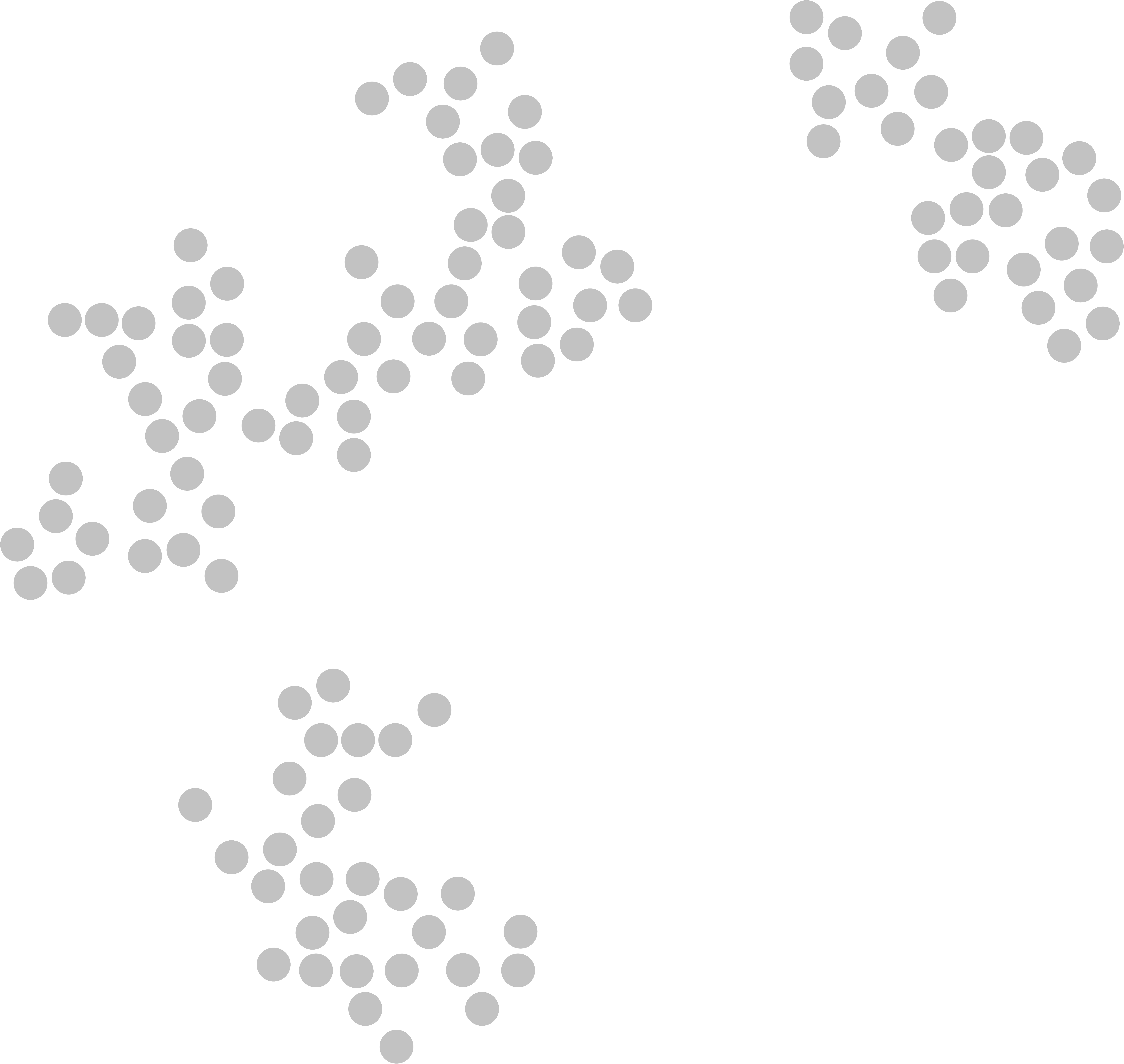}\\%
\includegraphics[width=\mylength,height=\mylength,keepaspectratio]{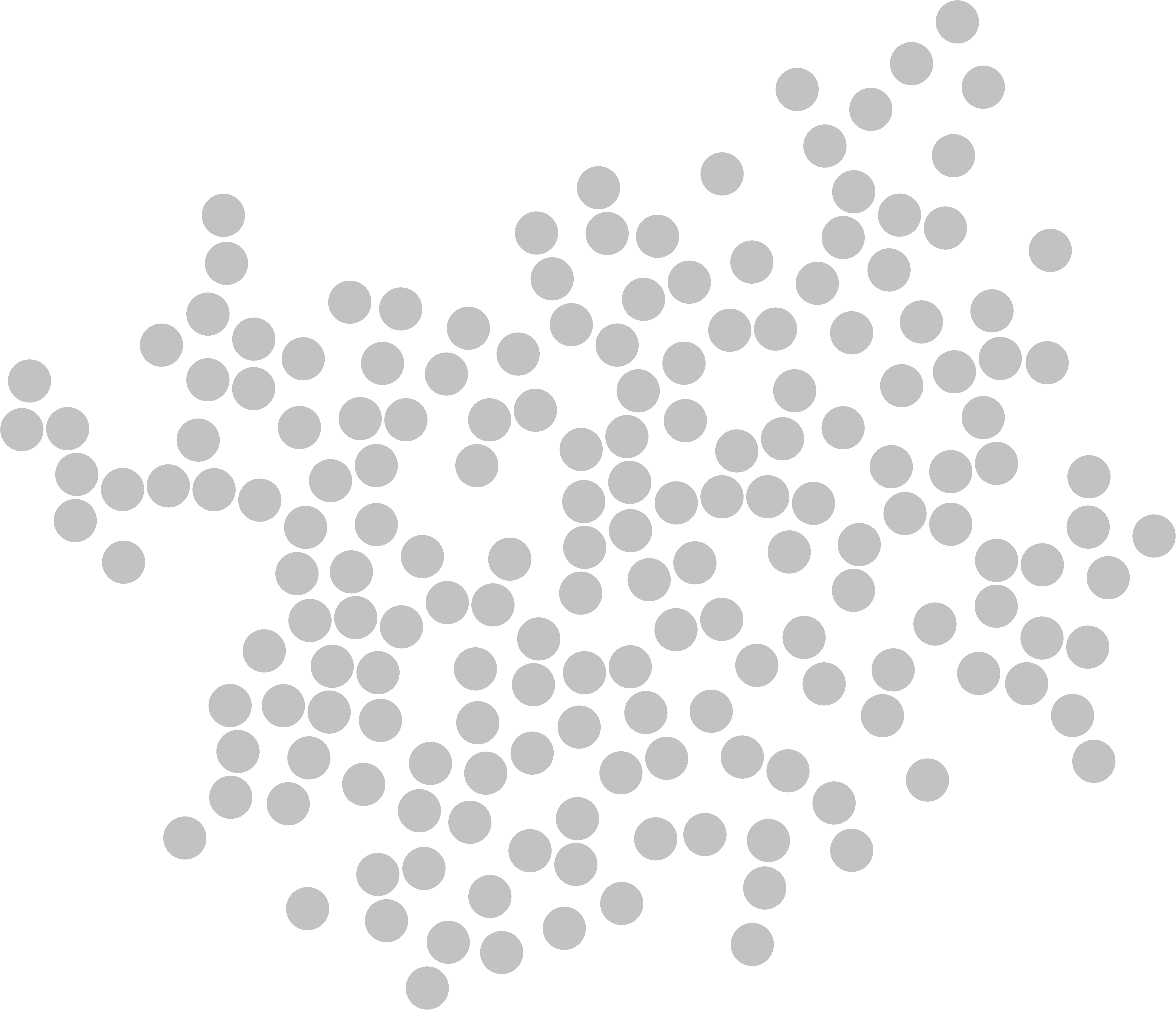}&  \includegraphics[width=\mylength,height=\mylength,keepaspectratio]{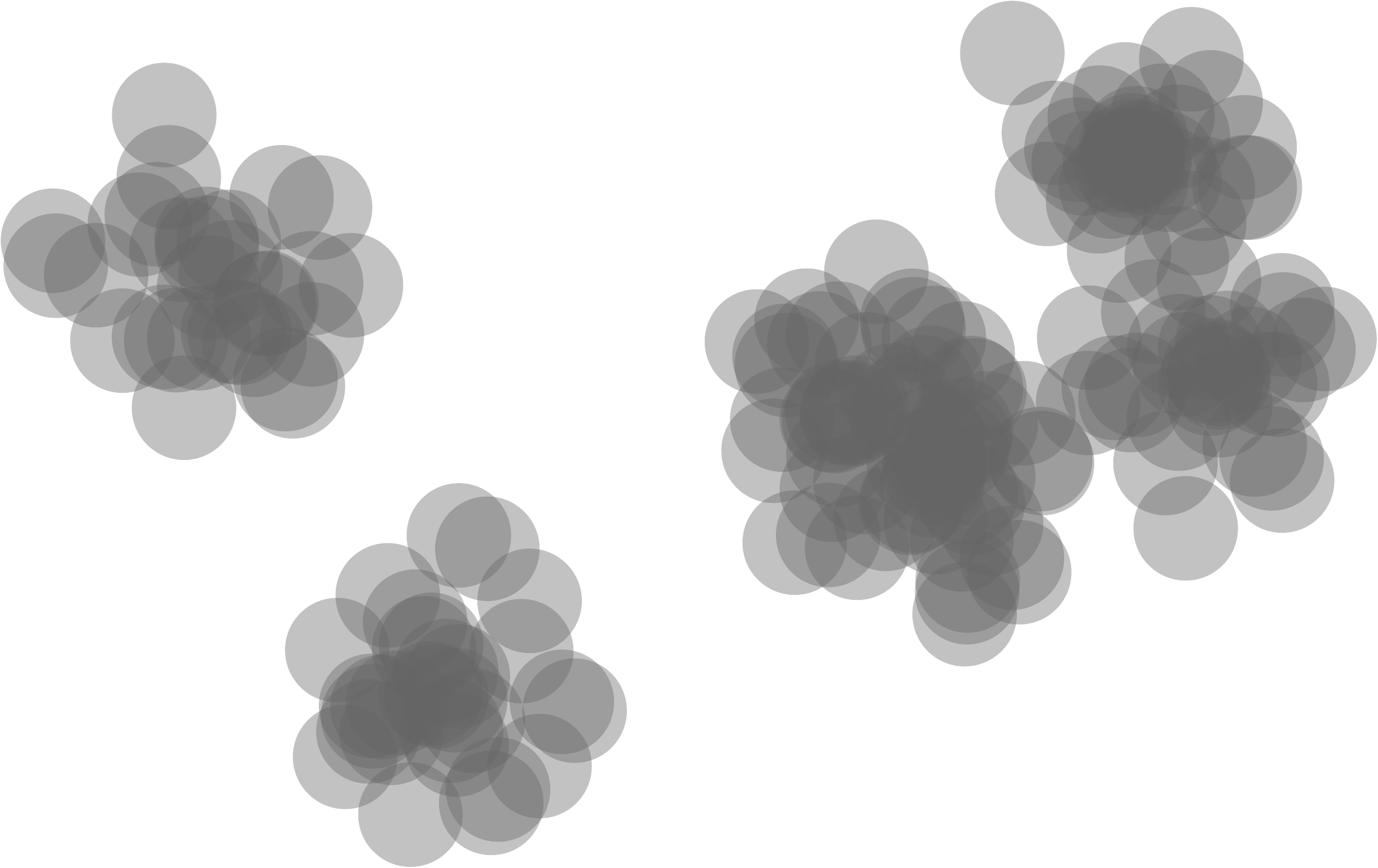}&  \includegraphics[width=1.8\mylength,height=1.8\mylength,keepaspectratio]{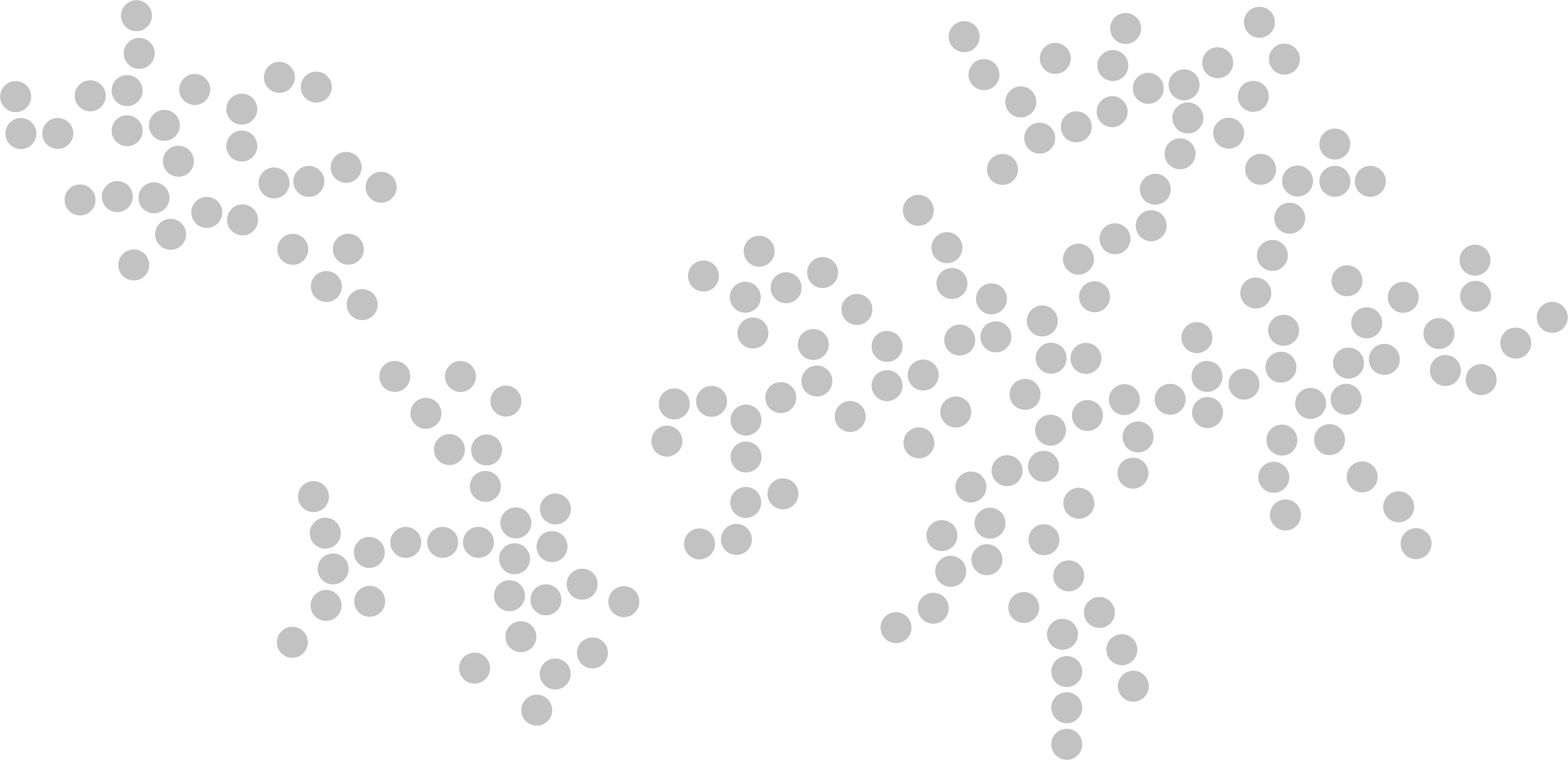}\\%
\includegraphics[width=\mylength,height=\mylength,keepaspectratio]{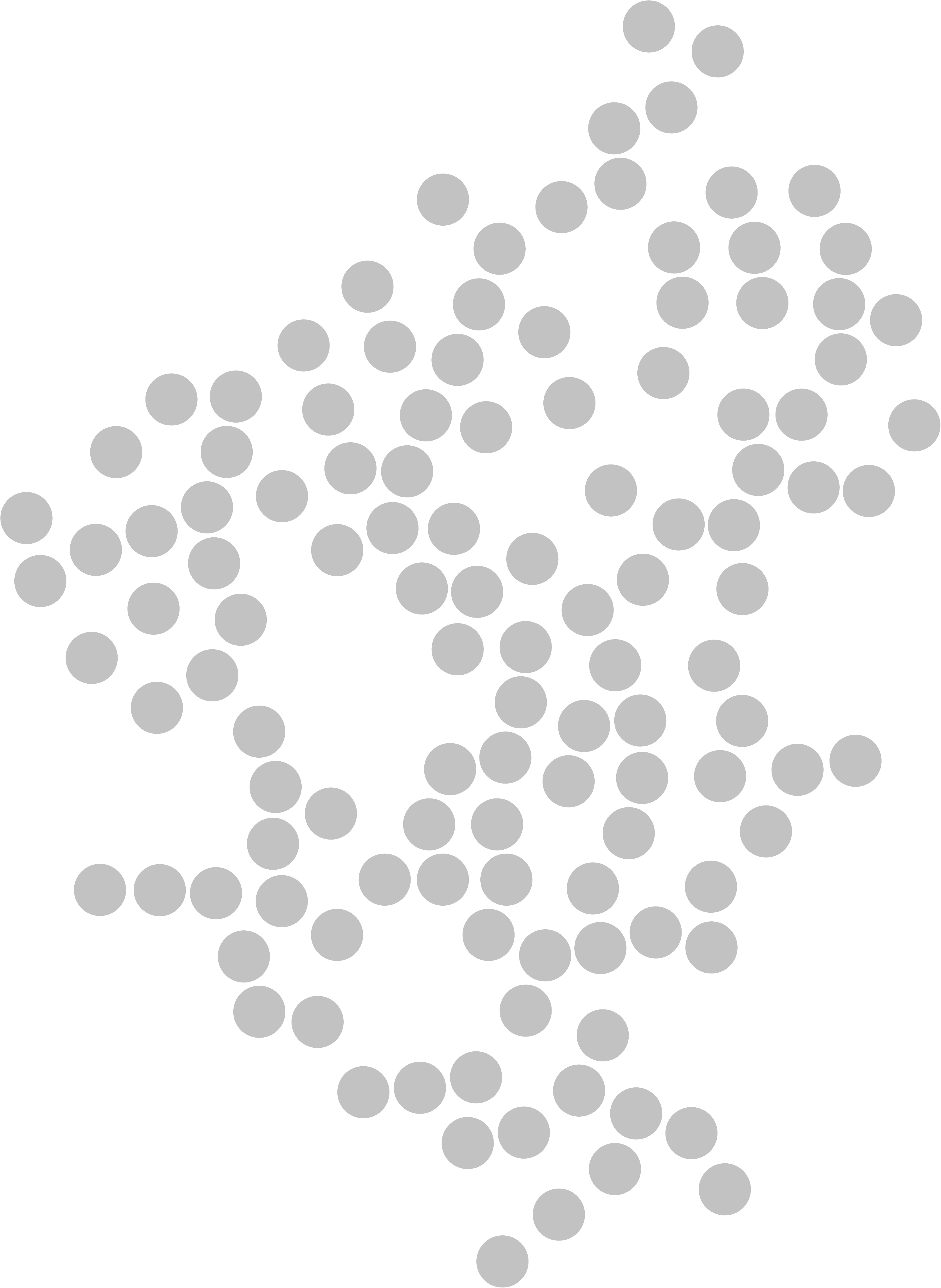}&  \includegraphics[width=\mylength,height=\mylength,keepaspectratio]{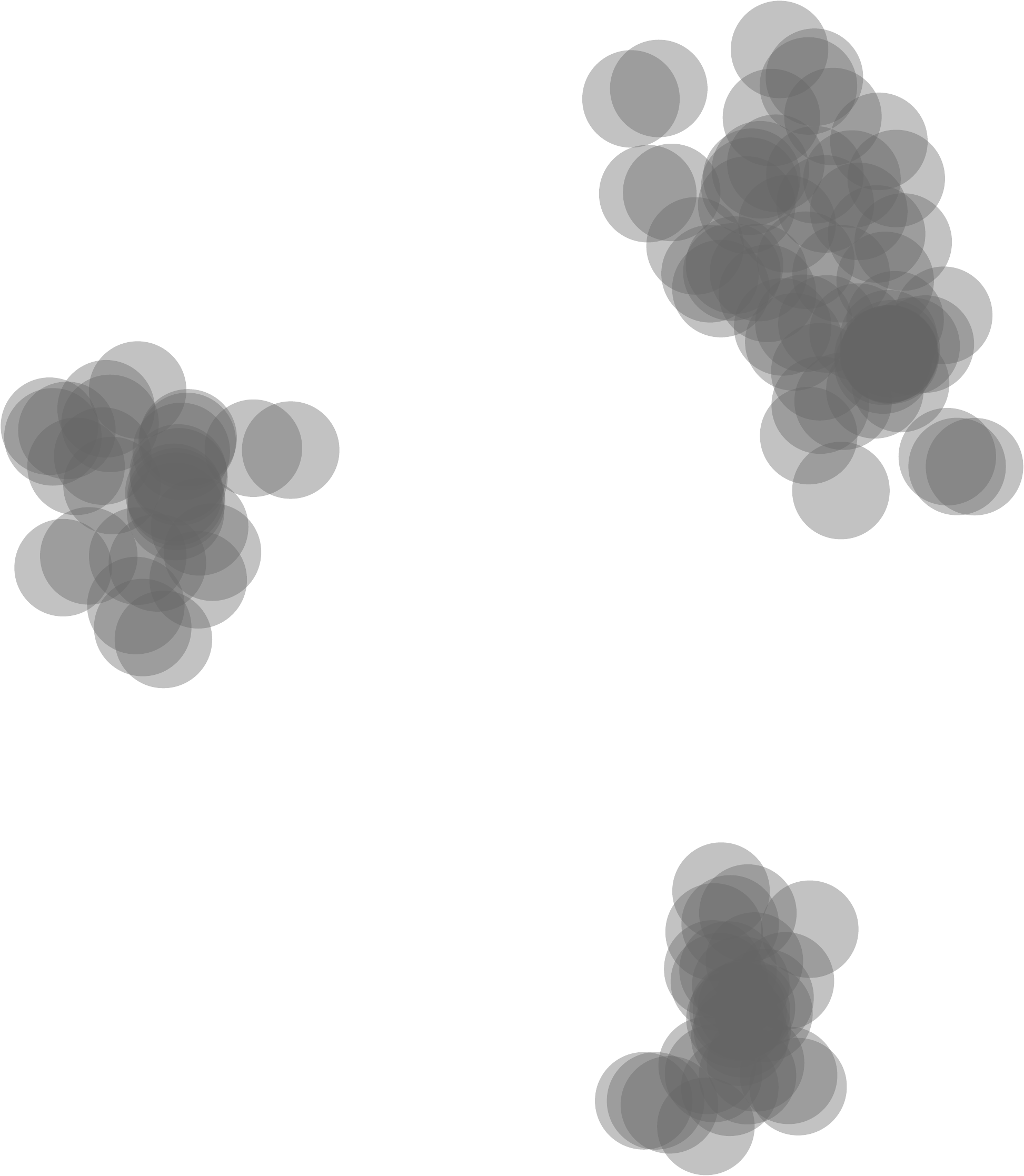}&  \includegraphics[width=1.6\mylength,height=1.6\mylength,keepaspectratio]{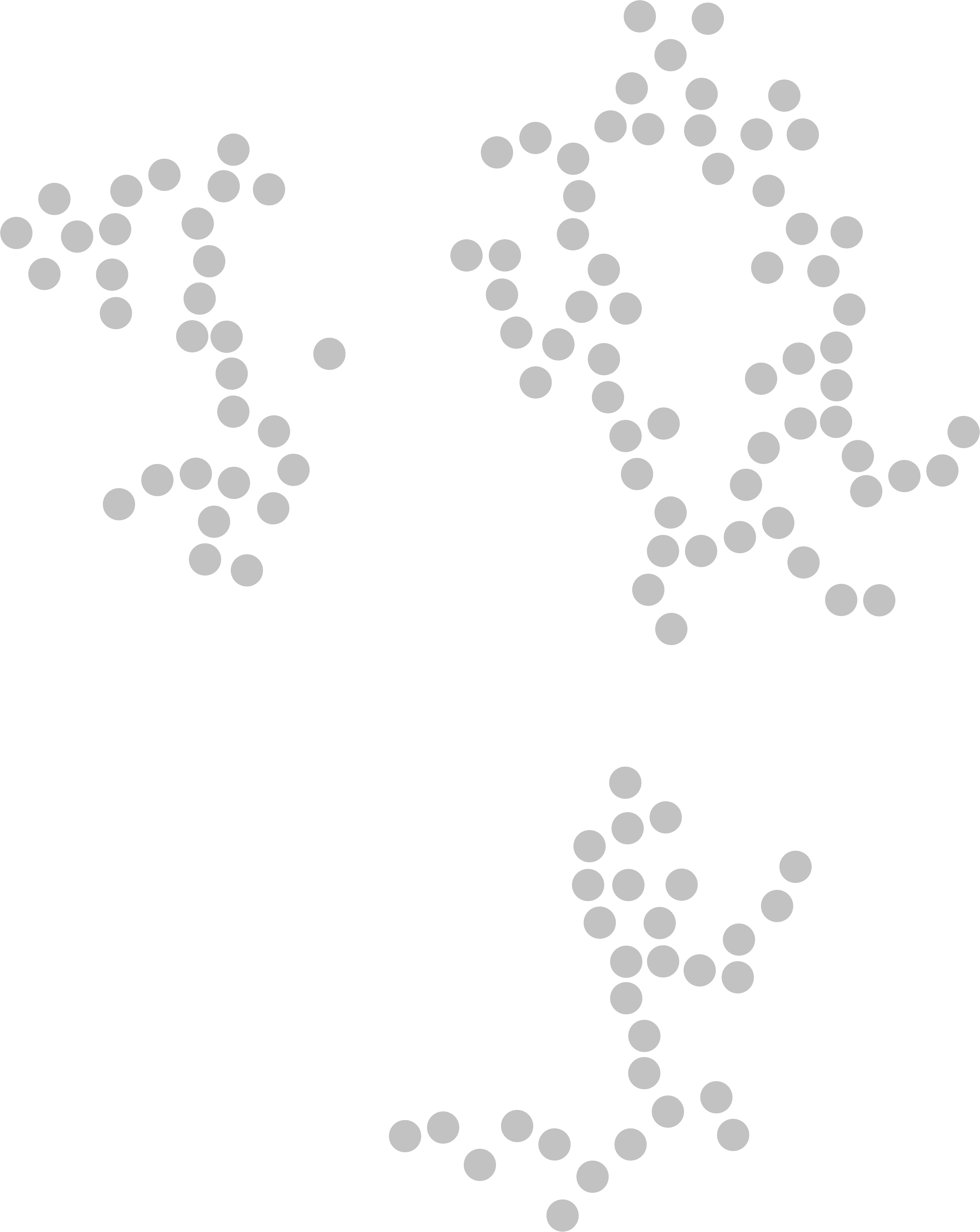}\\%
\includegraphics[width=\mylength,height=\mylength,keepaspectratio]{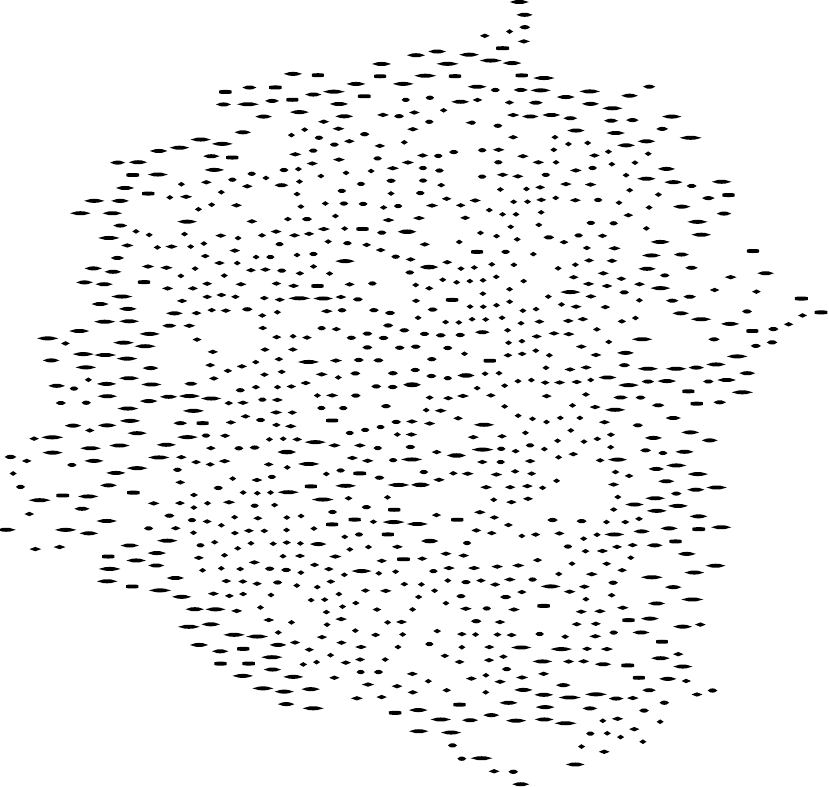}&  \includegraphics[width=\mylength,height=\mylength,keepaspectratio]{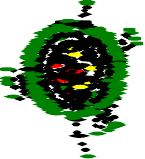}&  \includegraphics[width=1.6\mylength,height=1.6\mylength,keepaspectratio]{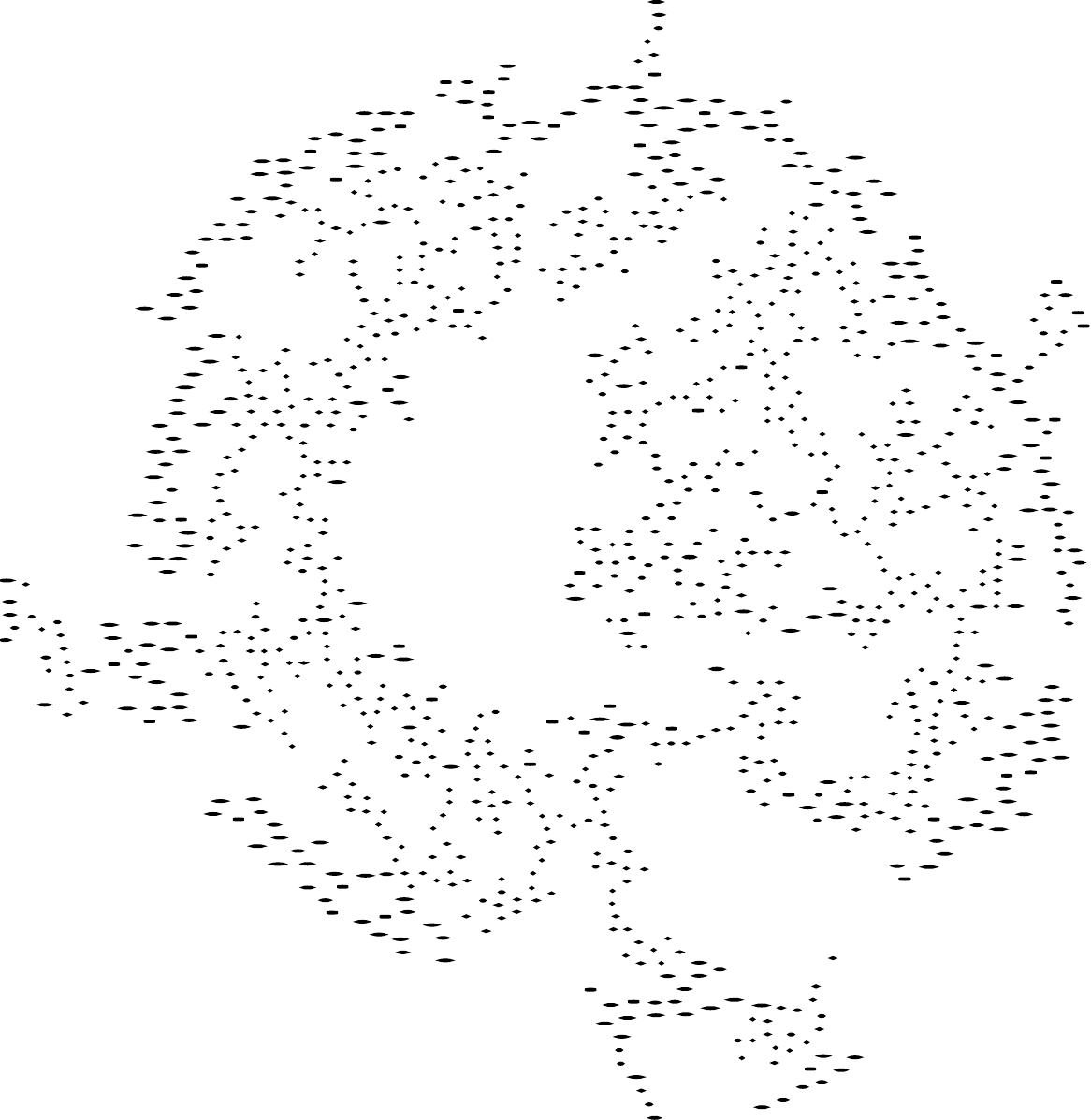}\\%
PRISM & original layout& GTree\\
\end{tabular}%
\caption{Comparison between PRISM, original, and GTree layouts. In four top rows the initial layouts were generated randomly. At the bottom are the drawings of nodes of graph "root" which was initially laid out by the Multi Dimensional Scaling algorithm of MSAGL. In our opinion, the initial structure is more preserved in the right column, containing the results of GTree.}
\label{fig:pdfs1}
\end{figure}

One can try to resolve overlap by scaling the node centers of the original layout. If there are no two coincident node centers this will work, but the resulting layout may require a huge area if some centers are close to each other.
We consider the \textit{area} of the final layout as one of the quality measures, and usually PRISM produces a smaller area than GTree, see Table~\ref{table:measures}.

In addition to comparing the areas,  we compare some other layout properties. Following Gansner and Hu~\cite{DBLP:journals/jgaa/GansnerH10}, we look at  \textit{edge length dissimilarity}, denoted as $\sigma_{\text{edge}}$. This measure reflects the relative change of the edge lengths of a Delaunay Triangulation on the node centers of the original layout. 

The other measure, which is denoted by $\sigma_{disp}$, is the Procrustean similarity~\cite{BorgMDS}. It shows how close the transformation of the original graph is to a combination of a scale, a rotation, and a shift transformation. PRISM and GTree performs similar in the last two measures as Table~\ref{table:measures} shows.

To distinguish the methods further, we measure the change in the set of $k$ closest neighbors of the nodes. Namely, let $p_1,\dots,p_n$ be the positions of the node centers, and let $k$ be an integer such that  $0 < k \le n$. 
Let $I = \{1,\dots,n\}$ be the set of node indices. For each $i \in I $ we define $N_k(i) \subset I\setminus \{i\}$,  such that $|N_k(p, i)|=k$, and for every $j \in I \setminus N_k(p, i)$ and for every $j' \in N_k(p, i)$ holds $\|p_j-p_i\| \ge \|p_{j'}-p_i\|$. In other words, $N_k(p, i)$ represents a set of $k$ closest neighbors of $i$, excluding $i$.
Let $p'_1,\dots,p'_n$ be transformed node centers. To see how much the layout is distorted nearby node $i$, we intersect $N_k(p,i)$ and $N_k(p',i)$. We measure the distortion as $(k-m)^2$, where $m$ is the number of elements in the intersection. One can see that if the node preserves its $k$ closest neighbors then the distortion is zero. 

Our experiments for $k$ from $8$ to $12$ show that under this measure
GTree produced a smaller error, showing less distortion, on 8 graphs from 14,
and on the rest PRISM produced a better result, see Table~\ref{tbl:kns}.
GTree produced a smaller error on all small random graphs from other
  collections\footnote{https://www.dropbox.com/sh/4q0k89yrv4x3ae3/AAA3xyKFRhLyyHXcG9jpcgata?dl=0}.
%from~\href{https://www.dropbox.com/sh/4q0k89yrv4x3ae3/AAA3xyKFRhLyyHXcG9jpcgata?dl=0}{other
  %collections}\footnote{https://www.dropbox.com/sh/4q0k89yrv4x3ae3/AAA3xyKFRhLyyHXcG9jpcgata?dl=0}.

\begin{table}[tbh]
%\begin{wraptable}{r}{7cm}
%\small
\caption{Similarity to the initial layout (left) and number of iterations for different graph sizes and different initialization methods (right). PR stands for PRISM % ($\sigma_{\text{edge}}$ and $\sigma_{\text{disp}}$) and the final layout area.
}
\label{table:measures} 
\centering
\resizebox{0.45\linewidth}{!}{%
\begin{tabular}{>{\raggedright\arraybackslash}p{1.25cm}
>{\raggedleft\arraybackslash}p{0.6cm}%
>{\raggedleft\arraybackslash}p{0.6cm}%
>{\raggedleft\arraybackslash}p{0.6cm}%
>{\raggedleft\arraybackslash}p{0.6cm}%
>{\raggedleft\arraybackslash}p{0.69cm}%
>{\raggedleft\arraybackslash}p{0.65cm}%
}%
\toprule%
\multicolumn{1}{c}{\bfseries }%
&\multicolumn{2}{>{\centering\arraybackslash}p{1.8cm}}{\bfseries  $\sigma_{\text{edge}}$}%
&\multicolumn{2}{>{\centering\arraybackslash}p{1.8cm}}{\bfseries $\sigma_{\text{disp}}$}%
&\multicolumn{2}{>{\centering\arraybackslash}p{1.8cm}}{\bfseries area}\tabularnewline%
%\cline{1-7}
%\midrule
%
% latex table generated in R 3.0.1 by xtable 1.7-1 package
% Thu Mar  6 12:57:04 2014
Graph & \scriptsize PR & \scriptsize GTree & \scriptsize PR & \scriptsize GTree & \scriptsize PR & \scriptsize GTree \\ 
  \midrule \midrule dpd & 0.34 & \cellcolor{colMst!25}0.28 & 0.37 & \cellcolor{colMst!25}0.36 & \cellcolor{colPrism!25} 0.82 &  0.84 \\ 
  unix & 0.22 & \cellcolor{colMst!25}0.19 & 0.24 & \cellcolor{colMst!25}0.20 & 2.38 &  2.38 \\ 
  rowe & 0.29 & \cellcolor{colMst!25}0.26 & \cellcolor{colPrism!25}0.23 & 0.24 & \cellcolor{colPrism!25} 0.68 &  0.73 \\ 
  size & 0.39 & \cellcolor{colMst!25}0.37 & \cellcolor{colPrism!25}0.24 & 0.26 & \cellcolor{colPrism!25} 1.09 &  1.28 \\ 
  ngk10\_4 & 0.30 & 0.30 & \cellcolor{colPrism!25}0.27 & 0.30 &  0.00 &  0.00 \\ 
  NaN & 0.56 & \cellcolor{colMst!25}0.44 & 0.73 & \cellcolor{colMst!25}0.51 & \cellcolor{colPrism!25} 4.03 &  4.34 \\ 
  b124 & 0.55 & \cellcolor{colMst!25}0.53 & 0.97 & \cellcolor{colMst!25}0.83 & \cellcolor{colPrism!25} 5.52 &  6.22 \\ 
  b143 & \cellcolor{colPrism!25}0.67 & 0.70 & 1.12 & \cellcolor{colMst!25}0.93 & \cellcolor{colPrism!25} 3.62 &  3.88 \\ 
  mode & 0.54 & \cellcolor{colMst!25}0.50 & 0.59 & \cellcolor{colMst!25}0.53 & \cellcolor{colPrism!25} 1.53 &  2.29 \\ 
  b102 & \cellcolor{colPrism!25}0.71 & 0.77 & 1.43 & \cellcolor{colMst!25}1.27 & \cellcolor{colPrism!25} 4.50 &  6.62 \\ 
  xx & 0.75 & \cellcolor{colMst!25}0.70 & 1.65 & \cellcolor{colMst!25}1.42 & \cellcolor{colPrism!25} 6.21 &  9.57 \\ 
  root & \cellcolor{colPrism!25}1.09 & 1.19 & 2.89 & \cellcolor{colMst!25}2.45 & \cellcolor{colPrism!25}34.58 & 91.87 \\ 
  badvoro & \cellcolor{colPrism!25}0.88 & 0.92 & \cellcolor{colPrism!25}2.27 & 2.42 & \cellcolor{colPrism!25}25.68 & 47.43 \\ 
  b100 & \cellcolor{colPrism!25}0.84 & 0.98 & \cellcolor{colPrism!25}3.08 & 3.14 & \cellcolor{colPrism!25}20.64 & 37.38 \\ 
   \bottomrule 
\end{tabular}%
}\qquad%
\resizebox{0.45\linewidth}{!}{%
\begin{tabular}{>{\raggedright\arraybackslash}p{1.05cm}%
>{\raggedleft\arraybackslash}p{0.6cm}%
>{\raggedleft\arraybackslash}p{0.6cm}%
>{\raggedleft\arraybackslash}p{0.6cm}%
>{\raggedleft\arraybackslash}p{0.5cm}%
>{\raggedleft\arraybackslash}p{0.6cm}%
>{\raggedleft\arraybackslash}p{0.5cm}%
}%
\toprule%
\multicolumn{2}{>{\centering\arraybackslash}m{2.5cm}}{\bfseries init.~layout:}%
&\multicolumn{3}{>{\raggedleft\arraybackslash}m{2.9cm}}{\bfseries \footnotesize neato}%
&\multicolumn{2}{>{\centering\arraybackslash}m{1.5cm}}{\bfseries \footnotesize SFDP}%
\tabularnewline%
%\cline{1-7}
%\midrule
%
% latex table generated in R 3.0.1 by xtable 1.7-1 package
% Thu Mar  6 15:27:54 2014
Graph & $|V|$ & $|E|$ & \scriptsize{PR} & \scriptsize{GTree} & \scriptsize{PR} & \scriptsize{GTree} \\ 
  \midrule \midrule dpd &   36 &  108 & \cellcolor{colPrism!25} 4 &  7 & \cellcolor{colPrism!25}  3 &  6 \\ 
  unix &   41 &   49 & \cellcolor{colPrism!25} 3 &  4 &  12 & \cellcolor{colMst!25} 5 \\ 
  rowe &   43 &   68 &  5 & \cellcolor{colMst!25} 4 &  13 & \cellcolor{colMst!25} 7 \\ 
  size &   47 &   55 &  7 & \cellcolor{colMst!25} 3 &   9 & \cellcolor{colMst!25} 5 \\ 
  ngk10\_4 &   50 &  100 &  6 & \cellcolor{colMst!25} 3 &  14 & \cellcolor{colMst!25} 7 \\ 
  NaN &   76 &  121 &  8 & \cellcolor{colMst!25} 3 &  24 & \cellcolor{colMst!25} 6 \\ 
  b124 &   79 &  281 & 14 & \cellcolor{colMst!25} 4 &  30 & \cellcolor{colMst!25}12 \\ 
  b143 &  135 &  366 & 21 & \cellcolor{colMst!25} 6 &  37 & \cellcolor{colMst!25}12 \\ 
  mode &  213 &  269 & 37 & \cellcolor{colMst!25} 8 &  11 & \cellcolor{colMst!25} 6 \\ 
  b102 &  302 &  611 & 60 & \cellcolor{colMst!25}24 & 113 & \cellcolor{colMst!25}19 \\ 
  xx &  302 &  611 & 83 & \cellcolor{colMst!25}18 &  50 & \cellcolor{colMst!25}19 \\ 
  root & 1054 & 1083 & 95 & \cellcolor{colMst!25}18 &  99 & \cellcolor{colMst!25}22 \\ 
  badvoro & 1235 & 1616 & 40 & \cellcolor{colMst!25}20 &  50 & \cellcolor{colMst!25}23 \\ 
  b100 & 1463 & 5806 & 80 & \cellcolor{colMst!25}24 & 136 & \cellcolor{colMst!25}28 \\ 
   \bottomrule 
\end{tabular}%
}%
\end{table}%
%\end{wraptable}
%
\begin{table}[tbh]
%\begin{wraptable}{r}{7cm}
%\small
\caption{k closest neighbors error, the Multi Dimensional Scaling algorithm of MSAGL was used for the initial layout. PR stands for PRISM. 
}
\label{tbl:kns} 
\centering
\resizebox{0.65\linewidth}{!}{%
\begin{tabular}{>{\raggedright\arraybackslash}p{1.25cm}
>{\raggedleft\arraybackslash}p{0.7cm}%
>{\raggedleft\arraybackslash}p{0.7cm}%
>{\raggedleft\arraybackslash}p{0.7cm}%
>{\raggedleft\arraybackslash}p{0.7cm}%
>{\raggedleft\arraybackslash}p{0.7cm}%
>{\raggedleft\arraybackslash}p{0.7cm}%
>{\raggedleft\arraybackslash}p{0.7cm}%
>{\raggedleft\arraybackslash}p{0.7cm}%
>{\raggedleft\arraybackslash}p{0.7cm}%
>{\raggedleft\arraybackslash}p{0.7cm}%
>{\raggedleft\arraybackslash}p{0.7cm}%
}%
\toprule%
\multicolumn{1}{c}{\bfseries }%
&\multicolumn{2}{>{\centering\arraybackslash}p{1.8cm}}{\bfseries  $k=8$}%
&\multicolumn{2}{>{\centering\arraybackslash}p{1.8cm}}{\bfseries $k =9$}%
&\multicolumn{2}{>{\centering\arraybackslash}p{1.8cm}}{\bfseries  $k=10$}%
&\multicolumn{2}{>{\centering\arraybackslash}p{1.8cm}}{\bfseries  $k=11$}%
&\multicolumn{2}{>{\centering\arraybackslash}p{1.8cm}}{\bfseries  $k=12$}%
\tabularnewline%

Graph & \scriptsize PR & \scriptsize GTree & \scriptsize PR& \scriptsize GTree & \scriptsize PR & \scriptsize GTree& \scriptsize PR &\scriptsize GTree & \scriptsize PR & \scriptsize GTree \\ 
  \midrule \midrule 
dpd & 7.75 & \cellcolor{colMst!25}6.06 & 9.61 & \cellcolor{colMst!25}7.36 & 9.5 & \cellcolor{colMst!25}8 & 10.14 & \cellcolor{colMst!25}8.5 & 9.97 & \cellcolor{colMst!25}7.64\\
unix & 8.56 & \cellcolor{colMst!25}7.05 & 10.51 & \cellcolor{colMst!25}8.8 & 10.95 & \cellcolor{colMst!25}10.02 & 11.66 & \cellcolor{colMst!25}10.54 & 13 & \cellcolor{colMst!25}11.41\\
rowe & \cellcolor{colPrism!25}6.28 & 8.09 & \cellcolor{colPrism!25}7.09 & 9.95 & \cellcolor{colPrism!25}7.49 & 10.49 & \cellcolor{colPrism!25}9.12 & 11.4 & \cellcolor{colPrism!25}11.05 & 12.51\\
size & \cellcolor{colPrism!25}4.68 & 6.09 & \cellcolor{colPrism!25}5.47 & 6.47 & \cellcolor{colPrism!25}6.28 & 7.57 & \cellcolor{colPrism!25}6.89 & 8.13 & \cellcolor{colPrism!25}8.26 & 10.02\\
ngk10\_4 & \cellcolor{colPrism!25}6.76 & 7.4 & \cellcolor{colPrism!25}7.52 & 9.26 & \cellcolor{colPrism!25}8.28 & 11.38 & \cellcolor{colPrism!25}10.72 & 13.74 & \cellcolor{colPrism!25}11.92 & 14.66\\
NaN & 11.83 & \cellcolor{colMst!25}8.95 & 14.46 & \cellcolor{colMst!25}11.5 & 17.32 & \cellcolor{colMst!25}13.88 & 19.88 & \cellcolor{colMst!25}16.37 & 22.17 & \cellcolor{colMst!25}19.7\\
b124 & \cellcolor{colPrism!25}11.03 & 11.44 & \cellcolor{colPrism!25}13.22 & 13.56 & \cellcolor{colPrism!25}14.76 & 15.54 & \cellcolor{colPrism!25}15.91 & 17.32 & \cellcolor{colPrism!25}18.23 & 20.04\\
b143 & 13.49 & \cellcolor{colMst!25}12.39 & 16.31 & \cellcolor{colMst!25}14.99 & 19.49 & \cellcolor{colMst!25}17.93 & 23.11 & \cellcolor{colMst!25}21.04 & 26.53 & \cellcolor{colMst!25}24.43\\
mode & 16.91 & \cellcolor{colMst!25}11.46 & 20.58 & \cellcolor{colMst!25}13.95 & 24.68 & \cellcolor{colMst!25}16.85 & 29.54 & \cellcolor{colMst!25}19.92 & 34.48 & \cellcolor{colMst!25}22.56\\
b102 & 15.99 & \cellcolor{colMst!25}14.62 & 19.61 & \cellcolor{colMst!25}18.78 & 23.38 & \cellcolor{colMst!25}22.77 & 27.28 & \cellcolor{colMst!25}26.77 & 32.15 & \cellcolor{colMst!25}31.45\\
xx & 15.68 & \cellcolor{colMst!25}15.65 & \cellcolor{colPrism!25}19.01 & 19.45 & \cellcolor{colPrism!25}23.05 & 23.37 & \cellcolor{colPrism!25}26.98 & 27.35 & \cellcolor{colPrism!25}31.29 & 32.47\\
root & 17.09 & \cellcolor{colMst!25}15.7 & 20.89 & \cellcolor{colMst!25}19.36 & 25.48 & \cellcolor{colMst!25}23.3 & 30.48 & \cellcolor{colMst!25}27.66 & 35.74 & \cellcolor{colMst!25}32.83\\
badvoro & 16.18 & \cellcolor{colMst!25}15.15 & 20.16 & \cellcolor{colMst!25}18.98 & 24.37 & \cellcolor{colMst!25}23.28 & 29.18 & \cellcolor{colMst!25}28.03 & 34.29 & \cellcolor{colMst!25}33.29\\
b100 & \cellcolor{colPrism!25}18 & 19.25 & \cellcolor{colPrism!25}22.11 & 23.65 & \cellcolor{colPrism!25}26.79 & 28.69 & \cellcolor{colPrism!25}32.03 & 34.46 & \cellcolor{colPrism!25}37.44 & 40.5\\
   \bottomrule 
\end{tabular}%
}%\qquad%
\end{table}%
We ran tests on the graphs from a subdirectory of the same site called ``dot\_files'', let us call this set of graphs collection $A$.
Each graph from $A$ represents the control flow of a method from a version of the .NET framework. $A$ contains 10077 graphs. The graph sizes do not exceed several thousands.
We used the Multi Dimensional Scaling algorithms of MSAGL for the initial layout in this test. The results of the run are summarized in Table~\ref{tbl:longrun}.
\begin{table}[tbh]
\caption{Statistics on collection $A$. Here k-cn stands for k-closest neighbors, and ``iters'' stands for the number of iterations. Each cell contains the number of graphs for the measure on which the method performed better. 
We can see that PRISM produced a layout of smaller area than the one of GTree on 8498 graph, against 1579 graphs where GTree  required less area.
 From the other side, GTree gives better results on all other measures.
 The columns of k-cn and ``iters'' do not sum to 10077, the number of graphs in $A$, because some of the results were equal for PRISM and GTree.}
\label{tbl:longrun} 
\centering
%\resizebox{0.25\linewidth}{!}{%
\begin{tabular}{>{\raggedright\arraybackslash}p{1.25cm}
>{\raggedleft\arraybackslash}p{0.7cm}%
>{\raggedleft\arraybackslash}p{0.7cm}%
>{\raggedleft\arraybackslash}p{0.7cm}%
>{\raggedleft\arraybackslash}p{0.7cm}%
>{\raggedleft\arraybackslash}p{0.9cm}%
>{\raggedleft\arraybackslash}p{0.7cm}%
}%
\tabularnewline%
Method & \scriptsize k-cn & \scriptsize $\sigma_{\text{edge}}$ & \scriptsize $\sigma_{\text{disp}}$& area & iters & time\\ 
  \midrule PRISM & 3237 & 4741 & 4114 & \cellcolor{colPrism!25}8498 & 46 & 7\\
GTree & \cellcolor{colMst!25}4088 & \cellcolor{colMst!25}5336 & \cellcolor{colMst!25}5963 & 1579 & \cellcolor{colMst!25}9986 & \cellcolor{colMst!25}10070\\
   \bottomrule 
\end{tabular}%
%}
\end{table}%
%8498 gtree wins 1579,
%on iter prism wins 46 gtree wins 9986, and we have 45 draws
%on time prism wins 7 gtree wins 10070, and we have 0 draws
\begin{figure}[bth]
\centering
\begin{knitrout}
\definecolor{shadecolor}{rgb}{0.969, 0.969, 0.969}\color{fgcolor}
\includegraphics[width=0.65\linewidth]{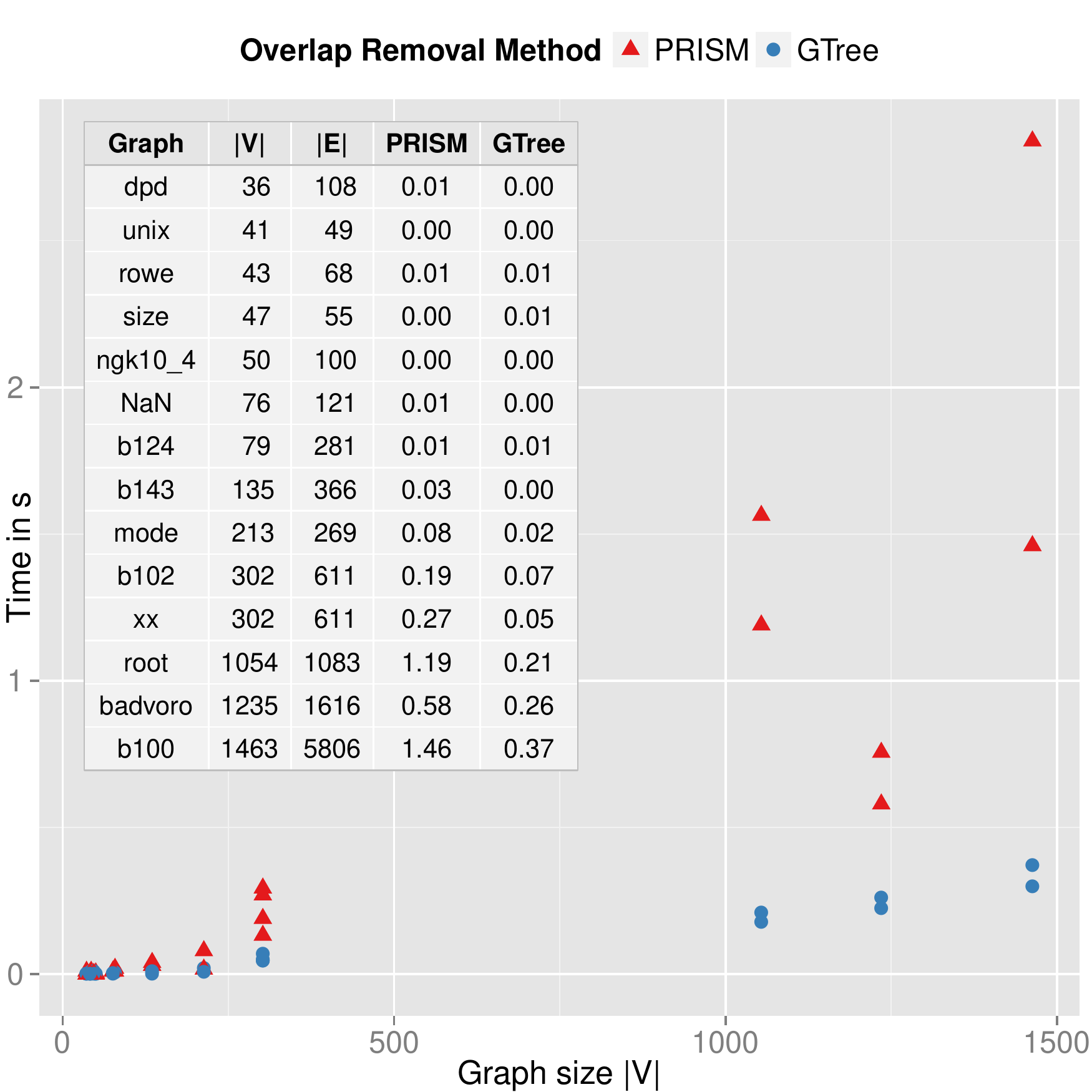} 
\end{knitrout}
\caption[Runtime Comparison PRISM-GTree]{ Runtimes for PRISM and GTree.}
\label{fig:runtimeOverview}
\end{figure}

\subsection*{Runtime Comparison}
Both methods remove the overlap iteratively using the proximity graph. However, while PRISM needs \mbox{$\mathcal{O}(|V|\cdot \sqrt{|V|})$} time to solve the stress model, GTree needs only $\mathcal{O}(|V|)$ time per iteration with the growing tree procedure. Therefore, GTree is asymptotically faster in a single iteration. In addition, as Table~\ref{table:measures} (right) shows,  GTree usually needs fewer iterations than PRISM, especially on larger graphs. 
The overall runtime can be seen in Figure \ref{fig:runtimeOverview}. It shows that GTree outperforms PRISM on larger graphs.

%\subsection{Maintaining the Initial Layout}
%
%To see how well the original graph structure is maintained by both
%methods, we compare the pairwise Euclidean node distances and the
%graph-theoretic shortest path distances of the graph. The quartile
%plots (\cref{fig:pathDistances}) show the distribution of these
%distances. It depends on the concrete graph which method represents 
%the shortest path distances better. Overall, the originalp
%graph structure is matched similarly well by PRISM and GTree. 
%
%For the initialization step of the layout we use two different
%algorithms of Graphviz; neato and SFDP. \Cref{fig:pdfs} shows
%results obtained by starting with neato. As the result shows, the layouts generated by both PRISM and GTree look fairly similar. And some layouts generated by PRISM appear to be more compact than corresponding layouts generated by GTree.   
%
%We further compared the ability of preserving original graph structure. We  generated a series of random layout, all with overlap nodes,  and applied both algorithm for overlap removal. Additional, we apply the same comparison on the ``root'' graph. Figure \ref{fig:pdfs1} shows the result.  In our opinion the structures are better preserved in the layouts generated by GTree.

%\clearpage
\begin{figure}[t]
        \centering
        \subcaptionbox{initial layout\label{fig:rootOriginal}}[0.2\linewidth]{
        \includegraphics[width=0.075\linewidth]{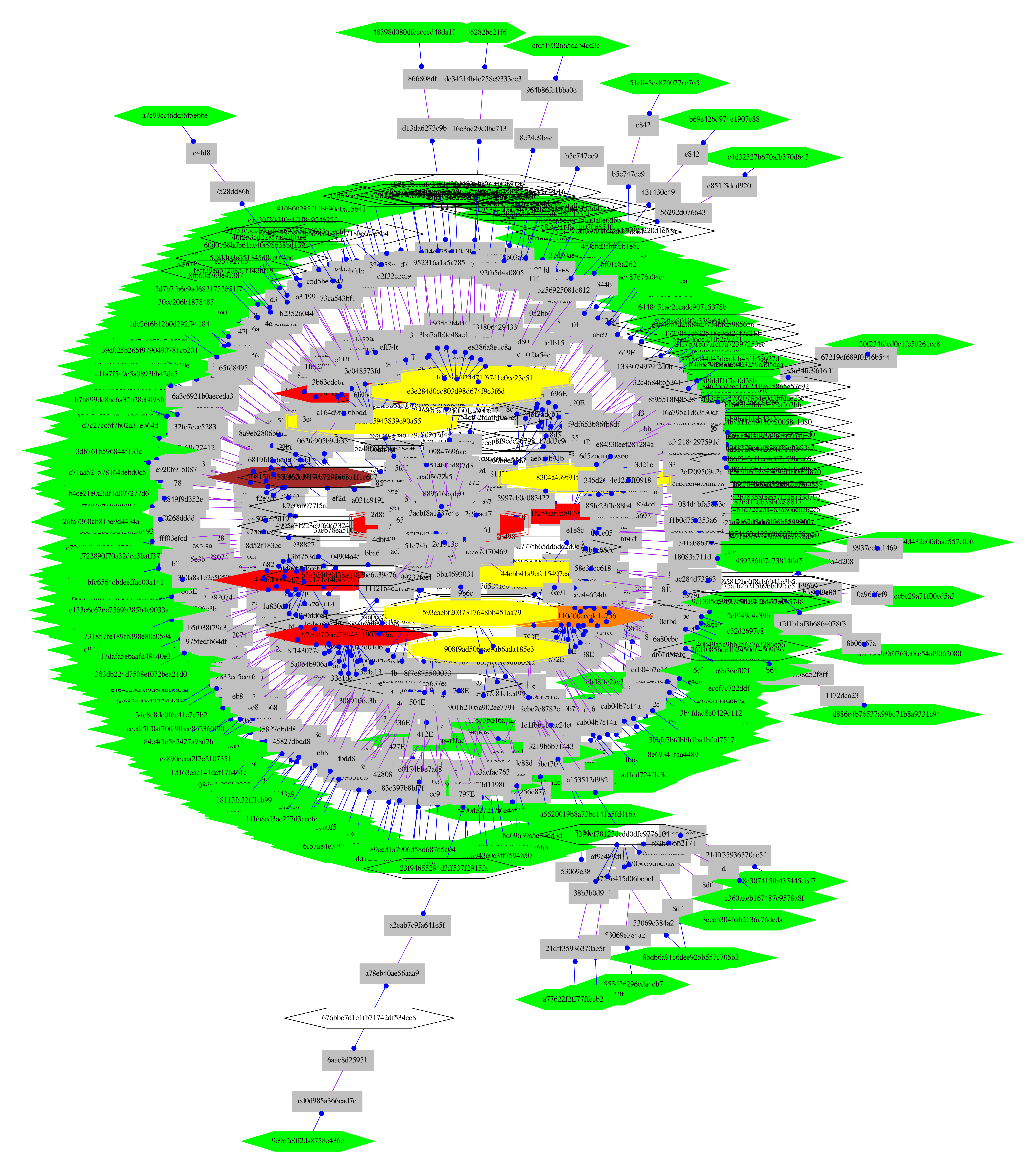}%
        }
        \subcaptionbox{PRISM\label{fig:rootPRISM}}{
        	\includegraphics[width=0.35\linewidth]{root-neato-prism.pdf}%
		}
		\subcaptionbox{GTree\label{fig:rootMST}}{
		\includegraphics[width=0.4\textwidth]{root-neato-gtree-exp.pdf}%
		}
\caption{root graph with 1054 nodes and 1083 edges.
(a) initial layout with NEATO, (b)
  applying PRISM, (c) applying GTree.}
\label{fig:root}
\end{figure}
\begin{figure}[t]
\centering
\resizebox{\linewidth}{!}{
\begin{tabular}{CCC}%
\includegraphics[width=\mylength,height=\mylength,keepaspectratio]{b100_gv-sfdpInit_dot-gtreeOverlap.pdf}&
\includegraphics[width=\mylength,height=\mylength,keepaspectratio]{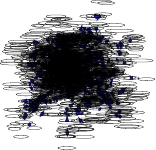}&
\includegraphics[width=\mylength,height=\mylength,keepaspectratio]{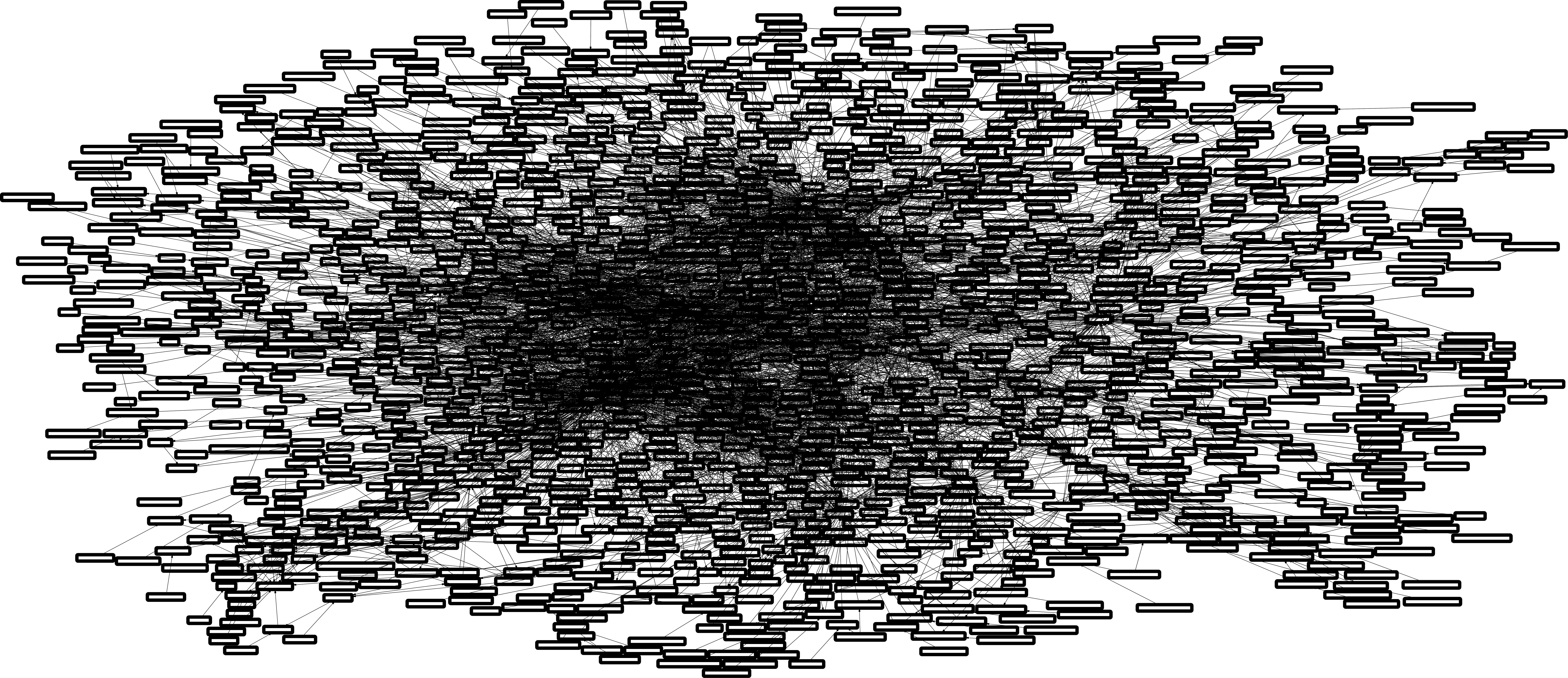}\\%
\includegraphics[width=\mylength,height=\mylength,keepaspectratio]{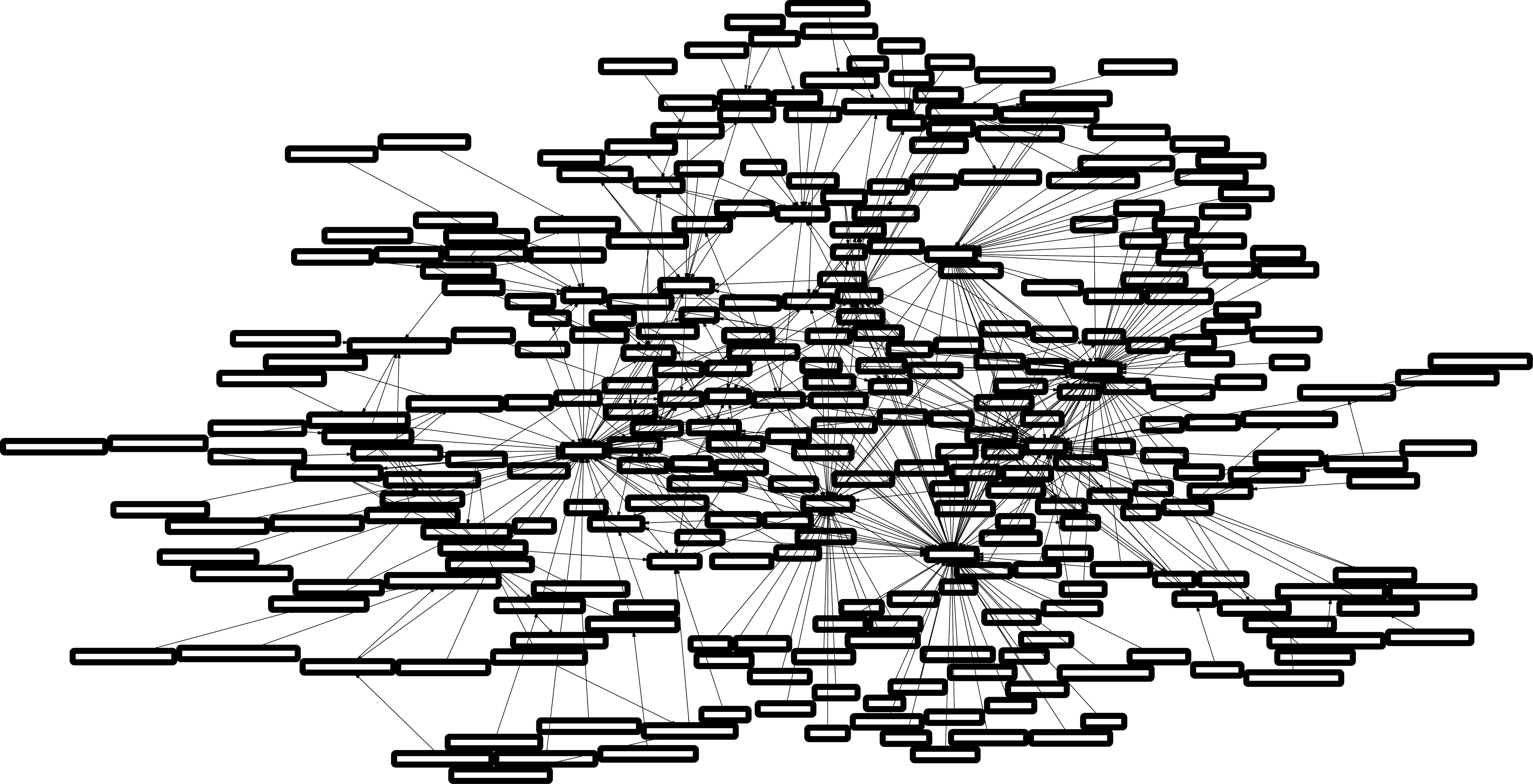}&
\includegraphics[width=\mylength,height=\mylength,keepaspectratio]{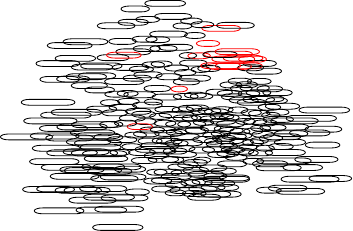}&
\includegraphics[width=\mylength,height=\mylength,keepaspectratio]{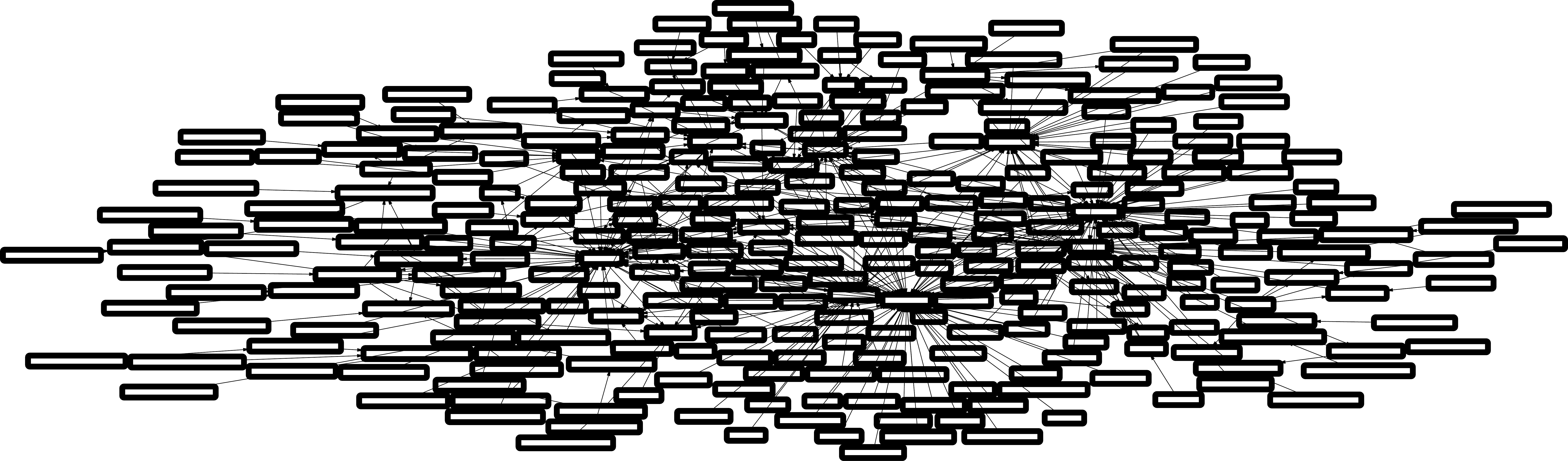}\\%
\includegraphics[width=\mylength,height=\mylength,keepaspectratio]{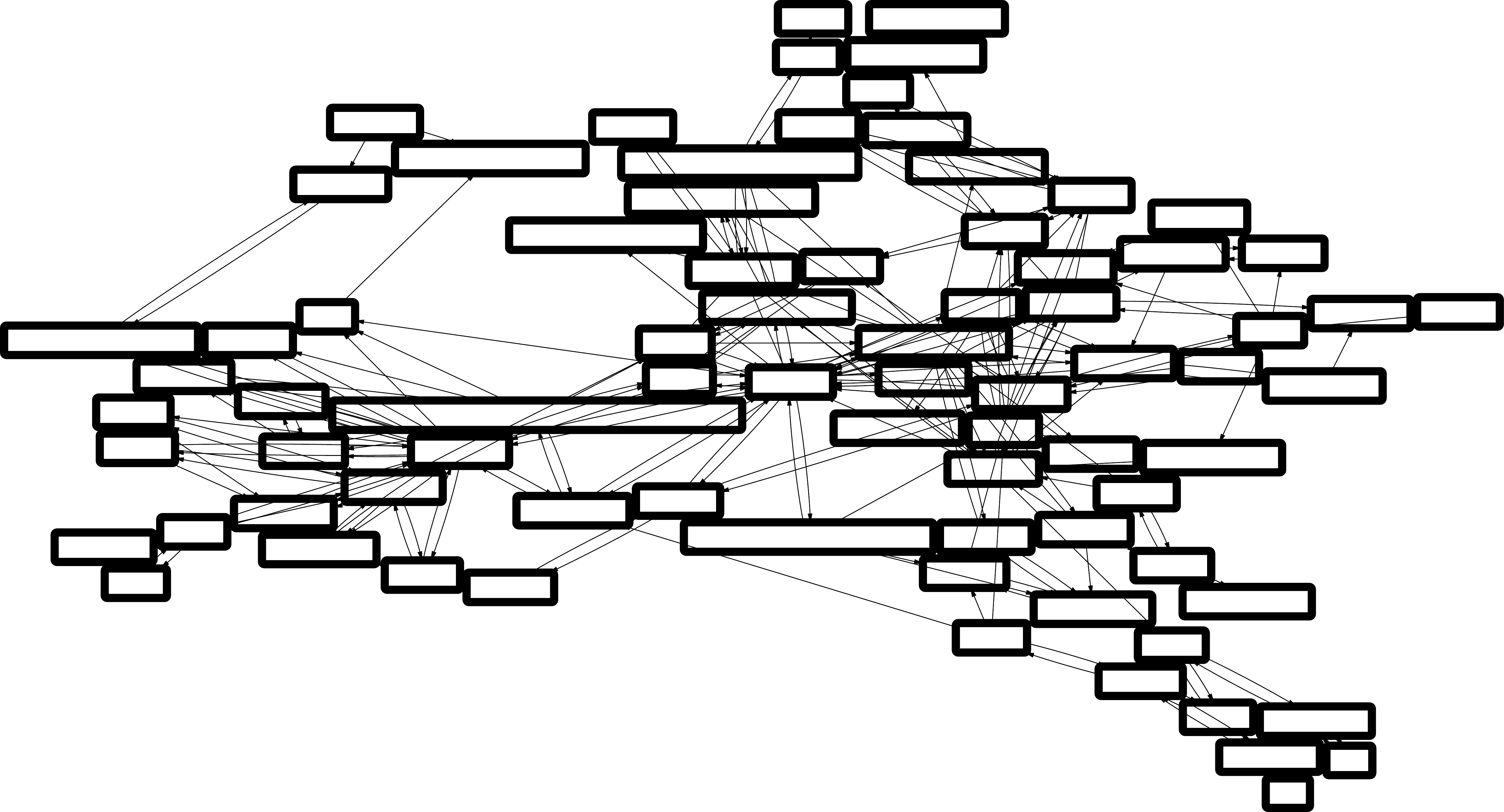}&
\includegraphics[width=\mylength,height=\mylength,keepaspectratio]{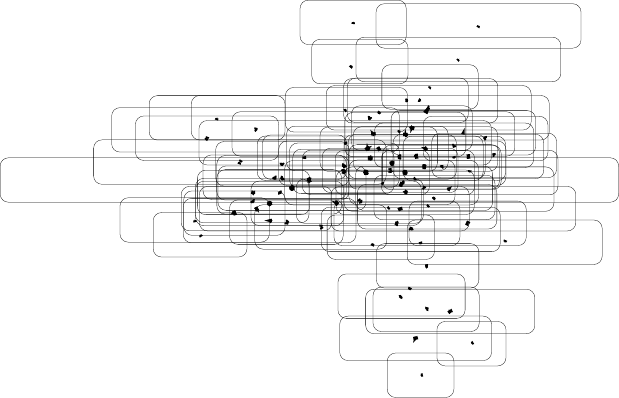}&
\includegraphics[width=\mylength,height=\mylength,keepaspectratio]{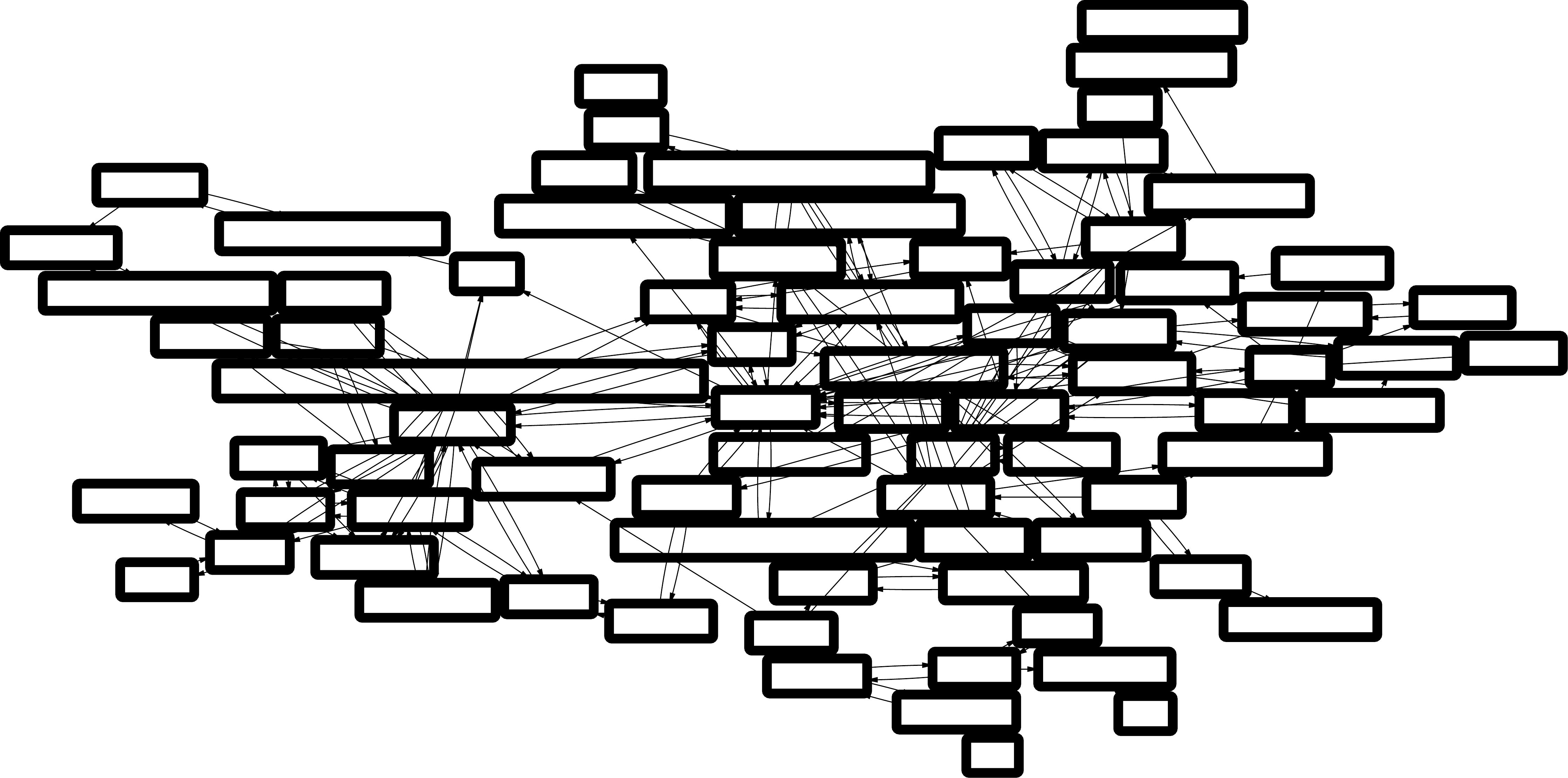}\\%
\includegraphics[width=\mylength,height=\mylength,keepaspectratio]{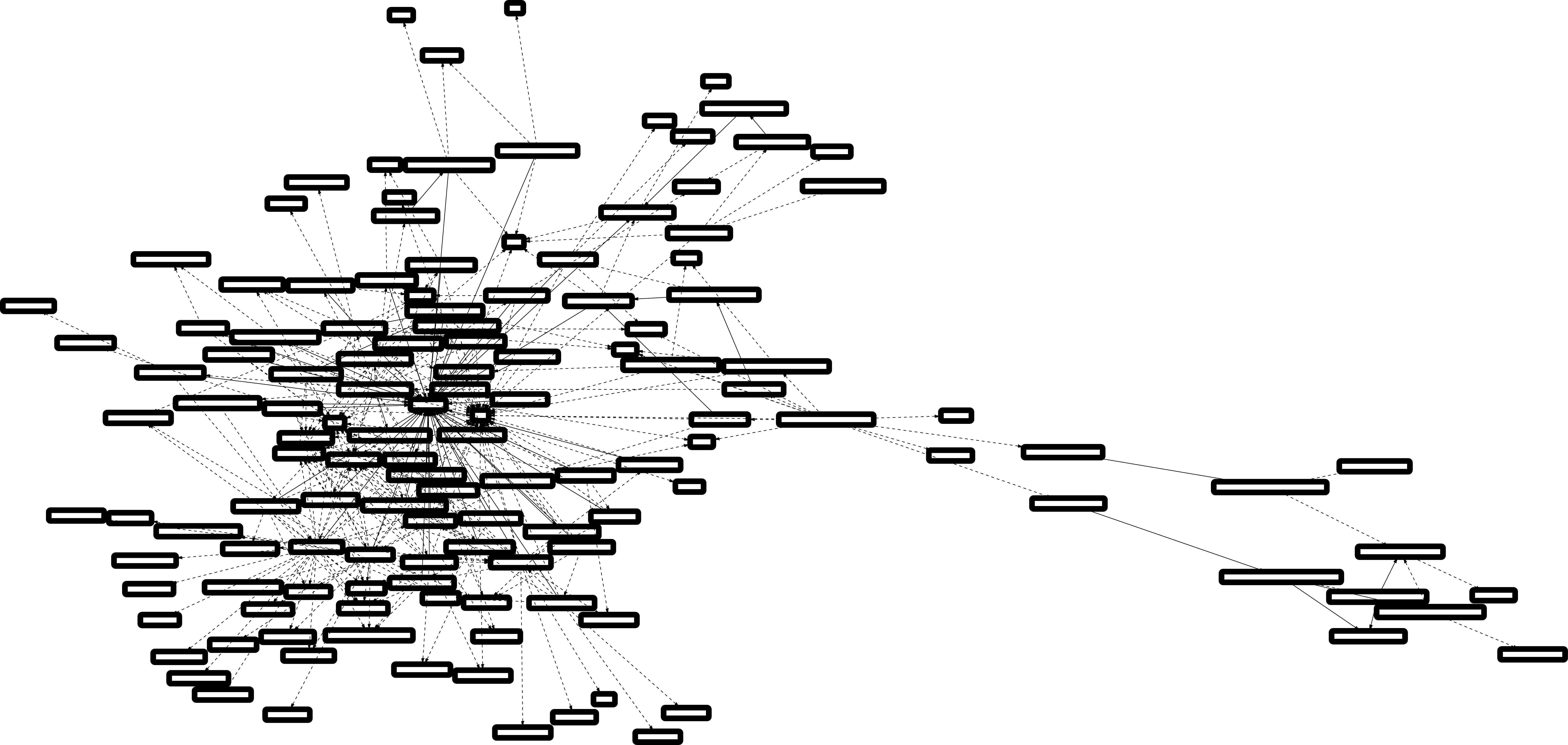}&
\includegraphics[width=\mylength,height=\mylength,keepaspectratio]{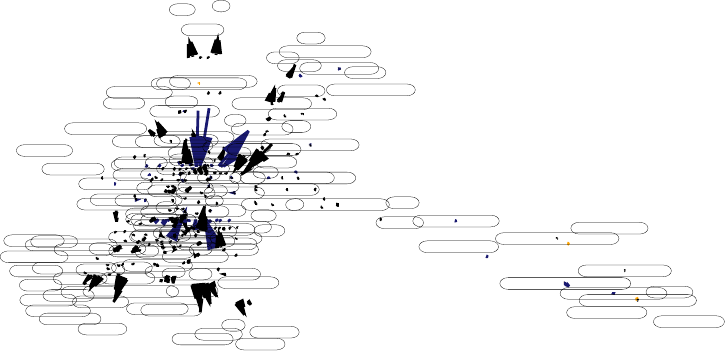}&
\includegraphics[width=\mylength,height=\mylength,keepaspectratio]{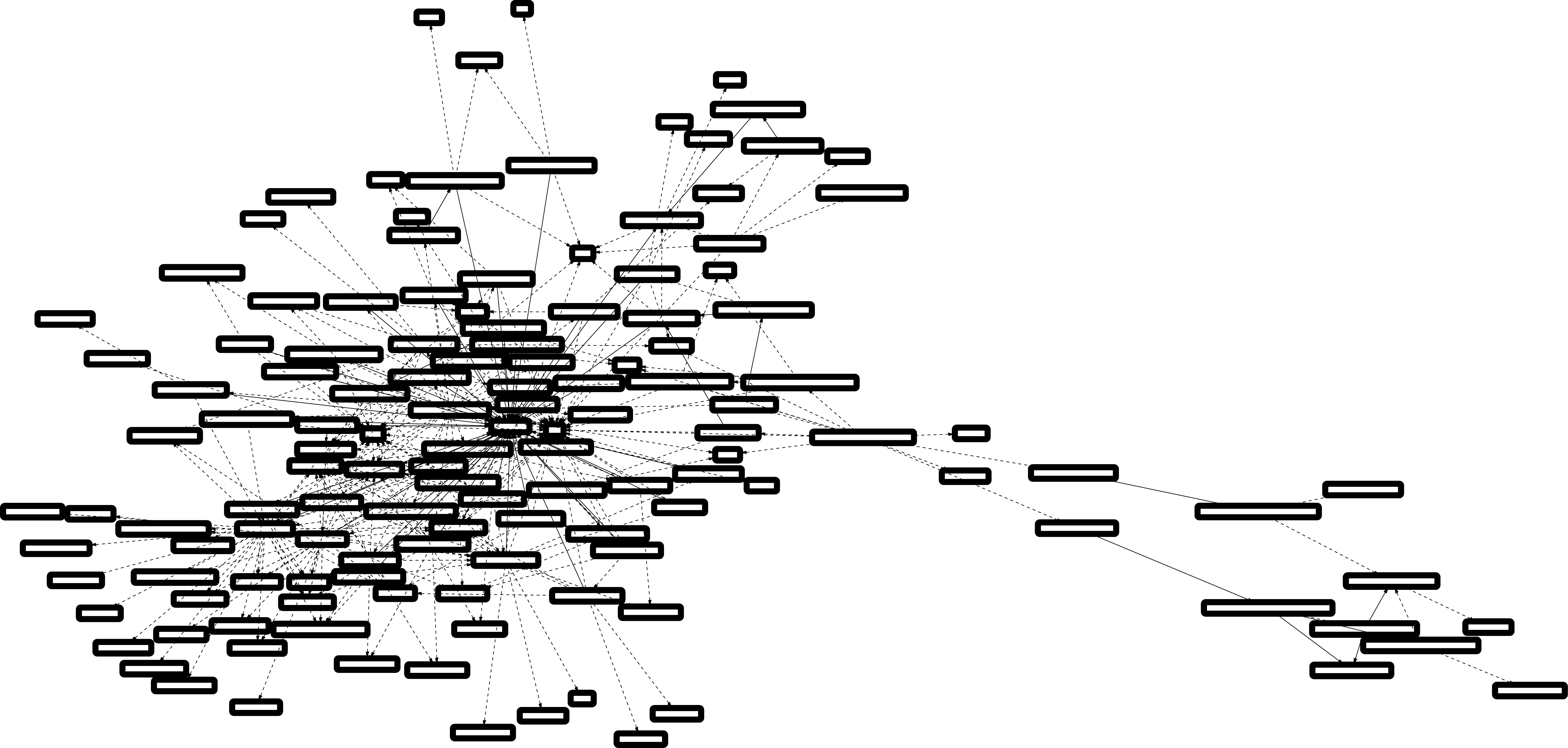}\\%
\includegraphics[width=\mylength,height=\mylength,keepaspectratio]{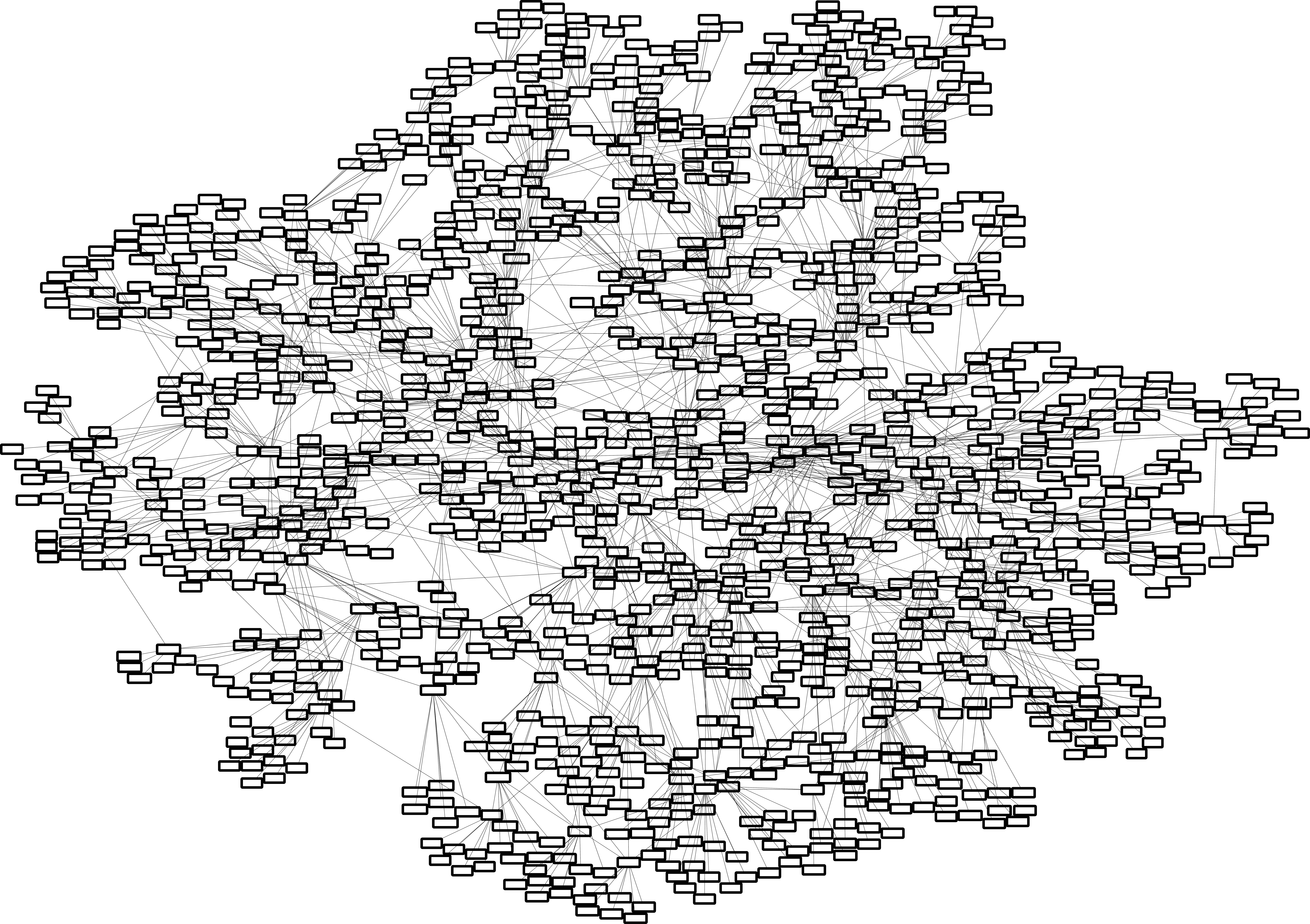}&
\includegraphics[width=\mylength,height=\mylength,keepaspectratio]{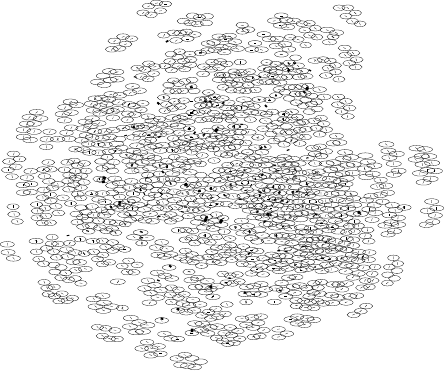}&
\includegraphics[width=\mylength,height=\mylength,keepaspectratio]{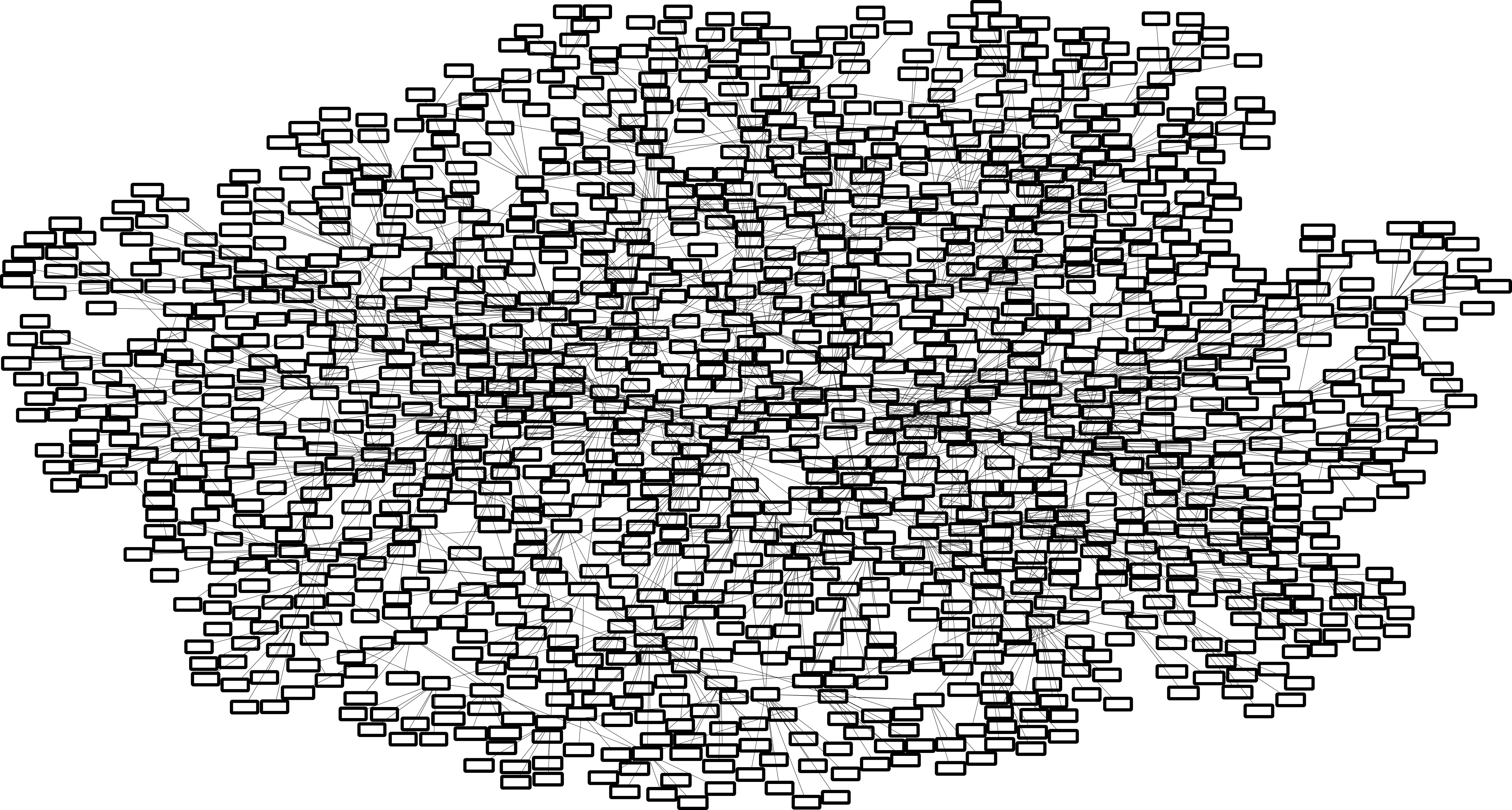}\\%
\includegraphics[width=\mylength,height=\mylength,keepaspectratio]{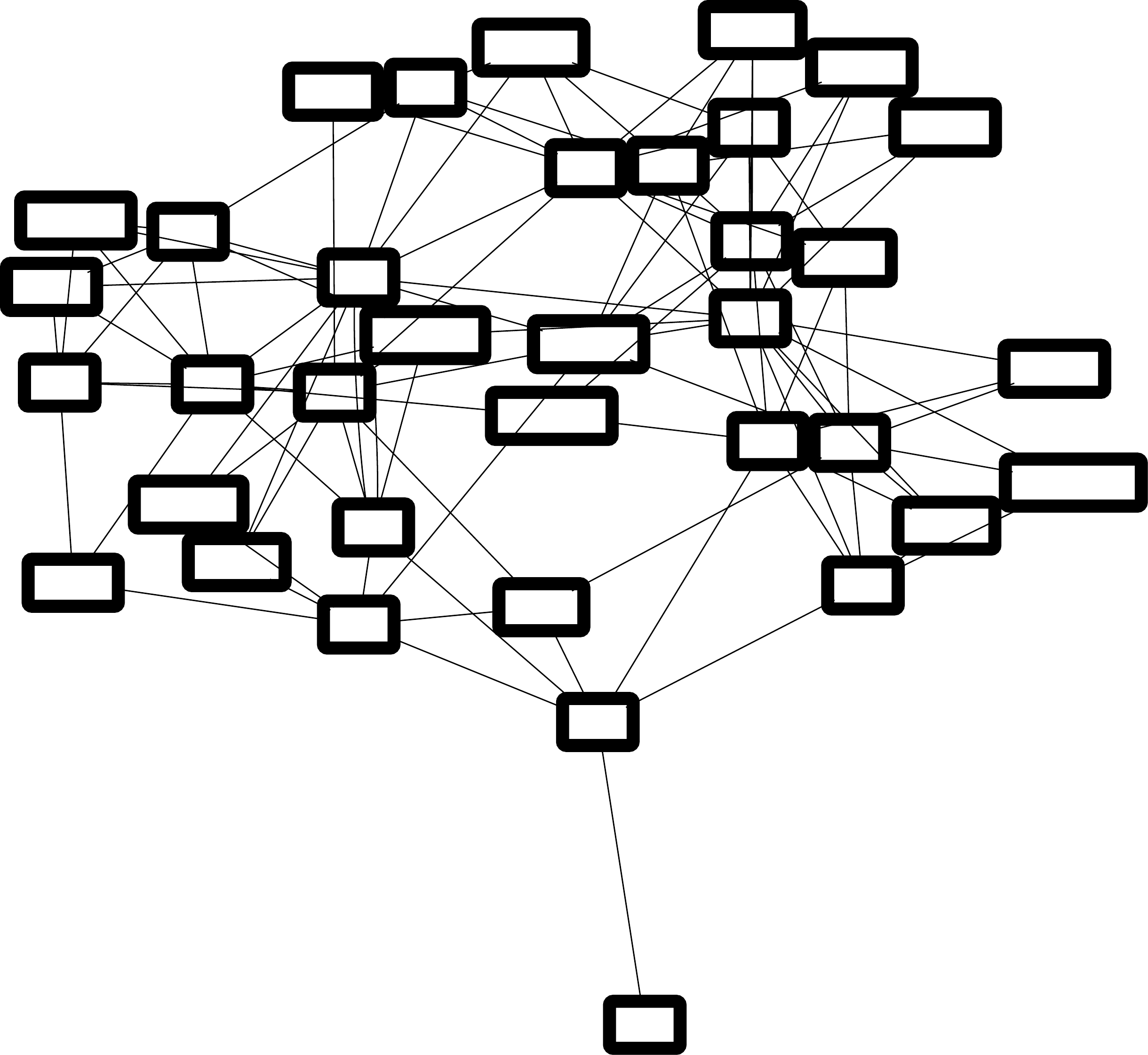}&
\includegraphics[width=\mylength,height=\mylength,keepaspectratio]{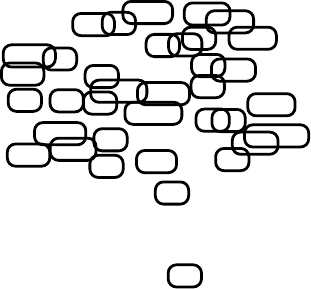}&
\includegraphics[width=\mylength,height=\mylength,keepaspectratio]{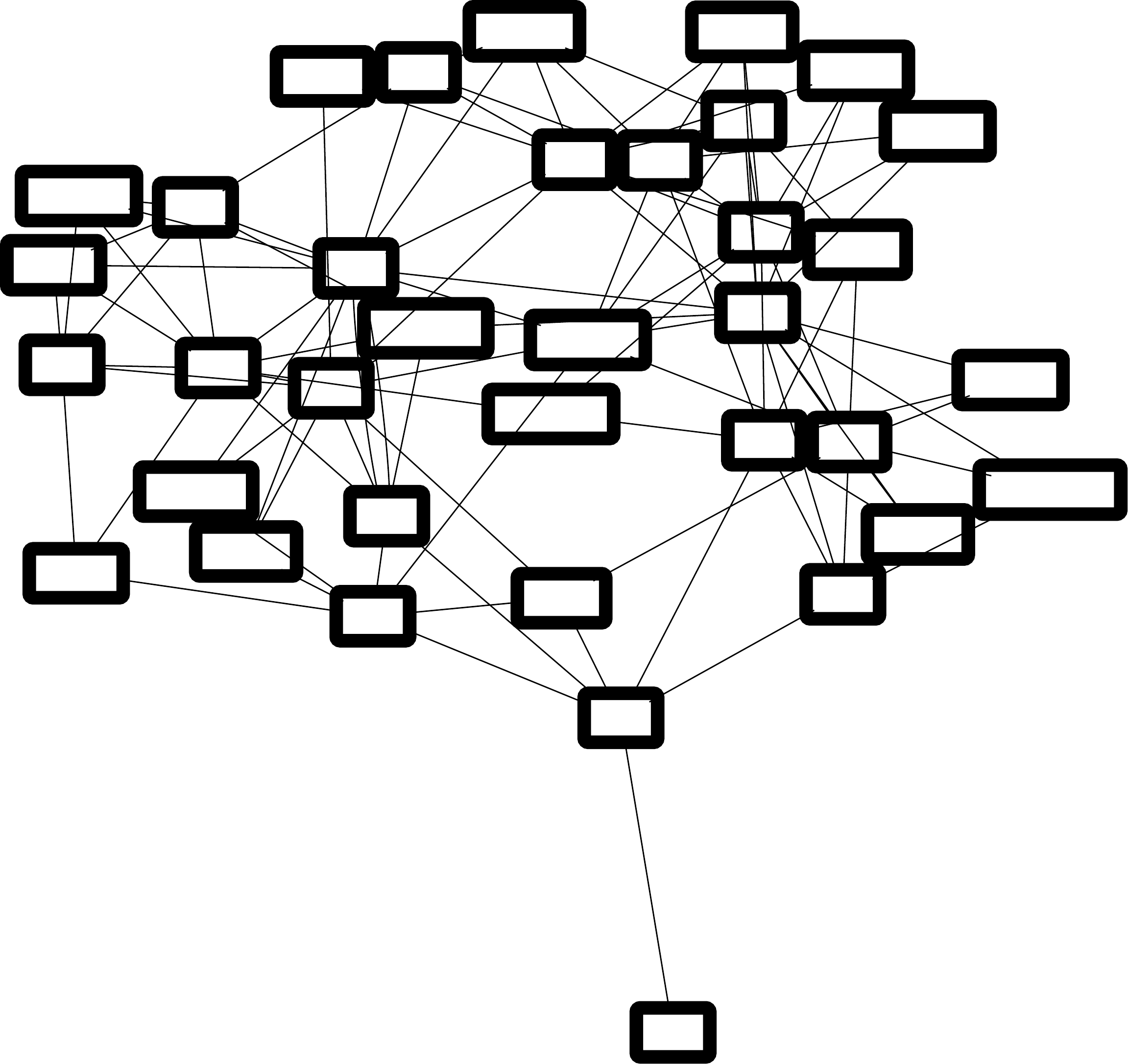}\\%
\includegraphics[width=\mylength,height=\mylength,keepaspectratio]{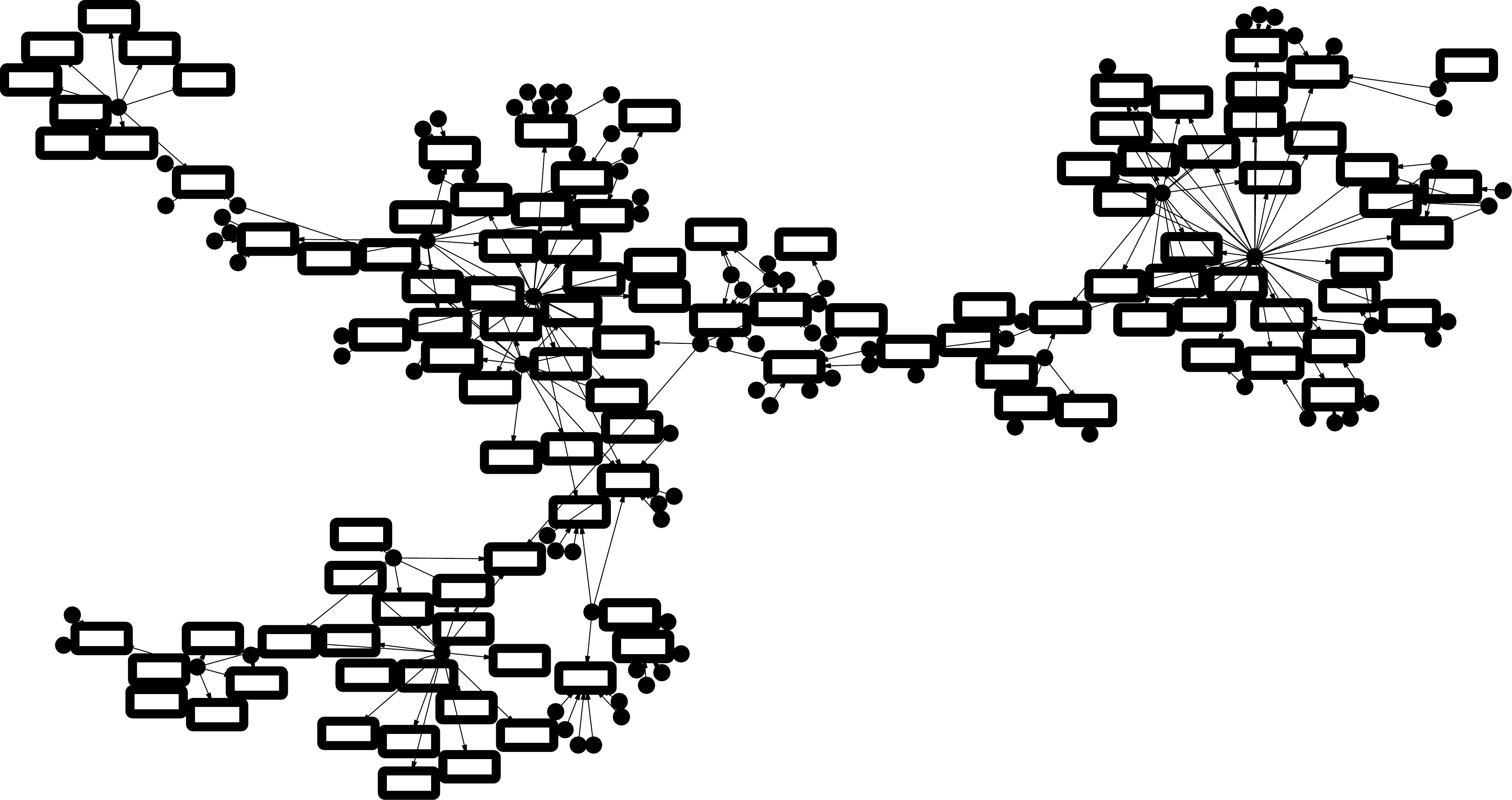}&
\includegraphics[width=\mylength,height=\mylength,keepaspectratio]{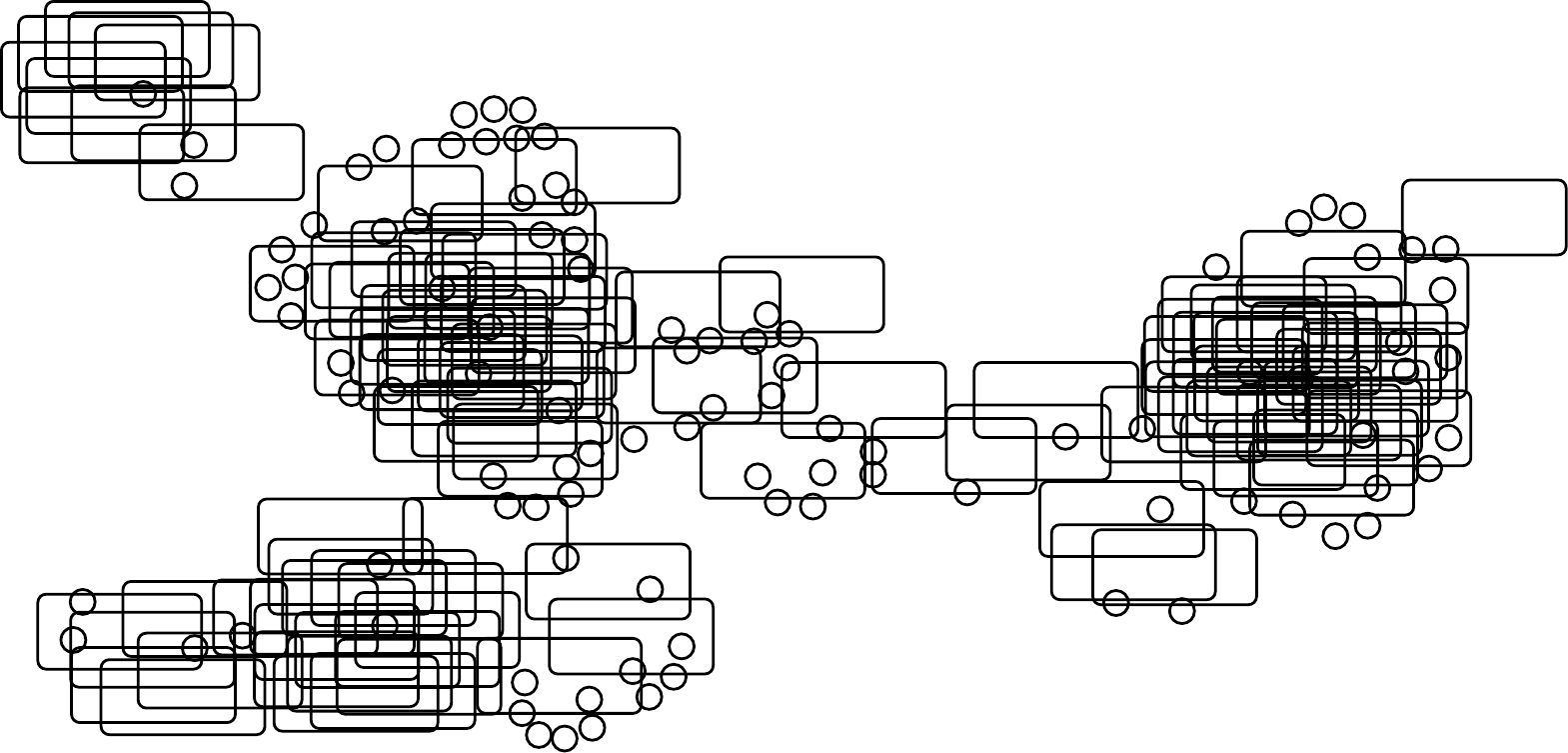}&
\includegraphics[width=\mylength,height=\mylength,keepaspectratio]{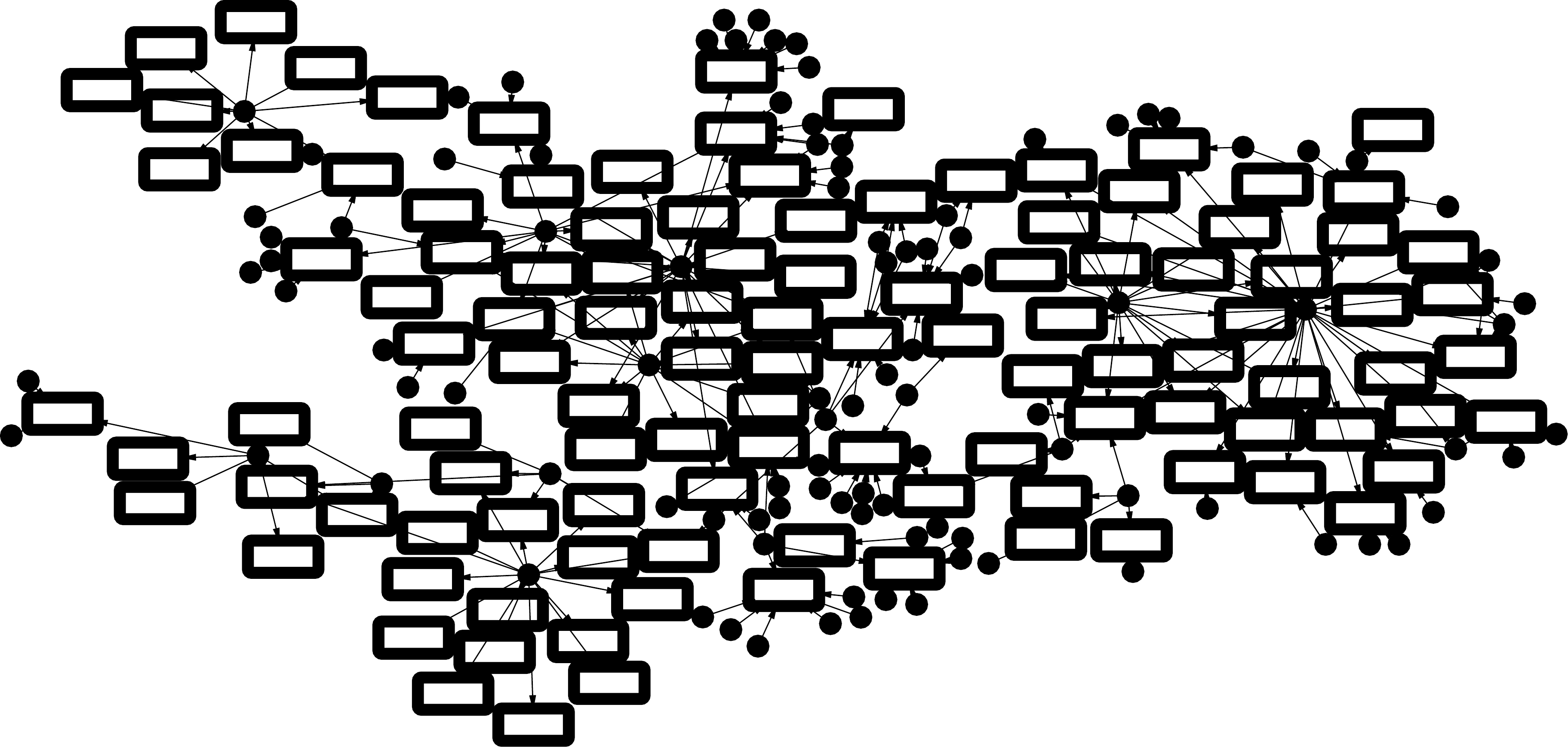}\\%
GTree & original & PRISM\\
\end{tabular}%
\begin{tabular}{CCC}
\includegraphics[width=\mylength,height=\mylength,angle=-90,keepaspectratio]{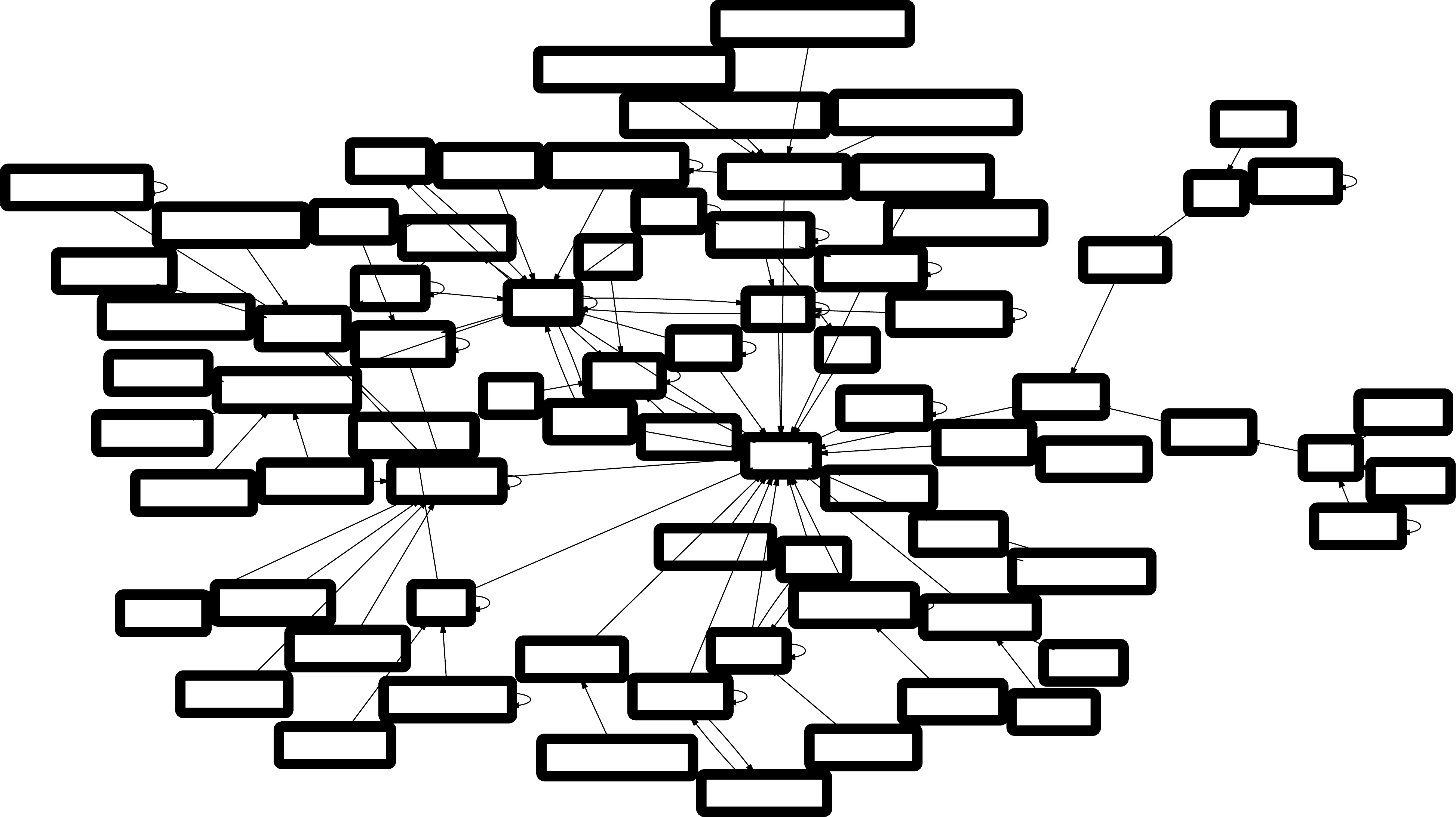}&  
\includegraphics[width=\mylength,height=\mylength,angle=-90,keepaspectratio]{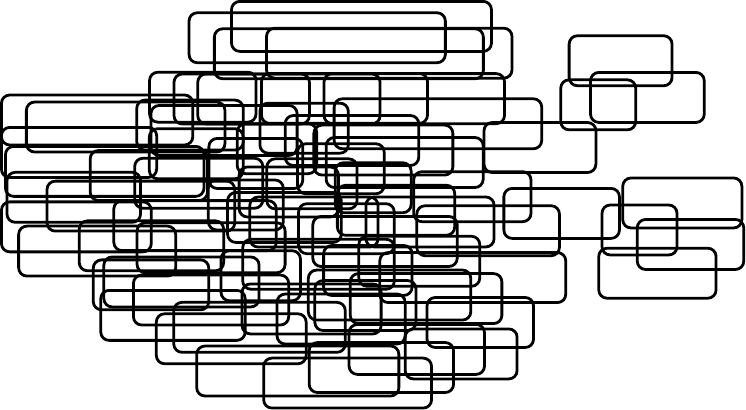}&  
\includegraphics[width=\mylength,height=\mylength,angle=-90,keepaspectratio]{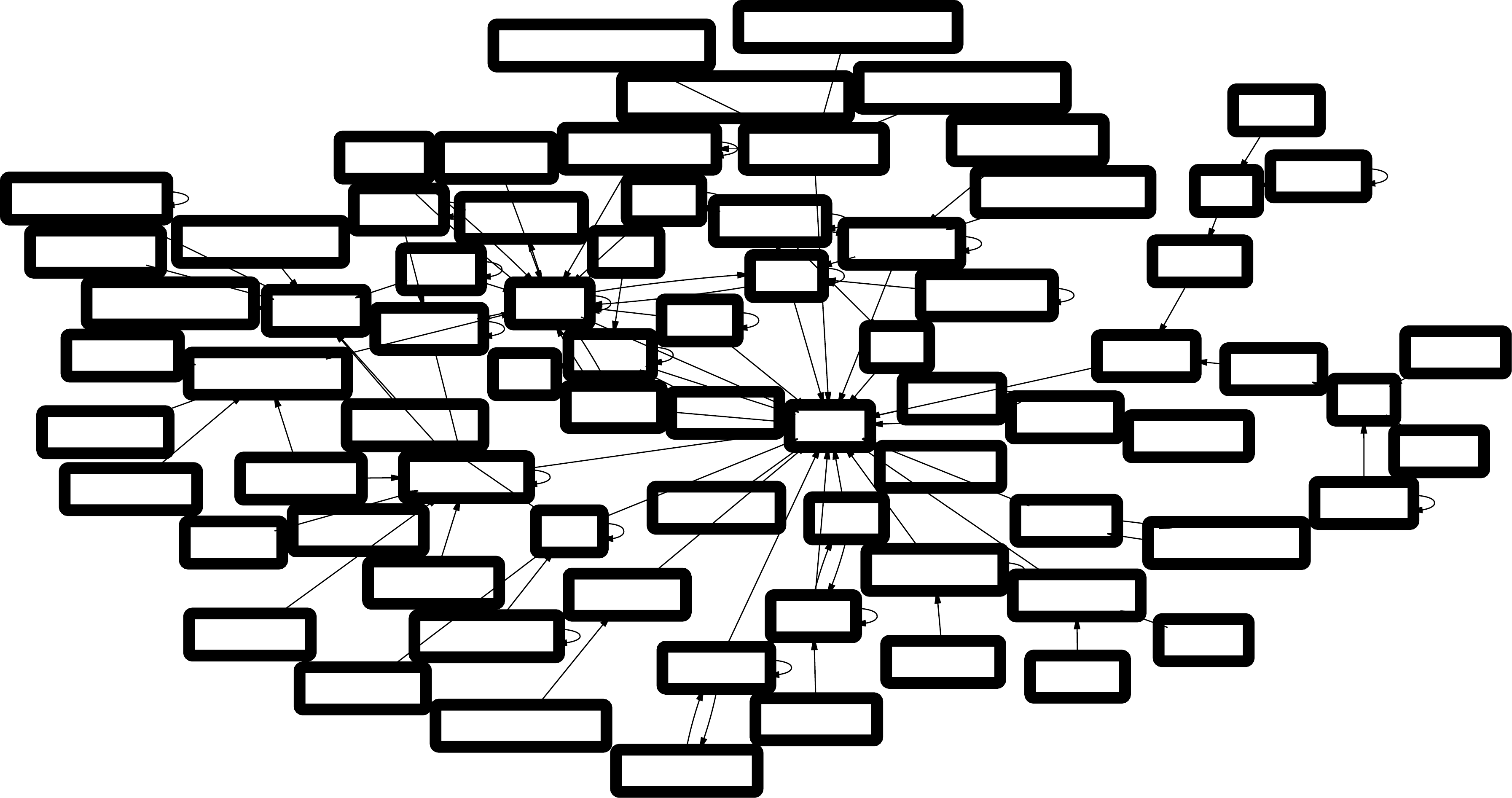}\\%
\includegraphics[width=\mylength,height=\mylength,keepaspectratio]{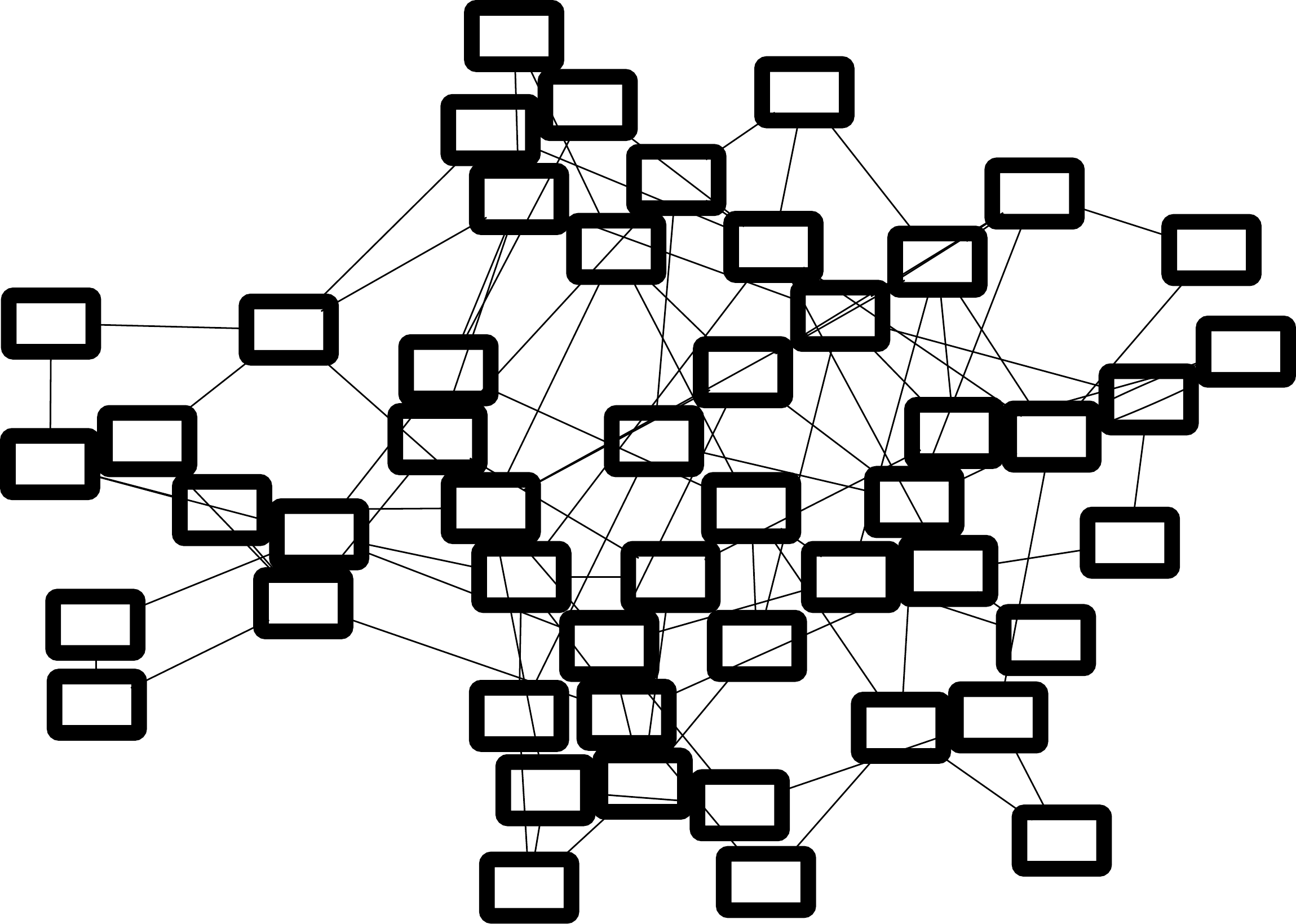}&
\includegraphics[width=\mylength,height=\mylength,keepaspectratio]{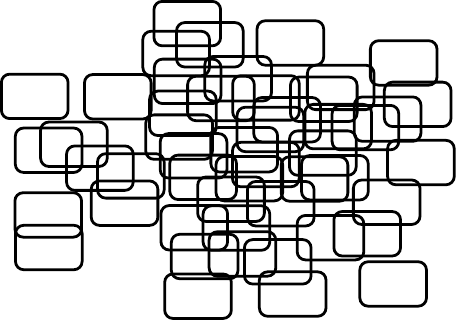}&
\includegraphics[width=\mylength,height=\mylength,keepaspectratio]{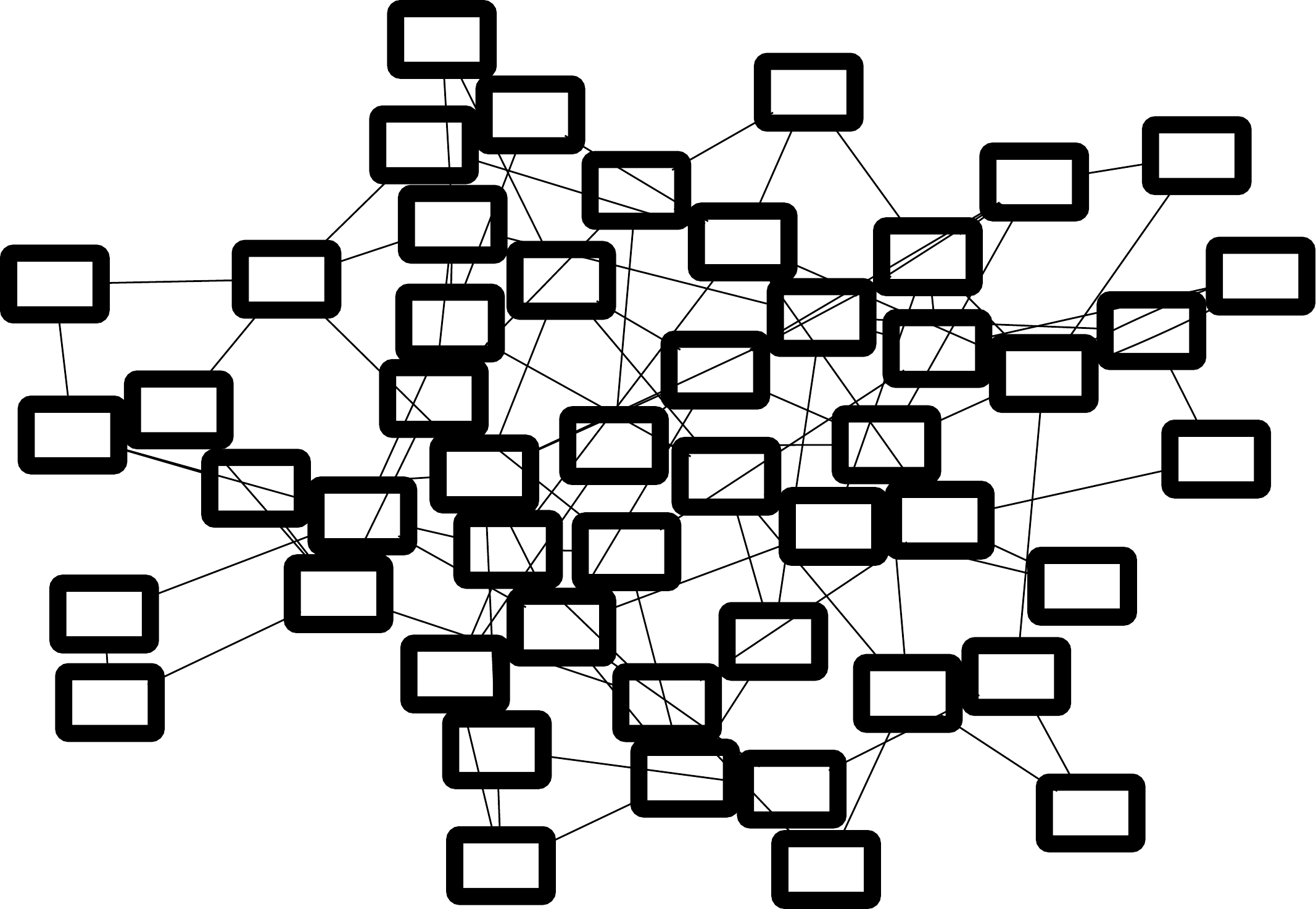}\\% 
\includegraphics[width=\mylength,height=\mylength,keepaspectratio]{root_gv-sfdpInit_dot-gtreeOverlap.pdf}&
\includegraphics[width=\mylength,height=\mylength,keepaspectratio]{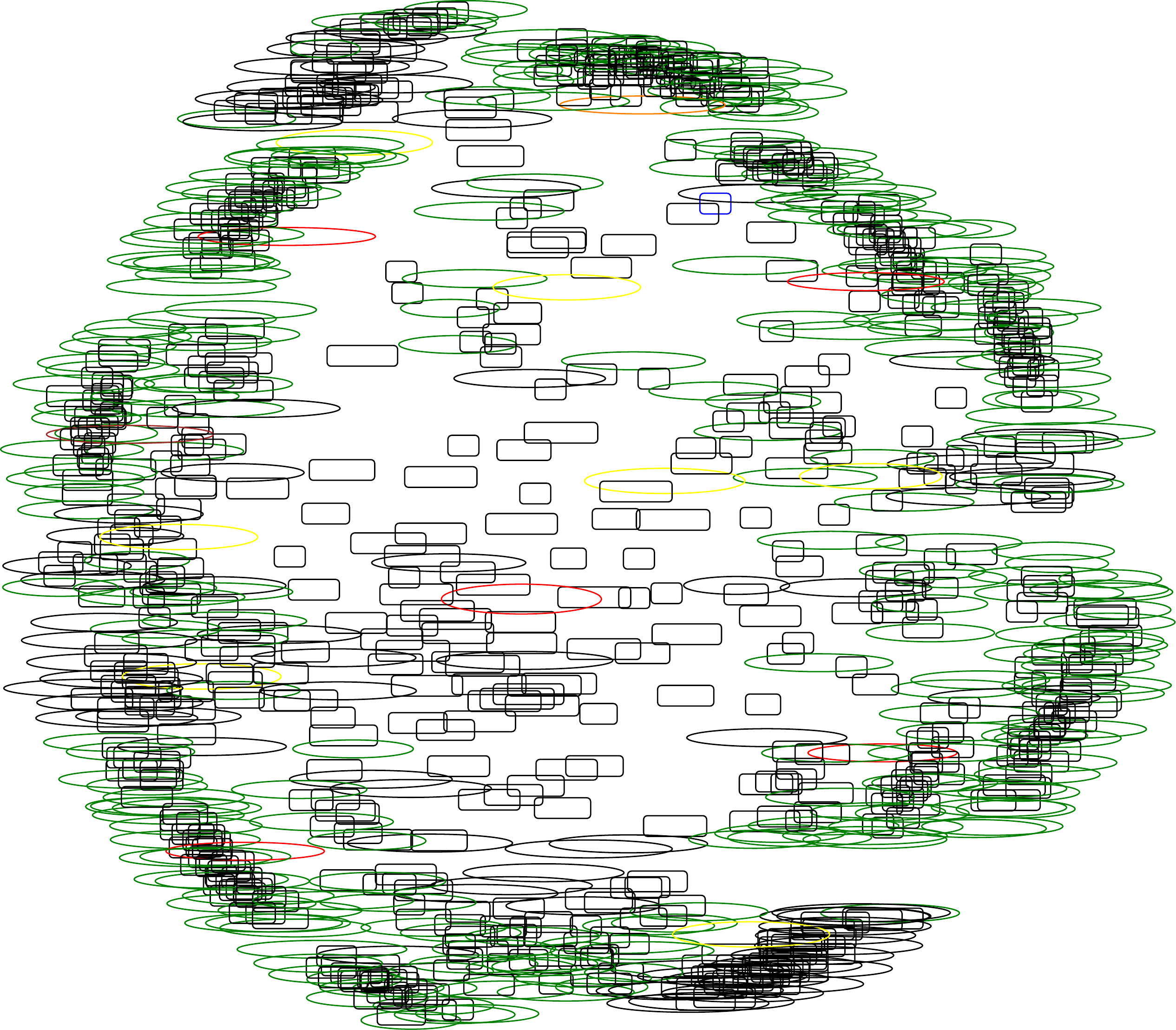}&
\includegraphics[width=\mylength,height=\mylength,keepaspectratio]{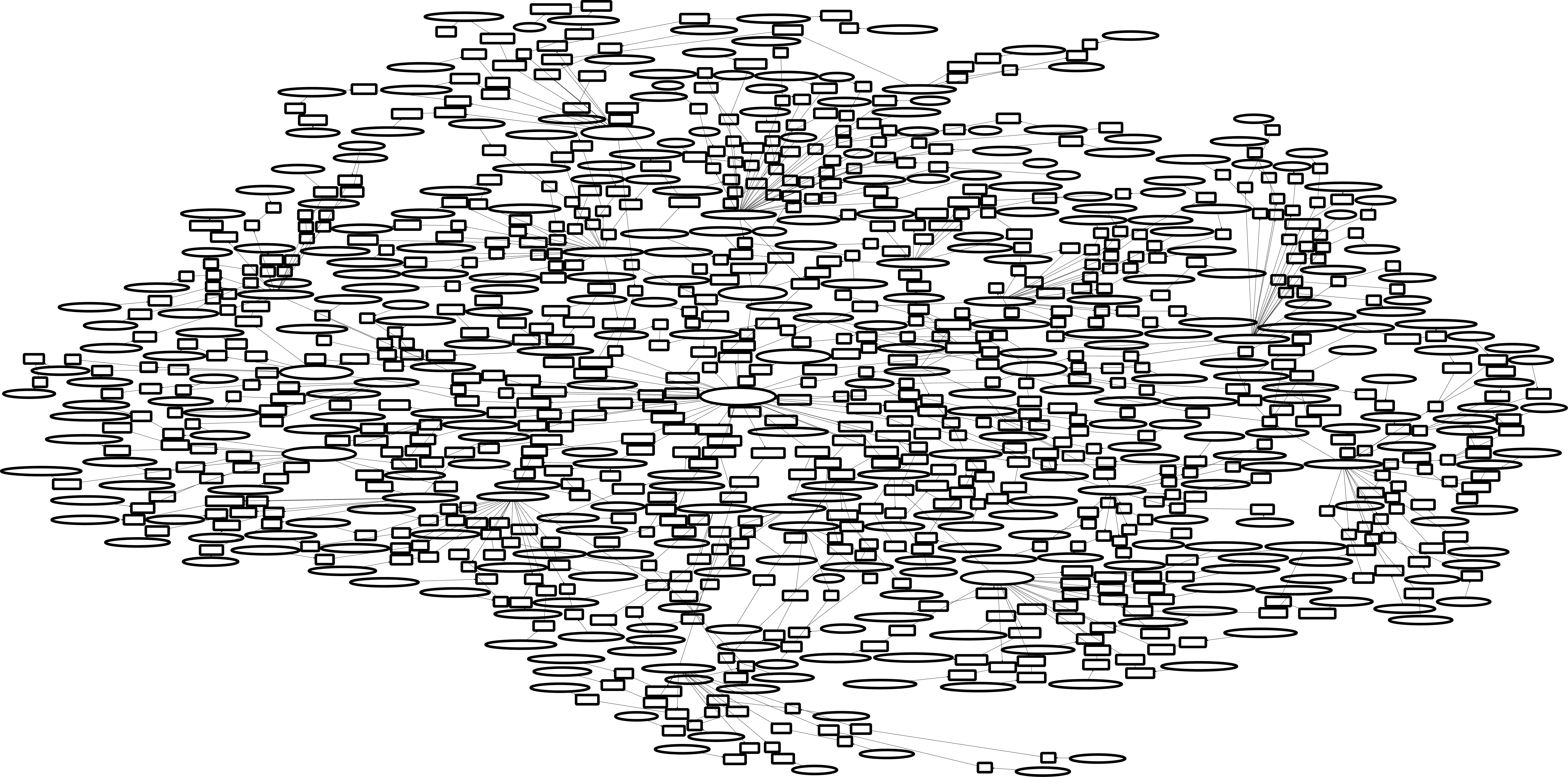}\\% 
\includegraphics[width=\mylength,height=\mylength,keepaspectratio]{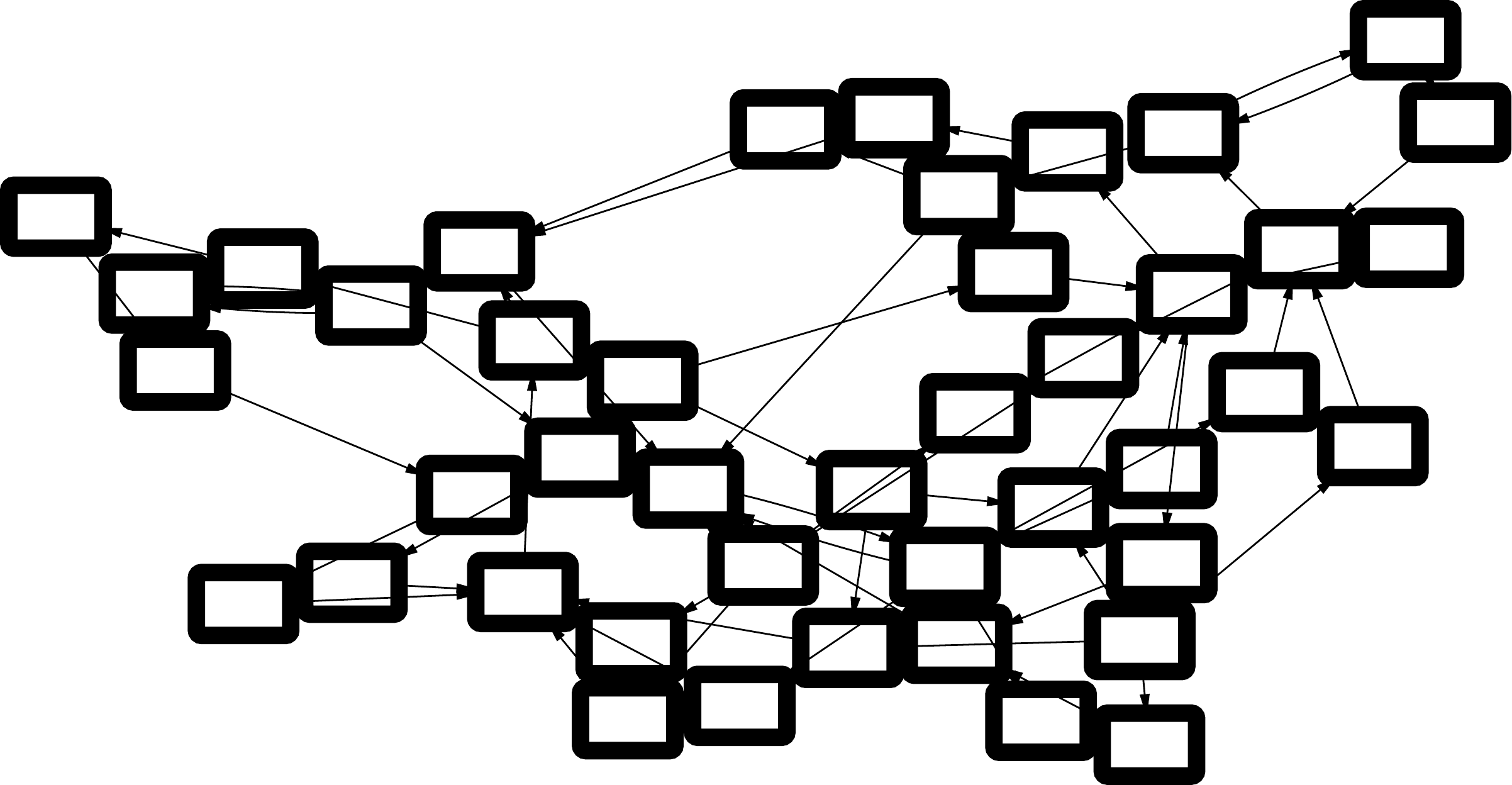}&
\includegraphics[width=\mylength,height=\mylength,keepaspectratio]{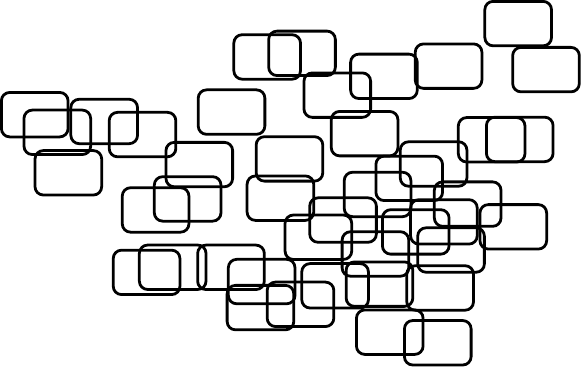}&
\includegraphics[width=\mylength,height=\mylength,keepaspectratio]{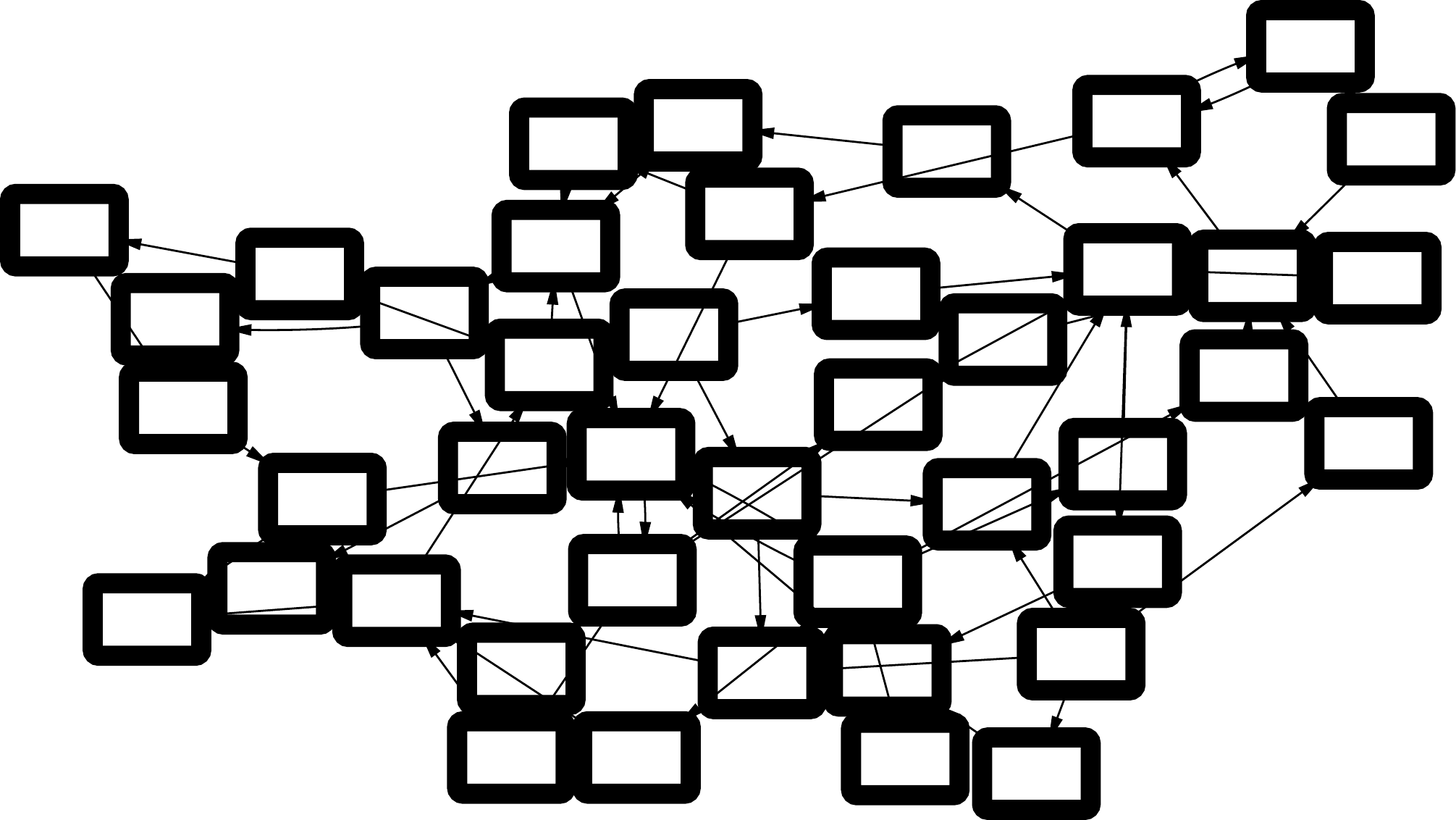}\\% 
\includegraphics[width=\mylength,height=\mylength,keepaspectratio]{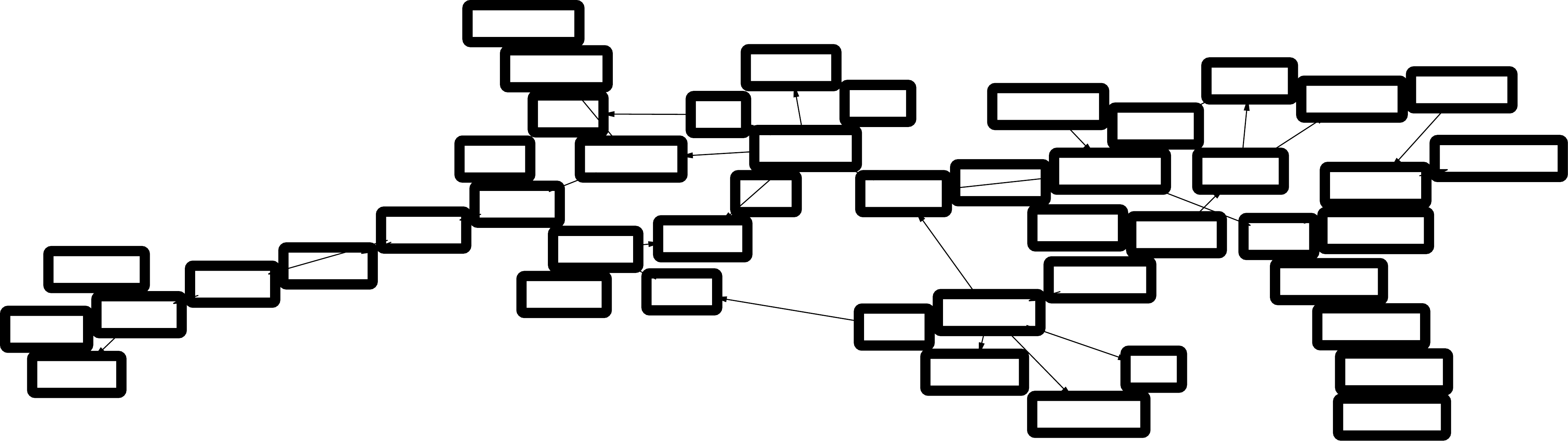}&
\includegraphics[width=\mylength,height=\mylength,keepaspectratio]{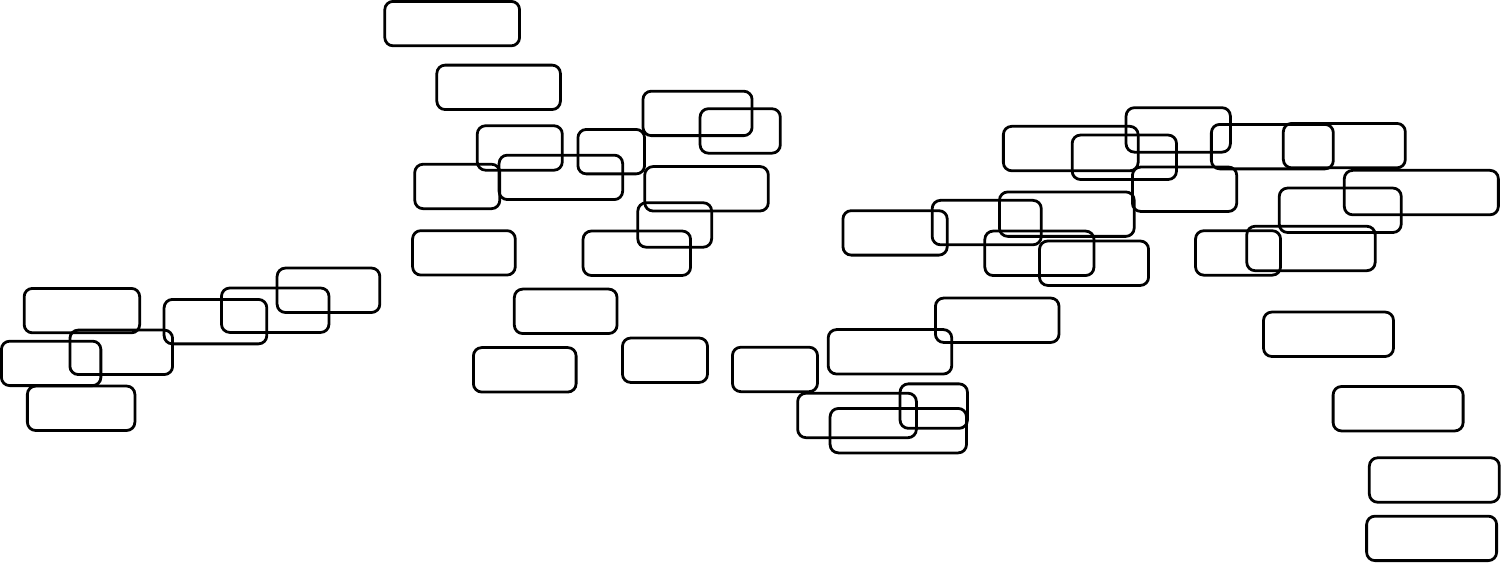}&
\includegraphics[width=\mylength,height=\mylength,keepaspectratio]{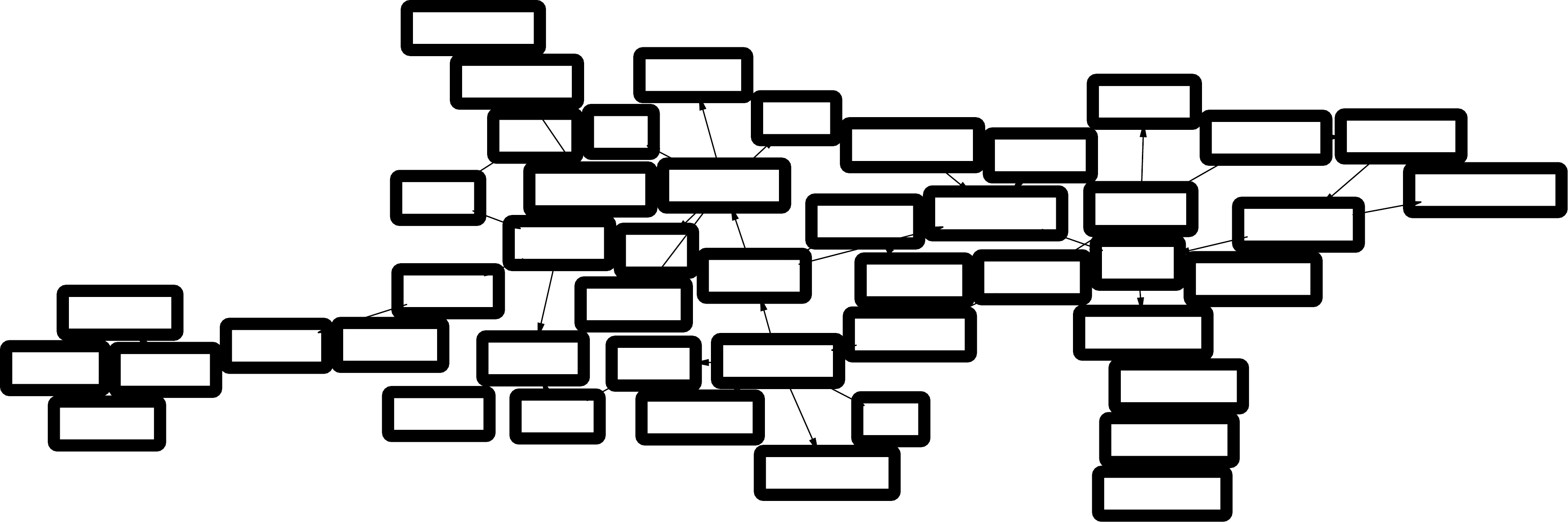}\\% 
\includegraphics[width=\mylength,height=\mylength,keepaspectratio]{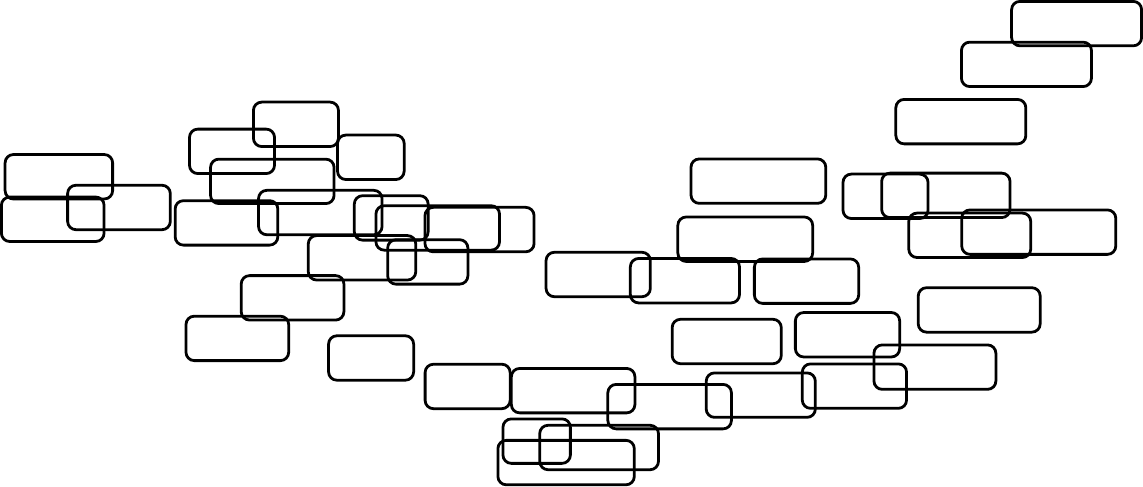}&
\includegraphics[width=\mylength,height=\mylength,keepaspectratio]{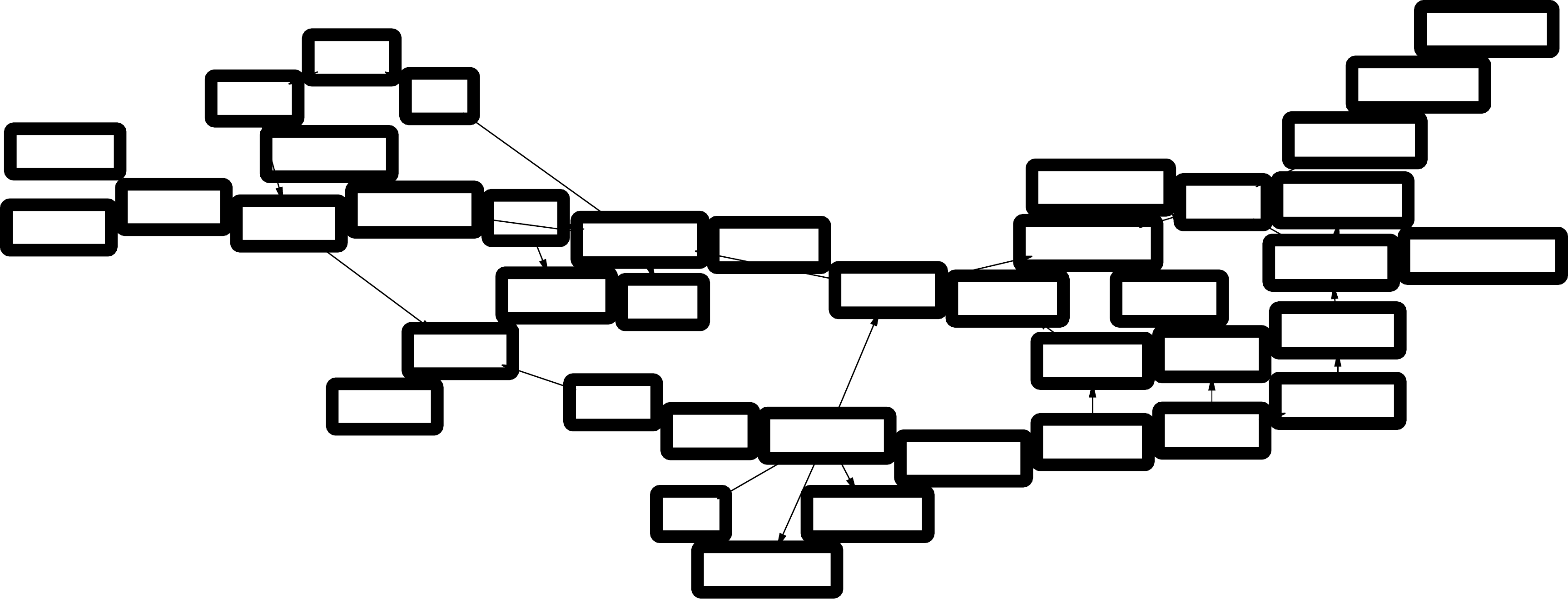}&
\includegraphics[width=\mylength,height=\mylength,keepaspectratio]{unix_gv-sfdpInit.pdf}\\% 
\includegraphics[width=\mylength,height=\mylength,keepaspectratio]{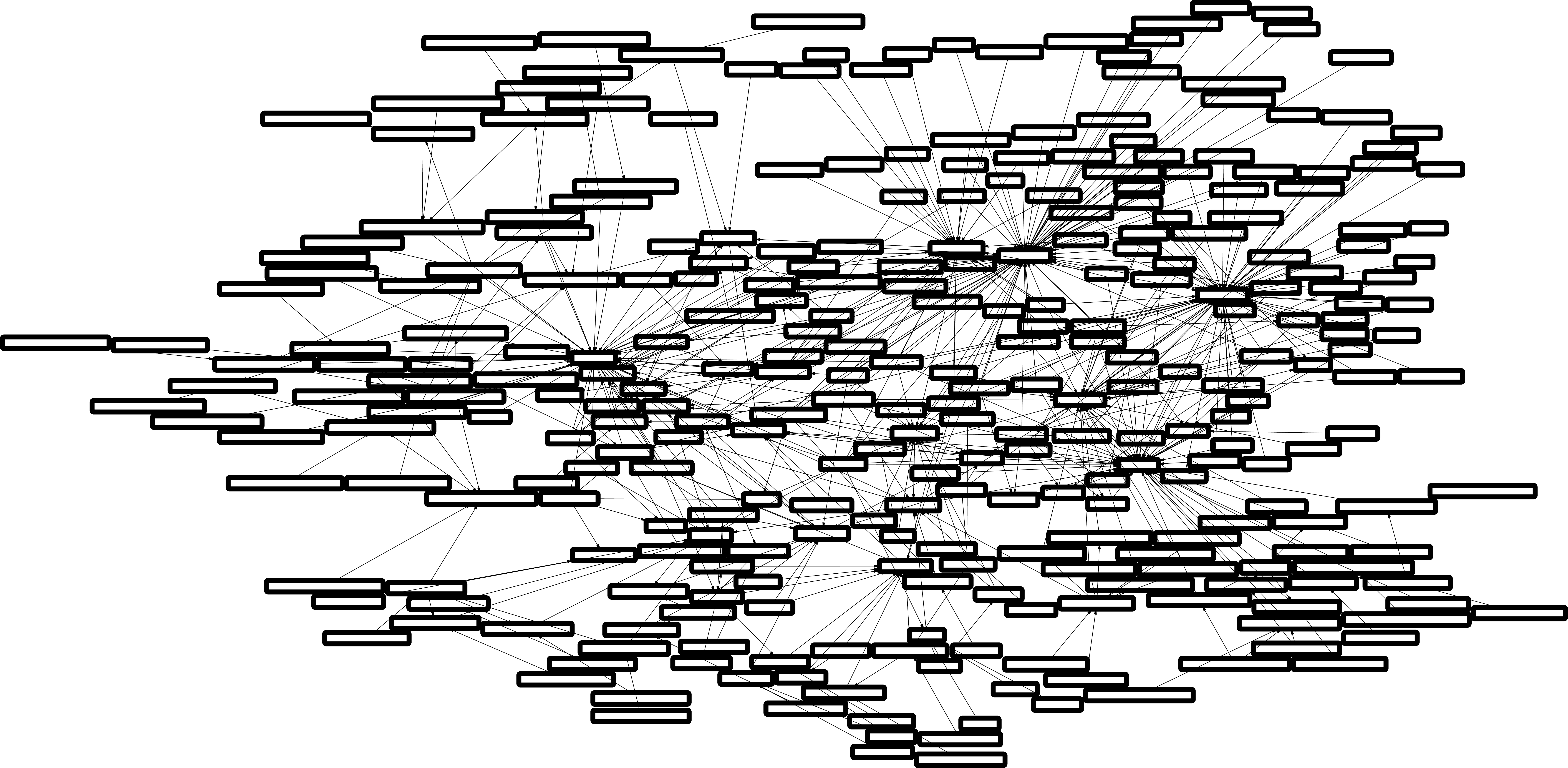}& 
\includegraphics[width=\mylength,height=\mylength,keepaspectratio]{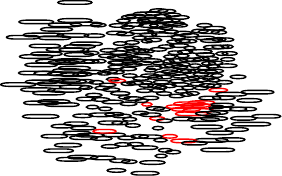}& 
\includegraphics[width=\mylength,height=\mylength,keepaspectratio]{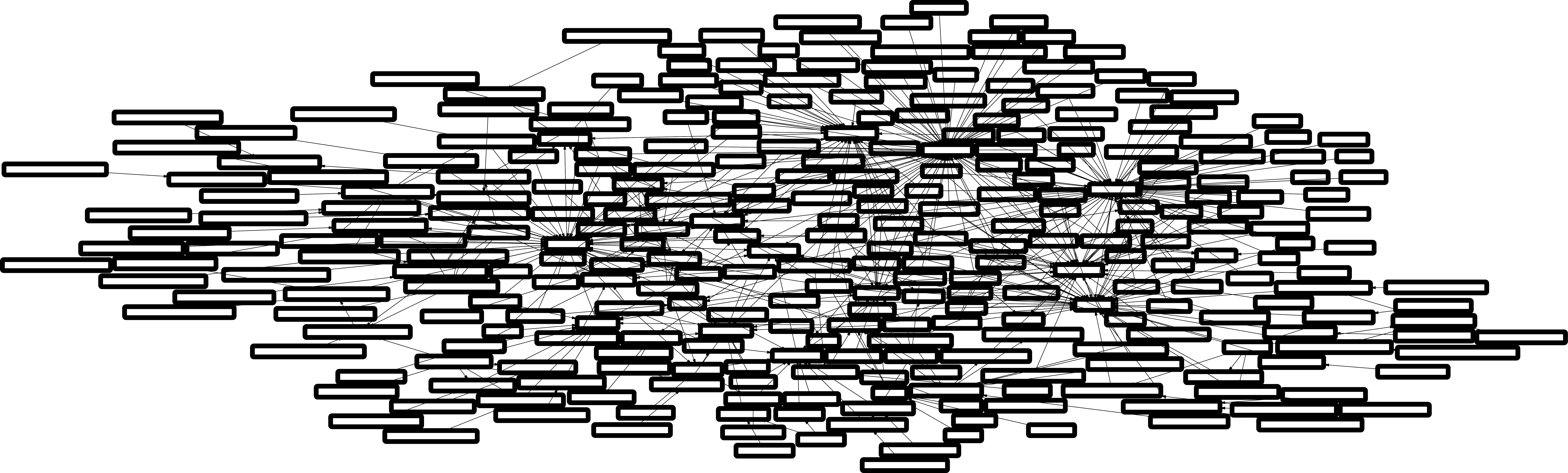}\\
GTree & original  &PRISM\\
\end{tabular} }
\caption{Results for GTree and PRISM initialized with SFDP. From top
  to bottom and left to right: b100, b102, b124, b143, badvoro, dpd,
  mode, - NaN, ngk10\_4, root, rowe, size, unix, and xx. To make the original drawings more readable they have been changed; In most cases the nodes were diminished and the edges removed. The drawings were scaled differently.}
\label{fig:pdfs}
\end{figure}

In Figure~\ref{fig:root} we experiment with the way we expand the
edges. Instead of the formula $p'_j = p'_i + t_{ij}(p_j-p_i)$, which
resolves the overlap between the nodes $i$ and $j$ immediately, we use
the update $p'_j = p'_i + \min(t_{ij},1.5)(p_j-p_i)$. As
a result, the algorithm runs a little bit slower but produces layouts
with smaller area. 

\section{Conclusion \& Future Work}
We proposed a new overlap removal algorithm that uses the minimum spanning tree. The algorithm is simple and easy to implement, and yet it preserves the initial layout well and is efficient. 

Although we introduced our approach in the context of graph visualization, our method can also be used for any other purpose where overlap needs to be resolved while maintaining the initial layout.
% As a byproduct we noticed that the used stress majorization solver, can make a huge difference in terms of runtime. This suggests that experimental studies comparing the various solvers, in the context of stress majorization for graph drawing, are necessary to develop more general guidelines. We are not aware of any such guidelines for stress majorization.
Finding a measure of how well an overlap removal algorithm preserves clusters of the initial layout seems to be an interesting challenge.

\bibliographystyle{abbrv} 
\bibliography{myReferences}

\end{document}